\documentclass[11pt,draftcls,onecolumn]{IEEEtran}
\usepackage{cite}

\usepackage{array,multirow}
\usepackage{graphicx}  
\usepackage{subfigure} 
\usepackage{amsmath}   
\usepackage{amssymb}   
\usepackage{eucal}     

\usepackage{pxfonts}
\usepackage{dsfont}

\newtheorem{theorem}{\textbf{Theorem}}
\newtheorem{corollary}{\textbf{Corollary}}

\newtheorem{definition}{\textbf{Definition}}
\newtheorem{lemma}{Lemma}
\newtheorem{remark}{Remark}
\newcommand{\dv}{\mathbf} 
\newcommand{\mc}{\mathcal} 
\newcommand{\qed}{\hfill \ensuremath{\Box}}

\DeclareMathAlphabet{\eurm}{U}{eur}{m}{n}
\DeclareMathAlphabet{\mathbsf}{OT1}{cmss}{bx}{n}
\DeclareMathAlphabet{\mathssf}{OT1}{cmss}{m}{sl}
\DeclareMathAlphabet{\mathcsf}{OT1}{cmss}{sbc}{n}

\DeclareSymbolFont{bsfletters}{OT1}{cmss}{bx}{n}  
\DeclareSymbolFont{ssfletters}{OT1}{cmss}{m}{n}
\DeclareMathSymbol{\bsfGamma}{0}{bsfletters}{'000}
\DeclareMathSymbol{\ssfGamma}{0}{ssfletters}{'000}
\DeclareMathSymbol{\bsfDelta}{0}{bsfletters}{'001}
\DeclareMathSymbol{\ssfDelta}{0}{ssfletters}{'001}
\DeclareMathSymbol{\bsfTheta}{0}{bsfletters}{'002}
\DeclareMathSymbol{\ssfTheta}{0}{ssfletters}{'002}
\DeclareMathSymbol{\bsfLambda}{0}{bsfletters}{'003}
\DeclareMathSymbol{\ssfLambda}{0}{ssfletters}{'003}
\DeclareMathSymbol{\bsfXi}{0}{bsfletters}{'004}
\DeclareMathSymbol{\ssfXi}{0}{ssfletters}{'004}
\DeclareMathSymbol{\bsfPi}{0}{bsfletters}{'005}
\DeclareMathSymbol{\ssfPi}{0}{ssfletters}{'005}
\DeclareMathSymbol{\bsfSigma}{0}{bsfletters}{'006}
\DeclareMathSymbol{\ssfSigma}{0}{ssfletters}{'006}
\DeclareMathSymbol{\bsfUpsilon}{0}{bsfletters}{'007}
\DeclareMathSymbol{\ssfUpsilon}{0}{ssfletters}{'007}
\DeclareMathSymbol{\bsfPhi}{0}{bsfletters}{'010}
\DeclareMathSymbol{\ssfPhi}{0}{ssfletters}{'010}
\DeclareMathSymbol{\bsfPsi}{0}{bsfletters}{'011}
\DeclareMathSymbol{\ssfPsi}{0}{ssfletters}{'011}
\DeclareMathSymbol{\bsfOmega}{0}{bsfletters}{'012}
\DeclareMathSymbol{\ssfOmega}{0}{ssfletters}{'012}

\newcommand{\calS}{{\mathcal{S}}}
\newcommand{\calU}{{\mathcal{U}}}
\newcommand{\calV}{{\mathcal{V}}}
\newcommand{\calX}{{\mathcal{X}}}
\newcommand{\calY}{{\mathcal{Y}}}

\begin{document}
{\fontencoding{OT1}\fontsize{9.4}{11.25pt}\selectfont
\title{Bounds on the Capacity of the Relay Channel with Noncausal State at Source\\}

\author{\vspace{0cm}
\authorblockN{ \small Abdellatif Zaidi \qquad Shlomo Shamai (Shitz) \qquad Pablo Piantanida \qquad Luc Vandendorpe\thanks{Abdellatif Zaidi is with Universit\'e Paris-Est Marne La Vall\'ee, 77454 Marne la Vall\'ee Cedex 2, France. Email: abdellatif.zaidi@univ-mlv.fr}
\thanks{Shlomo Shamai is with the Department of Electrical Engineering, Technion Institute of Technology, Technion City, Haifa 32000, Israel. Email: sshlomo@ee.technion.ac.il}
\thanks{Pablo Piantanida is with the Department of Telecommunications, SUPELEC, 91190 Gif-sur-Yvette, France. Email: pablo.piantanida@supelec.fr}
\thanks{Luc Vandendorpe is with \'{E}cole Polytechnique de Louvain, Universit\'e Catholique de Louvain, Louvain-la-Neuve 1348, Belgium. Email: luc.vandendorpe@uclouvain.be}
\thanks{This work has been supported by the European Commission in the framework of the FP7 Network of Excellence in Wireless Communications (NEWCOM++). The work of S. Shamai has also been supported by the CORNET consortium.}}}

\maketitle

\begin{abstract}
We consider a three-terminal state-dependent relay channel with the channel state available non-causally at only the source. Such a model may be of interest for node cooperation in the framework of cognition, i.e., collaborative signal transmission involving cognitive and non-cognitive radios. We study the capacity of this communication model. One principal problem is caused by the relay's not knowing the channel state. For the discrete memoryless (DM) model, we establish two lower bounds and an upper bound on channel capacity. The first lower bound is obtained by a coding scheme in which the source describes the state of the channel to the relay and destination, which then exploit the gained description for a better communication of the source's information message. The coding scheme for the second lower bound remedies the relay's not knowing the states of the channel by first computing, at the source, the appropriate input that the relay would send had the relay known the states of the channel, and then transmitting this appropriate input to the relay. The relay simply guesses the sent input and sends it in the next block. The upper bound is non trivial and it accounts for not knowing the state at the relay and destination. For the general Gaussian model, we derive lower bounds on the channel capacity by exploiting ideas in the spirit of those we use for the DM model; and we show that these bounds are optimal for small and large noise at the relay irrespective to the strength of the interference. Furthermore, we also consider a special case model in which the source input has two components one of which is independent of the state. We establish a better upper bound for both DM and Gaussian cases and we also characterize the capacity in a number of special cases. 
\end{abstract}


\section{Introduction}\label{secI}

We consider a three-terminal state-dependent relay channel (RC) in which, as shown in Figure \ref{StateDependentDiscreteMemorylessRelayChannel}, the source wants to communicate a message $W$ to the destination through the state-dependent RC in $n$ uses of the channel, with the help of the relay. The channel outputs $Y^n_2$ and $Y^n_3$ for  the relay and the destination, respectively, are controlled by the channel input $X^n_1$ from the source, the relay input $X^n_2$ and the channel state $S^n$, through a given memoryless probability law $W_{Y_2,Y_3|X_1,X_2,S}$. The channel state $S^n$ is generated according to the $n$-product of a given memoryless probability law $Q_S$. It is assumed that the channel state is known, noncausally, to only the source. The destination estimates the message sent by the source from the received channel output. In this paper we study the capacity of this communication system. We will refer to the model in Figure \ref{StateDependentDiscreteMemorylessRelayChannel} as \textit{general state-dependent RC with informed source}.

\begin{figure}[htpb]
\centering
        \includegraphics[width=0.8\linewidth]{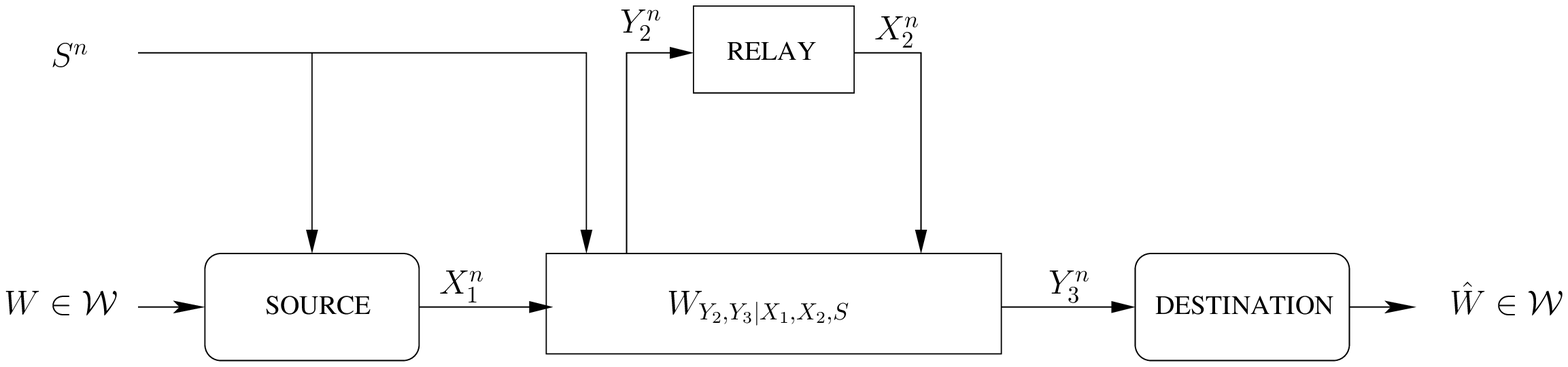}
\caption{General state-dependent relay channel with state information $S^n$ available non-causally at only the source.}
\label{StateDependentDiscreteMemorylessRelayChannel}
\end{figure}

We shall also study an important special case of the general model, shown in Figure~\ref{ModelWithHyperSource}. In this special model, the source alphabet $\mc X_1=\mc X_{1R} \times \mc X_{1D}$, $X^n_1=(X^n_{1R},X^n_{1D})$ and only the component $X^n_{1D}$ knows the states $S^n$. Furthermore, the memoryless conditional law $W_{Y_2,Y_3|X_{1R},X_{1D},X_2,S}$ factorizes as
\begin{equation}
W_{Y_2,Y_3|X_{1R},X_{1D},X_2,S}=W_{Y_2|X_{1R},S}W_{Y_3|X_{1D},X_2,S}.
\label{ConditionaProbability__ModelWithHyperSource}
\end{equation}

\noindent One can think of the two source encoder components in Figure~\ref{ModelWithHyperSource} as being two non-colocated base stations transmitting a common message to some destination with the help of a relay -- the common message may be obtained by means of message cognition at the encoder whose input is heard at the relay. 


\begin{figure}[htpb]
\centering
        \includegraphics[width=0.8\linewidth]{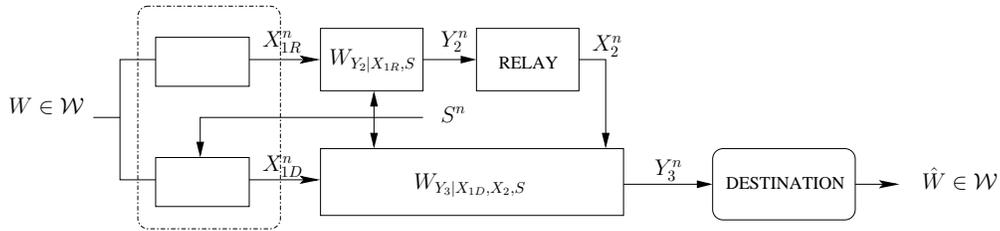}
\caption{State-dependent relay channel with the source input $X^n_1=(X^n_{1R},X^n_{1D})$, and only the component $X^n_{1D}$ knowing the states of the channel non-causally. \vspace{-0.7cm}}
\label{ModelWithHyperSource}
\end{figure}

\subsection{Background and Related Work}\label{secI_subsecA}
Channels whose probabilistic input-output relation  depends on random parameters, or channel states, have spurred much interest and can model a large variety of problems, each related to some physical situation of interest. The random state sequence may be known in a \textit{causal} or \textit{non-causal} manner. For single user models, the concept of channel state available at only the transmitter dates back to Shannon \cite{Sh58} for the causal channel state case, and to  Gel'fand and Pinsker \cite{GP80} for the non-causal channel state case. In \cite{HG83}, Heegard and El Gamal study a model in which the state sequence is known non-causally to only the encoder or to only the decoder. They also derive achievable rates for the case in which  partial channel state information (CSI) is given at varying rates to both the encoder and the decoder. In  \cite{C83}, Costa studies an additive Gaussian channel with additive Gaussian state known at only the encoder, and shows that Gel'fand-Pinsker coding with a specific auxiliary random variable, known as \textit{dirty paper coding} (DPC),  achieves the channel capacity. Interestingly, in this case, the DPC removes the effect of the additive channel state on the capacity as if there were no channel state present in the model or the channel state were known to the decoder as well. For a comprehensive review of state-dependent channels and related work, the reader may refer to \cite{KSM08}.

A growing body of work studies multi-user state-dependent models. Recent advances in this regard can be found in \cite{KSM08,SBSV07a,KL07a,ZKLV09a,KELW07,PKEZ07,ZKLV08a,ZKLV10,ZV09b,CS05,S05,LS10a,LS10b,PSS10,LSY10,LSY11,BL10,AMA09,KS09,CY11a,SCYA11a,K-FM11a}, and many other works. Key to the investigation of a state-dependent model is whether the parameters controlling the channel are known to \textit{all} or \textit{only some} of the users in the communication model. If the parameters of the channel are known to only some of the users, the problem exhibits some \textit{asymmetry} which makes its investigation more difficult in general. Also, in this case one has to expect some rate  penalty due to the lack of knowledge of the state at the uninformed encoders, relative to the case in which all encoders would be informed.

The state-dependent multiaccess channel (MAC) with only one informed encoder and degraded message sets is considered in \cite{KL04,KL07,KL07a,SBSV06,SBSV07,SBSV07a}; and the state-dependent relay channel (RC) with only informed relay is considered in \cite{ZKLV08a,ZKLV10}. For all these models, the authors develop non-trivial outer or upper bounds that permit to characterize the rate loss due to not knowing the state at the uninformed encoders. Key feature to the development of these outer or upper bounding techniques is that, in all these models, the uninformed encoder not only does not know the channel state but can learn no information about it. 

The model for the RC with informed source that we study in this paper seemingly exhibits some similarities with the RC with informed relay considered in \cite{ZKLV08a,ZKLV10}, and it also connects with the MAC with asymmetric channel state and degraded message sets considered in \cite{KL07a,SBSV07a,ZKLV09a}. However, establishing a non-trivial upper bound for the present model is more involved, comparatively. Partly, this is because, here, one uninformed encoder (the relay) is also a receiver; and, so, it can potentially get some information about the channel states from directly observing the past received sequence from the source. That is, at time $i$, the input $X_{2,i}$ of the relay can potentially depend on the channel states through its past output $Y_2^{i-1}=(Y_{2,1},\hdots,Y_{2,i-1})$. For the general model in Figure~\ref{StateDependentDiscreteMemorylessRelayChannel}, the relay can even know the states \textit{non-causally}, potentially. This is because $Y_2^{i-1}$ may depend on future values of the state through past source inputs  $X_{1,j}(W,S^n)$, $j=1,\hdots,i-1$. For the special case in Figure~\ref{ModelWithHyperSource}, the relay can know the states only \textit{strictly-causally}, but upper bounding the capacity seems still not easy. In our recent work \cite{ZPS11a, ZPS11b}, we have shown that, in a multiaccess channel, strictly causal knowledge of the state at one encoder can be beneficial in general for the other encoder \textit{even if the latter is informed non-causally}; but the capacity region is still to be characterized in general. Studying networks in which a subset of the nodes know the states non-causally and another subset know these states only strictly causally, i.e., networks with mixed -- noncausal and strictly causal, states appears to be more challenging in general, and is likely to capture additional interest, especially after  recent results on the utility of strictly causally known states in multiaccess channels \cite{LS10a,LS10b}.

\subsection{Main Contributions}\label{secI_subsecB}

For the general state-dependent RC with informed source shown in Figure~\ref{StateDependentDiscreteMemorylessRelayChannel}, we derive two lower bounds and an upper bound on the channel capacity. In the discrete memoryless (DM) case, the first lower bound is obtained by a block Markov coding scheme in which the source describes the channel state to the relay and destination \textit{ahead of time}. The source sends a two-layer description of the state consisting of two (possibly correlated) individual descriptions intended to be recovered at the relay and destination respectively. The relay recovers the individual description intended to it and then utilizes the estimated state as non-causal state information at the transmitter to implement collaborative source-relay binning in subsequent blocks, through a combined decode-and-forward \cite[Theorem 5]{CG79} and Gel'fand-Pinsker binning \cite{GP80}. The destination guesses the source's message sent cooperatively by the source and relay and the individual description which is intended to it from its output and the previously recovered state.  The rationale for the coding scheme which we use for the first lower bound is that, had the relay known the state with negligible distortion, then efficient cooperative source-relay binning in the spirit of \cite{KSS04} can be realized (recall that the model in \cite{KSS04} assumes availability of the state at both source and relay). 

We obtain the second lower bound by a block Markov coding scheme in which, rather than the channel state itself, the source describes to the relay the appropriate input that the relay would send had the relay known the channel states, assuming a decode-and-forward relaying strategy. The source sends this description to the relay ahead of time. The relay recovers the sent input and retransmits it in the appropriate subsequent block. The rationale for the coding scheme which we use for the second lower bound is that, if the input is produced at the source using binning against the known state and if the relay recovers it with negligible error, then all would appear as if the relay were informed of the channel state. This is because, from an operational point-of-view, the relay actually needs not know the channel state, but, rather, the appropriate input that it would send had it known this state. 

For the state-dependent general model, we also establish an upper bound on the capacity. This upper bound is non trivial and it accounts for not knowing the state at the relay and the destination. Then, considering the special model of Figure~\ref{ModelWithHyperSource}, we derive a better upper bound that accounts also for the loss incurred by not knowing the state at one of the source encoder components. We show that this upper bound is strictly tighter than the max-flow min cut or cut-set upper bound obtained by assuming that the state is available at all nodes. We note that upper-bounding techniques for related models with asymmetric channel states, i.e., models with states known only at some of the encoders have been developed recently in our previous work \cite{ZKLV10} for a relay channel with states known only at the relay, and in \cite{KL07a,SBSV07a,ZKLV09a} for a MAC with degraded message sets and states known only at one encoder. However, as we mentioned previously, the model that we study in this paper is more involved comparatively, essentially because, as a receiver the relay can get information about the unknown state. From this angle, our upper bounding techniques here are more linked to our recent works \cite{ZPS11a,ZPS11b}. 

Next, we also consider a memoryless Gaussian model in which the noise and the state are additive and Gaussian. The state represents an external interference and is known noncausally to only the source. We derive lower bounds on the capacity of the general Gaussian RC with informed source by applying the concepts that we develop for the DM case. Similar to the discrete case, one lower bound is based on the idea of describing the state to the relay beforehand; the relay recovers it and then utilizes it for collaborative binning in subsequent blocks. The other lower bound consists in transmitting to the relay a quantized version of the appropriate input that the relay would send had the relay known the channel state. We show that these lower bounds perform well in general and are optimal for large and small noise at the relay, respectively, irrespective to the strength of the interference. 

Furthermore, considering a Gaussian version of the special case model shown in Figure~\ref{ModelWithHyperSource}, we also develop a non-trivial upper bound on the capacity that is strictly better than the max-flow min cut or cut-set upper bound. We point out the rate loss in the upper bound incurred by the availability of the channel state at only the one source encoder component. Using this upper bound, we characterize the channel capacity in a number of cases, including when the interference corrupts transmission to the destination but not to the relay. 

\subsection{Outline and Notation}\label{secI_subsecC}

An outline of the remainder of this paper is as follows. Section \ref{secII} describes in more detail the communication models that we consider in this work.  Section \ref{secIII} provides lower and upper bounds on the capacity of the discrete memoryless model. Section \ref{secIV} provides lower and upper bound on the capacity of the Gaussian model; and characterizes the channel capacity in some cases. Section \ref{secV} contains some numerical results and discussions. Finally, Section \ref{secVI} concludes the paper.

We use the following notations throughout the paper. Upper case letters are used to denote random variables, e.g., $X$; lower case letters are used to denote realizations of random variables, e.g., $x$; and calligraphic letters designate alphabets, i.e., $\mc X$. The probability distribution of a random variable $X$ is denoted by $P_X(x)$. Sometimes, for convenience, we write it as $P_X$.  We use the notation $\mathbb{E}_{X}[\cdot]$ to denote the expectation of random variable $X$. A probability distribution of a random variable $Y$ given $X$ is denoted by $P_{Y|X}$. The set of probability distributions defined on an alphabet $\mc X$ is denoted by $\mc P(\mc X)$. The cardinality of a set $\mc X$ is denoted by $|\mc X|$. For convenience, the length $n$ vector $x^n$ will occasionally be denoted in boldface notation $\dv x$. The Gaussian distribution with mean $\mu$ and variance $\sigma^2$ is denoted by $\mathcal{N}(\mu,\sigma^2)$. Finally, throughout the paper, logarithms are taken to base $2$, and the complement to unity of a scalar $u \in [0,1]$ is denoted by $\bar{u}$, i.e., $\bar{u}=1-u$.

\section{System Model and Definitions}\label{secII}

In this section, we formally present our communication model and the related definitions. As shown in Figure \ref{StateDependentDiscreteMemorylessRelayChannel}, we consider a state-dependent relay channel denoted by $W_{Y_2,Y_3|X_1,X_2,S}$ whose outputs $Y^n_2 \in \mc Y^n_2$ and $Y^n_3 \in \mc Y^n_3$ for the relay and the destination, respectively, are controlled by the channel inputs $X^n_1 \in \mc X^n_1$ from the source and $X^n_2 \in \mc X^n_2$ from the relay, along with random states $S^n \in \mc S^n$. It is assumed that the channel state $S_i$ at time instant $i$ is independently drawn from a given distribution $Q_S$ and the channel states $S^n$ are non-causally known only at the source.

The source wants to transmit a message $W$ to the destination with the help of the relay, in $n$ channel uses. The message $W$ is assumed to be uniformly distributed over the set $\mc W=\{1,\hdots,M\}$. The information rate $R$ is defined as $n^{-1}\log M$ bits per transmission.

\noindent An $(M,n)$ code for the state-dependent relay channel with informed source consists of an encoding function at the source
$$\phi_1^n: \{1,\hdots,M\}\times \calS^n \rightarrow \mc X_1^n,$$ a sequence of encoding functions at the relay
$$\phi_{2,i}: \calY_{2,1}^{i-1} \rightarrow \calX_{2,i},$$
for $i=1,2,\ldots,n,$ and a decoding function at the destination
$$\psi^n: \mc Y_3^n \rightarrow \{1,\hdots,M\}.$$

Let a $(M,n)$ code be given. The sequences $X_{1}^n$ and $X_{2}^n$ from the source and the relay, respectively, are transmitted across a state-dependent relay channel modeled as a memoryless conditional probability distribution $W_{Y_2,Y_3|X_1,X_2,S}$. The joint probability mass function on ${\mc W}{\times}{\mc S^n}{\times}{\mc X^n_1}{\times}{\mc X^n_2}{\times}{\mc Y^n_2}{\times}{\mc Y^n_3}$ is given by
\begin{align}
P(w,s^n,x^n_1,x^n_2,y^n_2,y^n_3) &= P(w)\prod_{i=1}^{n}Q_S(s_i)P(x_{1,i}|w,s^n)P(x_{2,i}|y^{i-1}_2)\nonumber\\
&\hspace{1cm}{\cdot}W_{Y_2,Y_3|X_1,X_2,S}(y_{2,i},y_{3,i}|x_{1,i},x_{2,i},s_i).
\end{align}

The destination estimates the message sent by the source from the channel output $Y_{3}^n$. The average probability of error is defined as $P_e^n = \mathbb{E}_{S}\big[\mathrm{Pr}\big(\psi^n(Y_{3}^n)\neq W|S^n=s^n\big)\big].$

An  $(\epsilon,n,R)$ code for the state-dependent RC with informed source is an $(2^{nR},n)-$code $(\phi_1^n,\phi_2^n,\psi^n)$ having average probability of error $P_e^n$ not exceeding $\epsilon$.

A rate $R$ is said to be achievable if there exists a sequence of $(\epsilon_n,n,R)-$codes with $\lim_{n \rightarrow \infty} \epsilon_n=0$. The capacity $\mc C$ of the state-dependent RC with informed source is defined as the supremum of the set of achievable rates.

We shall also study the special case model shown in Figure~\ref{ModelWithHyperSource}, in which the source alphabet $\mc X_1=\mc X_{1R} {\times} \mc X_{1D}$, $X^n_1=(X^n_{1R},X^n_{1D})$ with the input component $X^n_{1R}$ function of only the message $W$ and the input component $X^n_{1D}$ function of $(W,S^n)$, i.e., $X^n_{1R}=\phi^n_{1R}(W)$ and $X^n_{1D}=\phi^n_{1D}(W,S^n)$ --- $\phi^n_{1R}$ and $\phi^n_{1D}$ are the source encoding functions, and the conditional distribution $W_{Y_2,Y_3|X_{1R},X_{1D},X_2,S}$ factorizing as \eqref{ConditionaProbability__ModelWithHyperSource}. 

\section{The Discrete Memoryless RC with Informed Source}\label{secIII}

In this section, we assume that the alphabets $\calS$, $\calX_1$, $\calX_2$, $\calY_2$, $\calY_3$ in the model are all discrete and finite.

\vspace{-0.2cm}

\subsection{Lower Bound on Channel Capacity: State Description}\label{secIII_subsecA}

The following theorem provides a lower bound on the capacity of the state-dependent general discrete memoryless RC with informed source.

\begin{theorem}\label{Theorem__LowerBound1__DiscreteChannel}
The capacity of the discrete memoryless state-dependent relay channel with informed source is lower bounded by
\begin{align}
R^{\text{lo}} = \max \: \min\: \{&I(U;Y_2|V,\hat{S}_R)-I(U;S,\hat{S}_D|V,\hat{S}_R),\nonumber\\
&I(U,V;Y_3|\hat{S}_D)-I(U,V;S,\hat{S}_R|\hat{S}_D)\}
\label{AchievableRate__Theorem__LowerBound1__DiscreteChannel}
\end{align}
subject to the constraints
\begin{subequations}
\begin{align}
I(S;\hat{S}_R) &\leq I(U_R;Y_2,\hat{S}_R|U,V)-I(U_R;S,\hat{S}_R,\hat{S}_D|U,V)\\
I(S;\hat{S}_D) &\leq I(U_D;Y_3,\hat{S}_D|U,V)-I(U_D;S,\hat{S}_R,\hat{S}_D|U,V)+[I(U;Y_3,\hat{S}_D|V)-I(U;S,\hat{S}_R,\hat{S}_D|V)]_{-}\\
I(S;\hat{S}_R,\hat{S}_D) + I(\hat{S}_R;\hat{S}_D) &\leq I(U_R;Y_2,\hat{S}_R|U,V)-I(U_R;S,\hat{S}_R,\hat{S}_D|U,V)\nonumber\\
&+I(U_D;Y_3,\hat{S}_D|U,V)-I(U_D;S,\hat{S}_R,\hat{S}_D|U,V)+[I(U;Y_3,\hat{S}_D|V)-I(U;S,\hat{S}_R,\hat{S}_D|V)]_{-}\nonumber\\
&-I(U_R;U_D|U,V,S,\hat{S}_R,\hat{S}_D)
\label{Constraints__AchievableRate__Theorem__LowerBound1__DiscreteChannel}
\end{align}
\end{subequations}
where $[x]_{-} \triangleq \min(x,0)$, and the maximization is over all joint measures on ${\mc S}\times\hat{\mc S}_R\times\hat{\mc S}_D\times{\mc U}_R\times{\mc U}_D\times{\mc U}\times{\mc V}\times{\mc X}_1\times{\mc X}_2\times{\mc Y}_2\times{\mc Y}_3$ of the form
\begin{align}
&P_{S,\hat{S}_R,\hat{S}_D,U_R,U_D,U,V,X_1,X_2,Y_2,Y_3}\nonumber\\
&\qquad = Q_SP_{\hat{S}_R,\hat{S}_D|S}P_{V|\hat{S}_R}P_{U|V,S,\hat{S}_R,\hat{S}_D}P_{U_R,U_D|V,U,S,\hat{S}_R,\hat{S}_D}P_{X_1|U_R,U_D,U,V,S,\hat{S}_R,\hat{S}_D}P_{X_2|V,\hat{S}_R}W_{Y_2,Y_3|X_1,X_2,S}.
\label{Measure__AchievableRate__Theorem__LowerBound1__DiscreteChannel}
\end{align}
and satisfying
\begin{align}
I(V;Y_3,\hat{S}_D) - I(V;\hat{S}_R) > 0.
\end{align}
\end{theorem}

\vspace{0.6cm}

\noindent \textbf{Proof:} An outline of the proof of Theorem \ref{Theorem__LowerBound1__DiscreteChannel} will follow, and complete error analysis appears in Appendix \ref{appendixTheorem__LowerBound1__DiscreteChannel}.

\vspace{0.2cm}

The following remarks are useful for a better understanding of the coding scheme which we use to achieve the lower bound in Theorem \ref{Theorem__LowerBound1__DiscreteChannel}. 

\begin{remark}\label{remark1}
The intuition for the coding scheme which we use to establish the lower bound in Theorem \ref{Theorem__LowerBound1__DiscreteChannel} is as follows. Had the relay known the state, the source and the relay could implement collaborative binning against that state for transmission to the destination \cite{KSS04}. Since the source knows the state of the channel non-causally, it can transmit a description of it to the relay \textit{ahead of time}. The relay recovers the state (with a certain distortion), and  then utilizes it in the relevant subsequent block through a collaborative binning scheme. The hope is that the benefit that the source can get from being assisted by a more capable relay will compensate the loss caused by the source's spending some of its resources to make the relay learn the state. 

\noindent In general, it may also turn to be useful to send a dedicated description of the state to the destination. The destination utilizes the recovered state as side information at the receiver. In the coding scheme that we employ to establish the lower bound in Theorem \ref{Theorem__LowerBound1__DiscreteChannel}, in addition to its message, the source also sends a two-layer description of the state to the relay and destination; one layer description dedicated for each. The two layers are possibly correlated. The relay guesses the source's message and the individual state description which is dedicated to it from the source transmission and the previously recovered state description. It then utilizes the new state estimate as non-causal state at the encoder for collaborative source-relay binning over the next block, through a combined decode-and-forward and Gel'fand-Pinsker binning. The destination guesses the source's message sent cooperatively by the source and relay and the individual state description which is dedicated to it from its output and the previously recovered state description.  
\end{remark}

\begin{remark}\label{remark2}
As it can be seen from the proof in Appendix \ref{appendixTheorem__LowerBound1__DiscreteChannel}, the source sends the descriptions intended to the relay and destination \textit{two blocks} ahead of time. That is, at the beginning of block $i$ the source describes the state vector $\dv s[i+2]$ to the relay and destination. While one block delay is sufficient to describe the state to the relay, a minimum of two blocks is necessary for the state reconstruction at the destination because of the used window decoding technique. In the following remark, we will comment onto the relevance of sliding window for decoding at the destination for our model.  
\end{remark}

\begin{remark}\label{remark3}
The coding scheme that we employ to prove the lower bound in Theorem \ref{Theorem__LowerBound1__DiscreteChannel} uses regular encoding sliding-window-decoding as a relaying strategy. Backward-decoding at the destination, which has been proved to sometimes offer rates higher than those of window-decoding for certain non-classic relaying models \cite{ZKLV10}, is also possible for our model. However, here, this would require sending \textit{independent} descriptions to the relay and destination. More specifically, with backward decoding, the individual description intended to the destination should correspond to the state sequence that affects block $i-1$, or an earlier block. That is, in this case the source would have to describe a "future" state vector to the relay and a "past" state vector to the destination. When the state sequence is i.i.d. across blocks, the two individual descriptions are independent, and, intuitively, this independence will cause the source to dedicate more of its rate to transmitting the state (in comparison to with sliding window), thus leaving a smaller rate for the transmission of the information message. 
\end{remark}

\textbf{Outline of Proof of Theorem \ref{Theorem__LowerBound1__DiscreteChannel}:} 

A formal proof of Theorem \ref{Theorem__LowerBound1__DiscreteChannel} with complete error analysis is given in Appendix \ref{appendixTheorem__LowerBound1__DiscreteChannel}. We now give a description of a random coding scheme which we use to obtain the lower bound given in Theorem \ref{Theorem__LowerBound1__DiscreteChannel}. This scheme is based on an appropriate combination of block Markov encoding \cite{CG79}, Gel'fand-Pinsker binning \cite{GP80}, multiple descriptions \cite{GC82} and Marton's coding for general broadcast channels \cite{M79,GM81,SS05}. Next, we outline the encoding and decoding procedures. 

 We transmit in $B+1$ blocks, each of length $n$. Let  $\dv s[i]$ denote the state sequence controlling the channel in block $i$, with $i=1,\hdots,B+1$. During each of the first $B$ blocks, the source encodes a message $w_i \in [1,2^{nR}]$ and sends it over the channel. In addition, during each of the first $B-1$ blocks, the source also sends two individual descriptions of $\dv s[i+2]$ intended to be recovered at the relay and destination, respectively. We denote by $\hat{\dv s}_R[\iota_{Ri}]$, $\iota_{Ri} \in [1,2^{n\hat{R}_R}]$, the description of $\dv s[i+2]$ intended to be recovered at the relay in block $i$, at rate $\hat{R}_R$, and by $\hat{\dv s}_D[\iota_{Di}]$, $\iota_{Di} \in [1,2^{n\hat{R}_D}]$, the description of $\dv s[i+2]$ intended to be recovered at the destination in block $i$, at rate $\hat{R}_D$. For the last two blocks, for convenience, we set $w_{B+1}=1$, $(\iota_{RB},\iota_{DB})=(1,1)$ and $(\iota_{RB+1},\iota_{DB+1})=(1,1)$. For fixed $n$, the average (channel coding) rate $R(B/(B+1))$ of the information message over $B+1$ blocks approaches $R$ as $B \longrightarrow +\infty$, and the average (source coding) rates $\hat{R}_R((B-1)/(B+1))$ and $\hat{R}_D((B-1)/(B+1))$ approach $\hat{R}_R$ and $\hat{R}_D$, respectively, as $B \longrightarrow +\infty$.

\textbf{Codebook generation:} Fix a measure $P_{S,\hat{S}_R,\hat{S}_D,U_R,U_D,U,V,X_1,X_2,Y_2,Y_3}$ of the form \eqref{Measure__AchievableRate__Theorem__LowerBound1__DiscreteChannel}. Calculate the marginals $P_{\hat{S}_R}$ and $P_{\hat{S}_D}$ induced by this measure. Fix $\epsilon > 0$, and let $M =2^{n[R-\epsilon]}$, 
\begin{align}
J_V &= 2^{n[I(V;\hat{S}_R)+\epsilon]} & M_R &= 2^{n[R_R-5\epsilon]}& J_R &=  2^{n[I(U_R;S,\hat{S}_R,\hat{S}_D|U,V)+\epsilon]}\nonumber\\
J_U &= 2^{n[I(U;S,\hat{S}_R,\hat{S}_D|V)+\epsilon]} & M_D &= 2^{n[R_D-5\epsilon]} & J_D &=  2^{n[I(U_D;S,\hat{S}_R,\hat{S}_D|U,V)+\epsilon]}
\label{ValuesForBinningVariablesInTheorem__LowerBound2__GeneralGaussianChannel}
\end{align}
with
\begin{align}
R_R &= I(U_R;Y_2,\hat{S}_R|U,V)-I(U_R;S,\hat{S}_R,\hat{S}_D|U,V)-\epsilon\nonumber\\
R_D &=I(U_D;Y_3,\hat{S}_D|U,V)-I(U_D;S,\hat{S}_R,\hat{S}_D|U,V)+[I(U;Y_3,\hat{S}_D|V)-I(U;S,\hat{S}_R,\hat{S}_D|V)]_{-}-\epsilon
\label{Choice__Rates__BC__Individual__Descriptions}
\end{align}
where $[x]_{-}$ denotes $\min(x,0)$.

\noindent We may assume that first term of the minimization in \eqref{AchievableRate__Theorem__LowerBound1__DiscreteChannel} is non-negative, i.e., $I(U;Y_2,\hat{S}_R|V)-I(U;S,\hat{S}_R,\hat{S}_D|V) \geq 0$.

\noindent We generate two statistically independent codebooks  (codebooks $1$ and $2$) by following the steps outlined below twice. We shall use  these codebooks for blocks with odd and even indices, respectively.

\begin{itemize}
\item[1)] Generate $2^{n\hat{R}_R}$ $n$-vectors $\hat{\dv s}_R[1],\hdots,\hat{\dv s}_R[2^{n\hat{R}_R}]$ independently according to a uniform distribution over the set $T^n_{\epsilon}(P_{\hat{S}_R})$ of $\epsilon-$typical $\hat{\dv S}_R$ $n-$ vectors.
\item[2)] Generate $2^{n\hat{R}_D}$ $n$-vectors $\hat{\dv s}_D[1],\hdots,\hat{\dv s}_D[2^{n\hat{R}_D}]$ independently according to a uniform distribution over the set $T^n_{\epsilon}(P_{\hat{S}_D})$ of $\epsilon-$typical $\hat{\dv S}_D$ $n-$ vectors.
\item[3)] Generate $J_VM$ independent and identically distributed (i.i.d.) codewords $\{\dv v(w',j_V)\}$ indexed by $w'=1,\hdots,M$, $j_V=1,\hdots,J_V$. Each codeword $\dv v(w',j_V)$ is with i.i.d. components drawn according to $P_V$.
\item[4)] For each codeword $\dv v(w',j_V)$, generate a collection of $J_UM$ codewords $\{\dv u(w',j_V,w,j_U)\}$ indexed by $w=1,\hdots,M$, $j_U=1,\hdots,J_U$. Each codeword $\dv u(w',j_V,w,j_U)$ is with i.i.d. components drawn according to $P_{U|V}$.
\item[5)] For each codeword $\dv v(w',j_V)$, for each codeword $\dv u(w',j_V,w,j_U)$, generate a collection of $J_RM_R$ codewords $\{\dv u_R(w',j_V,w,j_U,k,j_R)\}$ indexed by $k=1,\hdots,M_R$, $j_R=1,\hdots,J_R$. Each codeword $\dv u_R(w',j_V,w,j_U,k,j_R)$ is with i.i.d. components drawn according to $P_{U_R|V,U}$. 
\item[6)] For each codeword $\dv v(w',j_V)$, for each codeword $\dv u(w',j_V,w,j_U)$, generate a collection of $J_DM_D$ codewords $\{\dv u_D(w',j_V,w,j_U,l,j_D)\}$ indexed by $l=1,\hdots,M_D$, $j_D=1,\hdots,J_D$. Each codeword $\dv u_D(w',j_V,w,j_U,l,j_D)$ is with i.i.d. components drawn according to $P_{U_D|V,U}$. 
\item[7)] (Binning \`a-la Marton \cite{M79},\cite{GM81}): For $\iota_R \in [1,2^{n\hat{R}_R}]$, define the cells 
\begin{equation*}
\mc B_{\iota_R} = [(\iota_R-1)2^{n[R_R-\hat{R}_R-\epsilon]}+1,\iota_R2^{n[R_R-\hat{R}_R-\epsilon]}].
\end{equation*}
Similarly, for $\iota_D \in [1,2^{n\hat{R}_D}]$, define the cells
\begin{equation*}
\mc C_{\iota_D} = [(\iota_D-1)2^{n[R_D-\hat{R}_D-\epsilon]}+1,\iota_D2^{n[R_D-\hat{R}_D-\epsilon]}],
\end{equation*}
where without loss of generality $2^{n[R_R-\hat{R}_R-\epsilon]}$ and $2^{n[R_D-\hat{R}_D-\epsilon]}$ are considered to be integer valued.

\end{itemize}

\textbf{Encoding:} The encoders at the source and the relay encode messages using codebook 1 for blocks with odd indices, and codebook 2 for blocks with even indices. This is done because some of the decoding steps are performed jointly over two adjacent blocks, and so having independent codebooks makes the error events corresponding to these blocks independent and their probabilities easier to evaluate.

We pick up the story in block $i$. Let $w_i$ be the new message to be sent from the source node at the beginning of block $i$, and $w_{i-1}$ the message sent in the previous block $i-1$. The encoding at the beginning of block $i$ is as follows. 

\noindent The source finds, if possible, a pair $(\iota_{Ri},\iota_{Di}) \in [1,2^{n\hat{R}_R}]{\times}[1,2^{n\hat{R}_D}]$ such that $(\dv s[i+2],\hat{\dv s}_R[\iota_{Ri}],\hat{\dv s}_D[\iota_{Di}])$ are jointly typical. If such $(\iota_{Ri},\iota_{Di})$ does not exist, simply set $(\iota_{Ri},\iota_{Di})=(1,1)$. We shall show that a successful encoding of $\dv s[i+2]$ at the source is accomplished with high probability provided that $n$ is sufficiently large and
\begin{align}
\hat{R}_R &> I(S;\hat{S}_R)\nonumber\\
\hat{R}_D &> I(S;\hat{S}_D)\nonumber\\
\hat{R}_R+\hat{R}_D &> I(S;\hat{S}_R,\hat{S}_D)+I(\hat{S}_R;\hat{S}_D).
\label{Encoding__Step__Multiple__Description__Coding}
\end{align}

\noindent The source will send the quadruple $(w_{i-1},w_i,\iota_{Ri},\iota_{Di})$ over the channel. First, let us assume that the relay has decoded correctly message $w_{i-1}$ and the indices $(\iota_{Ri-2},\iota_{Ri-1})$, and the destination has decoded correctly message $w_{i-2}$ and the index $\iota_{Di-2}$. We shall show that our code construction allows the relay to decode correctly message $w_i$ and the index $\iota_{Ri}$ and the destination to decode correctly message $w_{i-1}$ and the index $\iota_{Di-1}$ at the end of block $i$ (with a probability of error $\leq \epsilon$). Thus, the information state $(w_{i-2},w_{i-1},\iota_{Ri-1},\iota_{Di-2})$ propagates forward and a recursive calculation of the probability of error can be made, yielding a probability of error $\leq (B+1)\epsilon$.


We continue with the strategy at the beginning of block $i$.
\begin{itemize}
\item[1)] The relay  knows $w_{i-1}$ and $\iota_{Ri-2}$ and searches for the smallest $j_V \in J_V$ such that $\dv v(w_{i-1},j_V)$ is jointly typical with $\hat{\dv s}_R[\iota_{Ri-2}]$ (the properties of jointly typical sequences guarantee that, with probability close to one, there exists one such $j_V$). Denote this $j_V$ by $j^{\star}_{Vi}=j_V(\hat{\dv s}_R[\iota_{Ri-2}],w_{i-1})$. Then the relay sends a vector $\dv x_2[i]$ with i.i.d. components given $\dv v(w_{i-1},j^{\star}_{Vi})$ and $\hat{\dv s}_R[\iota_{Ri-2}]$, drawn according to the marginal $P_{X_2|V,\hat{S}_R}$ induced by the distribution \eqref{Measure__AchievableRate__Theorem__LowerBound1__DiscreteChannel}. (For $i=1,2$, the relay does not know an estimate of the channel state and so it sends some default codeword).
\item[2)] The source first searches for the smallest $j_U \in J_U$ such that $\dv u(w_{i-1},j^{\star}_{Vi},w_i,j_U)$ is jointly typical with the vector $\dv s[i],\hat{\dv s}_R[\iota_{Ri-2}],\hat{\dv s}_D[\iota_{Di-2}])$ given $\dv v(w_{i-1},j^{\star}_{Vi})$. (Again, the properties of jointly typical sequences guarantee that there exists one such $j_U$). Denote this $j_U$ by $j^{\star}_{Ui}=j_U(\dv s[i],\hat{\dv s}_R[\iota_{Ri-2}],\hat{\dv s}_D[\iota_{Di-2}],w_{i-1},w_i)$. 
\item[3)] Next, the source searches for one pair $$\Big(\dv u_R(w_{i-1},j^{\star}_{Vi},w_i,j^{\star}_{Ui},k_i,j_{Ri}),\dv u_D(w_{i-1},j^{\star}_{Vi},w_i,j^{\star}_{Ui},l_i,j_{Di})\Big) \in \mc D_{\iota_{Ri}\iota_{Di}},$$
where
\begin{align}
\mc D_{\iota_{Ri}\iota_{Di}} = \Big\{\Big(\dv u_R(w_{i-1},j^{\star}_{Vi},w_i,j^{\star}_{Ui},&k_i,j_{Ri}),\dv u_D(w_{i-1},j^{\star}_{Vi},w_i,j^{\star}_{Ui},l_i,j_{Di})\Big)\:\:s.t.: \nonumber\\
& k_i \in \mc B_{\iota_{Ri}}, l_i \in \mc C_{\iota_{Di}}, j_{Ri} \in J_R, j_{Di} \in J_D\nonumber\\
& \Big(\dv u_R(w_{i-1},j^{\star}_{Vi},w_i,j^{\star}_{Ui},k_i,j_{Ri}),\dv s[i],\hat{\dv s}_R[\iota_{Ri-2}],\hat{\dv s}_D[\iota_{Di-2}]\Big) \in T^n_{\epsilon}(P_{U_RS\hat{S}_R\hat{S}_D|UV})\nonumber\\
& \Big(\dv u_D(w_{i-1},j^{\star}_{Vi},w_i,j^{\star}_{Ui},l_i,j_{Di}),\dv s[i],\hat{\dv s}_R[\iota_{Ri-2}],\hat{\dv s}_D[\iota_{Di-2}]\Big) \in T^n_{\epsilon}(P_{U_DS\hat{S}_R\hat{S}_D|UV})\nonumber\\
& \Big(\dv u_R(w_{i-1},j^{\star}_{Vi},w_i,j^{\star}_{Ui},k_i,j_{Ri}),\dv u_D(w_{i-1},j^{\star}_{Vi},w_i,j^{\star}_{Ui},l_i,j_{Di})\Big) \in T^n_{\epsilon}(P_{U_R,U_D|UVS\hat{S}_R\hat{S}_D})\Big\}.
\label{Encoding__Step__Marton__Binning}
\end{align}

We shall show that, with high probability, the source will find one such pair provided that $n$ is sufficiently large and 
\begin{align}
\hat{R}_R+\hat{R}_D &< R_R+R_D -I(U_R;U_D|U,V,S,\hat{S}_R,\hat{S}_D). 
\label{Condition__on__Sum__Rate__Encoding__Step__Marton__Binning}
\end{align}
Denote the found pair as $\Big(\dv u_R(w_{i-1},j^{\star}_{Vi},w_i,j^{\star}_{Ui},k_i,j^{\star}_{Ri}),\dv u_D(w_{i-1},j^{\star}_{Vi},w_i,j^{\star}_{Ui},l_i,j^{\star}_{Di})\Big)$.
\item[4)] The source then sends a vector $\dv x_1[i]$ with i.i.d. components given the vectors $\dv v(w_{i-1},j^{\star}_{Vi})$, $\dv u(w_{i-1},j^{\star}_{Vi},w_i,j^{\star}_{Ui})$, $\dv u_R(w_{i-1},j^{\star}_{Vi},w_i,j^{\star}_{Ui},k_i,j^{\star}_{Ri})$, $\dv u_D(w_{i-1},j^{\star}_{Vi},w_i,j^{\star}_{Ui},l_i,j^{\star}_{Di})$ and $(\dv s[i],\hat{\dv s}_R[\iota_{Ri-2}],\hat{\dv s}_D[\iota_{Di-2}])$, drawn according to the marginal $P_{X_1|V,U,U_R,U_D,S,\hat{S}_R,\hat{S}_D}$ induced by the distribution \eqref{Measure__AchievableRate__Theorem__LowerBound1__DiscreteChannel}. 

\end{itemize}

\textbf{Decoding:} Decoding and state reconstruction at the relay are based on classical joint typicality. Decoding and state reconstruction at the destination are based on joint typicality and window-decoding. The decoding and reconstruction procedures at the end of block $i$ are as follows.

\begin{itemize}
\item[1)] The relay knows $w_{i-1}$ and $\iota_{Ri-2}$ (in fact, the relay knows also $\iota_{Ri-1}$ but does not use it for decoding in this step). It declares that $(\hat{w}_i,\hat{\iota}_{Ri})$ are sent if there exists a unique triple $(\hat{w}_i,\hat{j}_{Ui},\hat{k}_i)$, $\hat{w}_i \in [1,M]$, $\hat{j}_{Ui} \in J_U$, $\hat{k}_i \in [1,M_R]$, such that $\dv u(w_{i-1},j^{\star}_{Vi},\hat{w}_i,\hat{j}_{Ui})$, $\dv u_R(w_{i-1},j^{\star}_{Vi},\hat{w}_i,\hat{j}_{Ui},\hat{k}_i,j_{Ri})$ are jointly typical with $(\dv y_2[i],\hat{\dv s}_R[\iota_{Ri-2}])$ given $\dv v(w_{i-1},j^{\star}_{Vi})$, for some $j_{Ri} \in J_R$, where $j^{\star}_{Vi}=j_V(\hat{\dv s}_R[\iota_{Ri-2}],w_{i-1})$. One can show that, with the choice \eqref{Choice__Rates__BC__Individual__Descriptions}, the decoding error in this step is small for sufficiently large $n$ if
\begin{align}
R &< I(U;Y_2,\hat{S}_R|V)-I(U;S,\hat{S}_R,\hat{S}_D|V).
\label{Bounds__on__Rates__Decoding__at__Relay__LowerBound2}
\end{align} 
If \eqref{Bounds__on__Rates__Decoding__at__Relay__LowerBound2} is satisfied, the estimate  $\hat{\iota}_{Ri}$ of $\iota_{Ri}$ at the relay is the index of the $\mc B_{\hat{\iota}_{Ri}}$ containing the found $\hat{k}_i$, i.e., $\hat{k}_i \in \mc B_{\hat{\iota}_{Ri}}$.
\item[2)] The destination knows the pair $(w_{i-2},l_{i-2})$ and the index $j^{\star}_{Vi-1}=j_V(\hat{\dv s}_R[\iota_{Ri-3}],w_{i-2})$ and decodes the pair $(w_{i-1},l_{i-1})$ based on the information received in block $i-1$ and block $i$. It declares that $(\hat{w}_{i-1},\hat{l}_{i-1})$ is sent if there is a unique triple $(\hat{w}_{i-1},\hat{j}_{Ui-1},\hat{l}_{i-1})$, $\hat{w}_{i-1} \in [1,M]$, $\hat{j}_{Ui-1} \in J_U$, $\hat{l}_{i-1} \in [1,M_D]$, and a unique $\hat{j}_{Vi} \in J_V$, such that $\dv u(w_{i-2},j^{\star}_{Vi-1},\hat{w}_{i-1},\hat{j}_{Ui-1})$, $\dv u_D(w_{i-2},j^{\star}_{Vi-1},\hat{w}_{i-1},\hat{j}_{Ui-1},\hat{l}_{i-1},j_{Di-1})$  are jointly typical with $(\dv y_3[i-1],\hat{\dv s}_D[\iota_{Di-3}])$ given $\dv v(w_{i-2},j^{\star}_{Vi-1})$ \underline{\textit{and}} $\dv v(\hat{w}_{i-1},\hat{j}_{Vi})$ is jointly typical with $(\dv y_3[i],\hat{\dv s}_D[\iota_{Di-2}])$. One can show that, with the choice \eqref{Choice__Rates__BC__Individual__Descriptions}, the decoding error in this step is small for sufficiently large $n$ if
\begin{align}
R  &< I(V,U;Y_3,\hat{S}_D)-I(V,U;S,\hat{S}_R,\hat{S}_D)\nonumber\\
0 &< I(V;Y_3,\hat{S}_D)-I(V;\hat{S}_R).
\label{Bounds__on__Rates__Decoding__at__Destination__LowerBound2}
\end{align}
If \eqref{Bounds__on__Rates__Decoding__at__Destination__LowerBound2} is satisfied, the estimate  $\hat{\iota}_{Di-1}$ of $\iota_{Di-1}$ at the destination is the index of the $\mc C_{\hat{\iota}_{Di-1}}$ containing the found $\hat{l}_{i-1}$, i.e., $\hat{l}_{i-1} \in \mc C_{\hat{\iota}_{Di-1}}$. Also, the destination obtains the correct index $j^{\star}_{Vi}=j_V(\hat{\dv s}_R[\iota_{Ri-2}],w_{i-1})$. \begin{flushright} \qed \end{flushright}
\end{itemize}

The achievable rate in Theorem \ref{Theorem__LowerBound1__DiscreteChannel} requires the relay to decode the message sent by the source \textit{fully}, and this can be rather a severe constraint.  We can generalize Theorem \ref{Theorem__LowerBound1__DiscreteChannel} by allowing the relay to decode the message sent by the source only \textit{partially} \cite{GA82}. This can be done by splitting the information sent by the source into two independent parts, one part is sent through the relay and the other part is sent directly to the destination. In the following corollary, the random variables $V$, $U$, $U_R$ and $U_D$ play the same roles as in Theorem \ref{Theorem__LowerBound1__DiscreteChannel} and $U_1$ is a new random variable that represents the information sent directly to the destination. 

\begin{corollary}\label{Corollary__LowerBound1__DiscreteChannel}
The capacity of the discrete memoryless state-dependent relay channel with informed source is lower bounded by
\begin{align}
R^{\text{lo}} = \max \: \min\: \{&I(U;Y_2|V,\hat{S}_R)-I(U;S,\hat{S}_D|V,\hat{S}_R),\nonumber\\
&I(U,V;Y_3|\hat{S}_D)-I(U,V;S,\hat{S}_R|\hat{S}_D)\}+ I(U_1;Y_3|U,V,\hat{S}_D)-I(U_1;S,\hat{S}_R|U,V,\hat{S}_D)
\label{AchievableRate__Theorem__LowerBound1__DiscreteChannel}
\end{align}
subject to the constraints
\begin{subequations}
\begin{align}
I(S;\hat{S}_R) &\leq I(U_R;Y_2,\hat{S}_R|U,V)-I(U_R;S,\hat{S}_R,\hat{S}_D|U,V)\\
I(S;\hat{S}_D) &\leq I(U_D;Y_3,\hat{S}_D|U_1,U,V)-I(U_D;S,\hat{S}_R,\hat{S}_D|U_1,U,V)+[I(U_1,U;Y_3,\hat{S}_D|V)-I(U_1,U;S,\hat{S}_R,\hat{S}_D|V)]_{-}\\
I(S;\hat{S}_R,\hat{S}_D) + I(\hat{S}_R;\hat{S}_D) &\leq I(U_R;Y_2,\hat{S}_R|U,V)-I(U_R;S,\hat{S}_R,\hat{S}_D|U,V)\nonumber\\
&+I(U_D;Y_3,\hat{S}_D|U_1,U,V)-I(U_D;S,\hat{S}_R,\hat{S}_D|U_1,U,V)+[I(U_1,U;Y_3,\hat{S}_D|V)-I(U_1,U;S,\hat{S}_R,\hat{S}_D|V)]_{-}\nonumber\\
&-I(U_R;U_D|U_1,U,V,S,\hat{S}_R,\hat{S}_D)
\label{Constraints__AchievableRate__Theorem__LowerBound1__DiscreteChannel}
\end{align}
\end{subequations}
where $[x]_{-} \triangleq \min(x,0)$, and the maximization is over all joint measures on ${\mc S}\times\hat{\mc S}_R\times\hat{\mc S}_D\times{\mc U}_R\times{\mc U}_D\times{\mc U}_1\times{\mc U}\times{\mc V}\times{\mc X}_1\times{\mc X}_2\times{\mc Y}_2\times{\mc Y}_3$ of the form
\begin{align}
&P_{S,\hat{S}_R,\hat{S}_D,U_R,U_D,U,V,X_1,X_2,Y_2,Y_3}\nonumber\\
&\qquad = Q_SP_{\hat{S}_R,\hat{S}_D|S}P_{V|\hat{S}_R}P_{U|V,S,\hat{S}_R,\hat{S}_D}P_{U_1|V,U,S,\hat{S}_R,\hat{S}_D}P_{U_R,U_D|V,U,U_1,S,\hat{S}_R,\hat{S}_D}P_{X_1|U_R,U_D,U,V,S,\hat{S}_R,\hat{S}_D}P_{X_2|V,\hat{S}_R}W_{Y_2,Y_3|X_1,X_2,S}
\end{align}
and satisfying $U_1 \leftrightarrow (V,U,S,\hat{S}_R,\hat{S}_D) \leftrightarrow U_R$ is a Markov chain and 
\begin{align}
0 &< I(V;Y_3,\hat{S}_D) - I(V;\hat{S}_R)\nonumber\\
0 &\leq I(U;Y_2|V,\hat{S}_R)-I(U;S,\hat{S}_D|V,\hat{S}_R)\nonumber\\
0 &\leq I(U_1;Y_3|U,V,\hat{S}_D)-I(U_1;S,\hat{S}_R|U,V,\hat{S}_D).
\end{align}
\end{corollary}

\vspace{0.4cm}

The proof of Corollary \ref{Corollary__LowerBound1__DiscreteChannel} follows by a fair extension of that of Theorem \ref{Theorem__LowerBound1__DiscreteChannel}, and so we omit it here for brevity.

\vspace{0.2cm}

\begin{remark}\label{remark__comparison-to-EURASIP}
In the coding scheme of Corollary~\ref{Corollary__LowerBound1__DiscreteChannel}, if the source sends no descriptions of the state to the relay and destination, i.e., $\hat{S}_R=\hat{S}_D=\O$, the coding scheme reduces to a generalized Gel'fand-Pinsker binning scheme at the source that is combined with partial DF. In this case, the relay sends codewords that carry part of the information message and are independent of the channel states. The following achievable rate\footnote{We note that the achievable rate \eqref{AchievabeRatePartialDecodeAndForwardDiscreteMemorylessChannel} is slightly larger than that of \cite[Theorem 1]{ZV09b}.} is obtained from Corollary~\ref{Corollary__LowerBound1__DiscreteChannel} by setting $\hat{S}_R=\hat{S}_D=\O$, $U_R=U_D=\O$ and $V=X_2$ independent of $S$, as
\begin{align}
 & R \: = \:\: \max \min \big\{\:I(U;Y_2|X_2)+ I(U_1;Y_3|U,X_2) -I(U,U_1;S|X_2),\:\: I(U,U_1,X_2;Y_3)-I(U,U_1;S|X_2)\:\big\}
 \label{AchievabeRatePartialDecodeAndForwardDiscreteMemorylessChannel}
 \end{align}
 with the maximization over joint measures of the form
 \begin{align}
 &P_{S,U,U_1,X_1,X_2,Y_2,Y_3}=Q_{S}P_{X_2}P_{U|S,X_2}P_{U_1,X_1|U,S,X_2}W_{Y_2,Y_3|X_1,X_2,S}
 \label{MeasureForAchievabeRatePartialDecodeAndForwardDiscreteMemorylessChannel}
 \end{align}
 and satisfying
 \begin{align}
 0 &\leq I(U;Y_2|X_2)-I(U;S|X_2)\nonumber\\
 0 &\leq I(U_1;Y_3|U,X_2)-I(U_1;S|U,X_2)\nonumber\\
 0 &\leq I(U,U_1;Y_3|X_2)-I(U,U_1;S|X_2).
 \label{Theprem1__MeasureConstraints}
 \end{align}

\end{remark}

\subsection{Lower Bound on Channel Capacity: Analog Input Description}\label{secIII_subsecB}

The following theorem provides a lower bound on the capacity of the state-dependent general discrete memoryless RC with informed source.


\begin{theorem}\label{Theorem__LowerBound2__DiscreteChannel}
The capacity of the discrete memoryless state-dependent relay channel with informed source is lower bounded by
\begin{align}
R^{\text{lo}} = \max \: I(U;Y_3)-I(U;S)
\label{AchievableRate__Theorem__LowerBound2__DiscreteChannel}
\end{align}
subject to the constraint
\begin{align}
I(X;\hat{X}) &< I(U_R;Y_2)-I(U_R;S)-I(U_R;U|S)
\label{Constraint__AchievableRate__Theorem__LowerBound2__DiscreteChannel}
\end{align}
where maximization is over all joint measures on ${\mc S}\times{\mc U}\times{\mc U_R}\times{\mc X}_1\times{\mc X}_2\times{\mc X}\times\hat{\mc X}\times{\mc Y}_2\times{\mc Y}_3$ of the form
\begin{align}
&P_{S,U,U_R,X_1,X_2,X,\hat{X},Y_2,Y_3}\nonumber\\
&\qquad = Q_SP_{U,U_R|S}P_{X_1|U,U_R,S}P_{X|U,S}P_{\hat{X}|X}\mathds{1}_{X_2=\hat{X}}W_{Y_2,Y_3|X_1,X_2,S}.
\label{Measure__AchievableRate__Theorem__LowerBound2__DiscreteChannel}
\end{align}
\end{theorem}

\vspace{0.6cm}

\textbf{Proof:} The proof of Theorem \ref{Theorem__LowerBound2__DiscreteChannel} appears in Appendix \ref{appendixTheorem__LowerBound2__DiscreteChannel}.

\vspace{0.2cm}

\begin{remark}\label{remark5}
The rationale for the coding scheme which we use to obtain the lower bound in Theorem \ref{Theorem__LowerBound2__DiscreteChannel} is as follows. Had the relay known the message to be sent in each block and the state that corrupts the transmission in that block, then the relay generates its input using a collaborative Gel'fand-Pinsker scheme as in \cite{KSS04}.

\noindent For our model, the source knows the message that the relay should optimally send in each block (if the relay gets the message correctly). It also knows the state sequence that corrupts the transmission in that block. It can then generate the appropriate relay input vector that the relay would send had the relay known the message and the state. The source can send this vector to the relay \textit{ahead of time}, and if the relay can estimate it to high accuracy, then collaborative source-relay binning in the sense of \cite{KSS04} is readily realized for transmission from the source and relay to the destination.

\noindent More precisely, let us consider transmission in two adjacent blocks $i$ and $i+1$. In the beginning of block $i$, the source sends the information $w_i$ of the current block, and, in addition, describes to the relay the input $\dv x[i+1]$ that the relay should send in the next block $i+1$ had the relay known the message $w_{i+1}$ and the state $\dv s[i+1]$. Let $\hat{\dv x}[m_i]$ be a description of $\dv x[i+1]$. The information of block $i$ and the index $m_i$ which the source sends in block $i$ are precoded using binning against the state that controls transmission in the current block $i$. The vector $\dv x[i+1]$, however, is the input that the relay would send in the next block $i+1$ had the relay known the state $\dv s[i+1]$, and so is generated at the source using binning against the state $\dv s[i+1]$. The vector $\dv x[i+1]$, and its description which is sent to the relay during block $i$, are intended to combine coherently with the source transmission in block $i+1$.
\end{remark}

\begin{remark}\label{remark6}
In the scheme we described briefly in Remark \ref{remark5}, the relay needs only estimate the code vector $\dv x[i+1]$ sent by the source in block $i$, and transmit the obtained estimate in the next block $i+1$. For instance, the relay does not need know the information message $w_{i+1}$ that the estimated vector actually carries, let alone the state sequence $\dv s[i+1]$ that controls the channel in block $i+1$. Thus, from a practical viewpoint, this may be particularly convenient for communication with an oblivious relay. Transmission from the source terminal to the relay terminal can be regarded as that of an analog source which, in block $i$, produces a sequence $\dv x[i+1]$. This source has to be transmitted by the source terminal over a state-dependent channel and reconstructed at the relay terminal. The reconstruction error at the relay terminal influences the rate at which information can be decoded reliably at the destination by acting as an additional noise term. 

\end{remark}

\begin{remark}\label{remark7}
A block Markov encoding is used to establish Theorem \ref{Theorem__LowerBound2__DiscreteChannel}. In block $i$, the source transmits the message $w_i$ and the index $m_i$ of a description $\hat{\dv x}[m_i]$ of the input $\dv x[i+1]$. In Theorem \ref{Theorem__LowerBound2__DiscreteChannel}, the auxiliary random vector $U^n$ represents the Gel'fand-Pinsker vector associated with the information message and is binned against the state $S^n$; and the auxiliary random vector $U^n_R$ represents the Gel'fand-Pinsker vector associated with the description information ans is binned against $(U^n,S^n)$.  

\end{remark}

\subsection{Upper Bounds on Channel Capacity}\label{secIII_subsecC}

As we mentioned in Section~\ref{secI}, the relay does not know the states of the channel directly in our model, but it can potentially get some information about $S^n$ from the past received sequence from the informed source. More precisely, the input of the relay $X_{2,i}$ at time $i$ depends on the channel states through $Y_2^{i-1}=(Y_{2,1},\hdots,Y_{2,i-1})$ which in turn depends on these states through $S^{i-1}$ and the past source inputs  $X_{1,j}(W,S^n)$, $j=1,\hdots,i-1$. Further, because the source knows the states non-causally this dependence may even be non-causal. This aspect makes establishing non-trivial upper bounds on the capacity, i.e., bounds that are strictly better than the cut-set upper bound
\begin{align}
R^{\text{up}}_{\text{triv}} &= \max_{p(x_1,x_2|s)} \min \Big\{ I(X_1;Y_2,Y_3|S,X_2),\:\: I(X_1,X_2;Y_3|S)\Big\}
\label{CutSetUpperBound__GeneralDiscreteChannel}
\end{align}
not easy.

The following theorem provides an upper bound on the capacity of the state-dependent general discrete memoryless RC with informed source. 

\begin{theorem}\label{Theorem_UpperBound__DiscreteChannel}
The capacity of the discrete memoryless state-dependent relay channel with informed source is upper-bounded by
\begin{equation}
R^{\text{up}} = \max \: \min\: \{I(V;Y_2,Y_3|U,X_2)-I(V;S|U,X_2),\:\: I(V;Y_3)-I(V;S)\}
\label{UpperBound__DiscreteChannel}
\end{equation}
where the maximization is over measures of the form
\begin{align}
P_{S,U,V,X_1,X_2,Y_2,Y_3} &= Q_SP_{U|S}P_{X_2|U,S}P_{V,X_1|U,S}W_{Y_2,Y_3|X_1,X_2,S}.
\label{Measure__UpperBound__DiscreteChannel}
\end{align}
and $U \in \calU$, $V \in \calV$ are auxiliary random variables with
\begin{subequations}
\begin{align}
\label{BoundsOnCardinalityOfAuxiliaryRandonVariableU__UpperBound__DiscreteMemorylessChannel}
&|\mc U| \leq |\mc S||\mc X_1||\mc X_2|\\
&|\mc V| \leq \Big(|\mc S||\mc X_1||\mc X_2|\Big)^2,
\label{BoundsOnCardinalityOfAuxiliaryRandonVariableV__UpperBound__DiscreteMemorylessChannel}
\end{align}
\label{BoundsOnCardinalityOfAuxiliaryRandonVariables__UpperBound__DiscreteMemorylessChannel}
\end{subequations}
respectively.
\end{theorem}

\vspace{0.6cm}

\textbf{Proof:} The proof of Theorem~\ref{Theorem_UpperBound__DiscreteChannel} appears in Appendix~\ref{appendixTheorem__UpperBound__DiscreteChannel}.

\vspace{0.2cm}

\noindent Note that the relay input $X_2$ depends on the state $S$ in the measure \eqref{Measure__UpperBound__DiscreteChannel}, and this reflects our discussion above. Also, one can specialize the upper bound \eqref{UpperBound__DiscreteChannel} to the special model of Figure~\ref{ModelWithHyperSource} using the channel structure \eqref{ConditionaProbability__ModelWithHyperSource} and obtain an upper bound on the capacity of this model. Instead, we establish a better upper bound, by better exploiting the fact that the input component $X^n_{1R}$ that is heard at the relay does not know the state $S^n$ at all in this model, and that the relay output $Y^{i-1}_2$ is function of only the strictly causal part $S^{i-1}$ of the state in this case. The result is stated in the following theorem.

\begin{theorem}\label{Theorem__UpperBound__DiscreteModelWithHyperSource}
The capacity of the discrete memoryless state-dependent relay model of Figure~\ref{ModelWithHyperSource} is upper-bounded by
\begin{align}
R^{\text{up}} &= \max \min \:\Big\{I(X_{1R};Y_2|X_2,S),\: I(X_2;Y_3)\Big\}+I(X_{1D};Y_3|X_2,S)\Big\}
\label{UpperBound__DiscreteModelWithHyperSource}
\end{align}
where the maximization is over all joint measures of the form
\begin{align}
P_{S,X_{1R},X_{1D},X_2,Y_2,Y_3} &= Q_SP_{X_2}P_{X_{1R}|X_2}P_{X_{1D}|X_2,S}W_{Y_2|X_{1R},S}W_{Y_3|X_{1D},X_2,S}
\label{MeasureForUpperBound__DiscreteModelWithHyperSource}
\end{align}
\end{theorem}

\vspace{0.6cm}

\textbf{Proof:} The proof of Theorem \ref{Theorem__UpperBound__DiscreteModelWithHyperSource} appears in Appendix \ref{appendixTheorem__UpperBound__DiscreteModelWithHyperSource}.

\vspace{0.2cm}

\begin{remark}
We note that although the output $Y^{i-1}_2$ at the relay at time $i$ in the special case model of Figure~\ref{ModelWithHyperSource} can convey information only about the strictly causal part $S^{i-1}$ of the state, upper bounding the channel capacity is not trivial \textit{even} in this case. For a related somewhat simpler model, a two-user multiaccess channel with common and one individual messages, we have shown recently in \cite{ZPS11a, ZPS11b} that  strictly causal knowledge of the state at the encoder that sends only the common message \textit{can increase} the transmission rate of the other encoder in general \textit{even if this one knows the states non-causally} --- however, the capacity region is still to be characterized in general. For the special case model in Figure~\ref{ModelWithHyperSource}, it is not clear yet how the relay could exploit optimally the information about the state $S^{i-1}$ that is contained in $Y^{i-1}_2$. The second term of the minimization in \eqref{UpperBound__DiscreteModelWithHyperSource} upper-bounds the information that the source and the relay can send to the destination by 
\begin{equation}
I(X_2;Y_3)+I(X_{1D};Y_3|X_2,S)=I(X_{1D},X_2;Y_3|S)-I(X_2;S|Y_3),
\end{equation}
which is strictly better than the corresponding term in the cut-set upper bound \eqref{CutSetUpperBound__GeneralDiscreteChannel}.

\end{remark}

\section{The Gaussian RC with Informed Source}\label{secIV}

\subsection{System Model}\label{secIV_subsecA}

In this section, we consider a full-duplex state-dependent RC informed source in which the channel state and the noise are additive and Gaussian. In this model, the channel state can model an additive Gaussian interference which is assumed to be known (non-causally) to only the source. The channel outputs $Y_{2,i}$ and $Y_{3,i}$ at time instant $i$ for the relay and the destination, respectively, are related to the channel input $X_{1,i}$ from the source and $X_{2,i}$ from the relay, and the channel state $S_i$, by
\begin{subequations}
\begin{align}
Y_{2,i}&=X_{1,i}+S_i+Z_{2,i}\\
Y_{3,i}&=X_{1,i}+X_{2,i}+S_i+Z_{3,i}.
\end{align}
\label{ChannelModel__StateDependent__GaussianRC}
\end{subequations}
The channel state $S_i$ is zero mean Gaussian random variable with variance $Q$; and only the source knows the state sequence $S^n$ (non-causally). The noises $Z_{2,i}$ and $Z_{3,i}$ are zero mean Gaussian random variables with variances $N_2$ and $N_3$, respectively; and are mutually independent and independent from the state sequence $S^n$ and the channel inputs $(X^n_1,X^n_2)$.

We shall also consider an important special case of the general Gaussian model \eqref{ChannelModel__StateDependent__GaussianRC} for which our bounds will be more tight. In this special case, the source input $X_{1,i}=(X_{1R,i},X_{1D,i})$ with $X_{1R,i}$ independent of the channel state $S^n$, and the channel outputs $Y_{2,i}$ and $Y_{3,i}$ at time instant $i$ for the relay and the destination, respectively, are related to the channel inputs from the source and relay and the channel state $S_i$ by
\begin{subequations}
\begin{align}
Y_{2,i}&=X_{1R,i}+S_i+Z_{2,i}\\
Y_{3,i}&=X_{1D,i}+X_{2,i}+S_i+Z_{3,i}.
\end{align}
\label{GaussianChannelModel__RCwithHyperSource}
\end{subequations}

For the general model \eqref{ChannelModel__StateDependent__GaussianRC}, we consider the following individual power constraints on the average transmitted power at the source and the relay,
\begin{equation}
\sum_{i=1}^{n}X_{1,i}^2 \leq nP_1, \qquad \sum_{i=1}^{n}X_{2,i}^2 \leq nP_2.
\label{IndividualPowerConstraintsFullDuplexRegime}
\end{equation}
For the special case model \eqref{GaussianChannelModel__RCwithHyperSource}, we consider separate power constraints on the average transmitted power at the encoder components,
\begin{equation}
\sum_{i=1}^{n}X_{1R,i}^2 \leq nP_{1R}, \quad \sum_{i=1}^{n}X_{1D,i}^2 \leq nP_{1D}, \quad \sum_{i=1}^{n}X_{2,i}^2 \leq nP_2.
\label{IndividualPowerConstraintsModelwithHyperSource}
\end{equation}

The definition of a code for the Gaussian model is the same as that given in the discrete case, with the additional constraint that the channel inputs should satisfy the appropriate power constraint, \eqref{IndividualPowerConstraintsFullDuplexRegime} or \eqref{IndividualPowerConstraintsModelwithHyperSource}.

\subsection{Lower Bounds on Channel Capacity}\label{secIV_subsecB}

The following theorem provides a lower bound on the capacity of the state-dependent general Gaussian RC with informed source.

\begin{theorem}\label{Theorem__LowerBound2__GeneralGaussianChannel}
The capacity of the state-dependent Gaussian RC with informed source is lower-bounded by
\begin{align}
R^{\text{lo}}_{\text{G}} = \max \frac{1}{2}\log\Big(1+\frac{(\sqrt{\bar{\gamma}P_1}+\sqrt{P_2-D})^2}{N_3+D+{\gamma}P_1}\Big),
\label{LowerBound2__GeneralGaussianChannel}
\end{align}
where
\begin{align}
D &:= P_2\frac{N_2}{N_2+{\gamma}P_1}
\label{Distortion__LowerBound2__GeneralGaussianChannel}
\end{align}
and the maximization is over $\gamma \in [0,1]$.
\end{theorem}

\vspace{0.6cm}

\begin{remark}\label{remar11}
It is insightful to observe that the rate in Theorem \ref{Theorem__LowerBound2__GeneralGaussianChannel} does not depend on the strength of the state $S$. This makes the coding scheme appreciable, particularly for the case of arbitrary strong interference in which classical coding schemes suffer greatly from the strong interference unknown at the relay.\\
\end{remark}

\textbf{Outline of Proof of Theorem \ref{Theorem__LowerBound2__GeneralGaussianChannel}:} The result in Theorem \ref{Theorem__LowerBound2__DiscreteChannel} for the DM case can be extended to memoryless channels with discrete time and continuous alphabets using standard techniques \cite[Chapter 7]{G68}. The proof of Theorem \ref{Theorem__LowerBound2__GeneralGaussianChannel} follows through evaluation of the lower bound of Theorem \ref{Theorem__LowerBound2__DiscreteChannel} using the following jointly Gaussian input distribution. For $0 \leq \gamma \leq 1$, we let $X \sim \mc N(0,P_2)$ and $X_{1R} \sim \mc N(0,{\gamma}P_1)$, with $X$ jointly Gaussian with $S$ with $\mathbb{E}[XS]=0$; and $X_{1R}$ jointly Gaussian with $(S,X)$, with $\mathbb{E}[X_{1R}S]=\mathbb{E}[X_{1R}X]=0$. Also, for $0 \leq D \leq P_2$ given, we consider the test channel $\hat{X}=aX+\tilde{X}$, where  $a:=1-D/P_2$ and $\tilde{X}$ is a Gaussian random variable with zero mean and variance $\tilde{P}_2=D(1-D/P_2)$, independent from $X$ and $S$. Using this test channel, we calculate $\mathbb{E}[(X-\hat{X})^2]=D$ and $\mathbb{E}[\hat{X}^2]=P_2-D$.

\noindent We use the following choices of the auxiliary random variables in Theorem \ref{Theorem__LowerBound2__DiscreteChannel},

\begin{align}
U &= \Big(\sqrt{\frac{\bar{\gamma}P_1}{P_2}}+\sqrt{\frac{P_2-D}{P_2}}\Big)X+{\alpha}S\\
U_R &= X_{1R}+\alpha_R\Big(S+\frac{\sqrt{\bar{\gamma}P_1}}{\sqrt{\bar{\gamma}P_1}+\sqrt{P_2-D}}X\Big),
\label{JointSourceRelay__DPC__LowerBound2}
\end{align}
 where
\begin{align}
\alpha =& \frac{(\sqrt{\bar{\gamma}P_1}+\sqrt{P_2-D})^2}{(\sqrt{\bar{\gamma}P_1}+\sqrt{P_2-D})^2+(N_3+D+{\gamma}P_1)} \quad \text{and} \quad \alpha_R =\frac{{\gamma}P_1}{{\gamma}P_1+N_2}.
\end{align}

Through straightforward algebra, which we omit here for brevity, it can be shown that the evaluation of the lower bound of Theorem \ref{Theorem__LowerBound2__DiscreteChannel} using the above choice gives the lower bound in Theorem \ref{Theorem__LowerBound2__GeneralGaussianChannel}.

\textbf{Alternative Proof of Theorem \ref{Theorem__LowerBound2__GeneralGaussianChannel}:} The encoding and transmission scheme is as follows. For $0 \leq \gamma \leq 1$, let $X \sim \mc N(0,P_2)$ and $X_{1R} \sim \mc N(0,{\gamma}P_1)$, with $X$ jointly Gaussian with $S$ with $\mathbb{E}[XS]=0$; and $X_{1R}$ jointly Gaussian with $(S,X)$, with $\mathbb{E}[X_{1R}S]=\mathbb{E}[X_{1R}X]=0$. Also, let $0 \leq D \leq P_2$ be given, and consider the test channel $\hat{X}=aX+\tilde{X}$, where  $a:=1-D/P_2$ and $\tilde{X}$ is a Gaussian random variable with zero mean and variance $\tilde{P}_2=D(1-D/P_2)$, independent from $X$ and $S$. Using this test channel, we calculate $\mathbb{E}[(X-\hat{X})^2]=D$ and $\mathbb{E}[\hat{X}^2]=P_2-D$.

\noindent We use the two random variables $U$ and $U_R$ given by \eqref{JointSourceRelay__DPC__LowerBound2} to generate the auxiliary codewords $U_i$ and $U_{R,i}$ which we will use in the sequel.

\noindent As in the discrete case, a block Markov encoding is used. For each block $i$, let $\dv x[i]$ be a Gaussian signal which carries message $w_i \in [1,2^{nR}]$ and is obtained via a DPC considering $\dv s[i]$ as noncausal channel state information, as
\begin{equation}
\Big(\sqrt{\frac{\bar{\gamma}P_1}{P_2}}+\sqrt{\frac{P_2-D}{P_2}}\Big) \dv x[i] = \dv u[i] - {\alpha}\dv s[i],
\end{equation}
where the components of $\dv u[i]$ are generated i.i.d. using the auxiliary random variable $U$.

For every block $i$, the source quantizes $\dv x[w_i]$ into $\hat{\dv x}[m_i]$, where $m_i \in [1,2^{n\hat{R}}]$. Using the above test channel, the source can encode $\dv x[w_i]$ successfully at the quantization rate
\begin{align}
\hat{R} &=I(X;\hat{X})\nonumber\\
        &= \frac{1}{2}\log(\frac{P_2}{D}).
\label{QuantizationRate__LowerBound2__GeneralGaussianChannel__SourceCodingConstraint}
\end{align}

Let $m_i$ be the index associated with $\dv x[w_{i+1}]$. In the beginning of block $i$, the source sends a superposition of two Gaussian vectors,
\begin{equation}
\dv x_1[i]=\dv x_{1R}[m_i]+\sqrt{\frac{\bar{\gamma}P_1}{P_2}}\dv x[w_i].
\label{SourceInput__LowerBound2__GeneralGaussianChannel}
\end{equation}
In equation \eqref{SourceInput__LowerBound2__GeneralGaussianChannel}, the signal $\dv x_{1R}[m_i]$ carries message $m_{i}$ and is obtained via a DPC considering $(s[i],\dv x[w_i])$ as noncausal channel state information, as
\begin{align}
\dv x_{1R}[m_i] &=\dv u_R[i]-\alpha_R\Big(\dv s[i]+\sqrt{\frac{\bar{\gamma}P_1}{P_2}}\dv x[w_i]\Big),\label{JointSourceRelay__DPC__LowerBound2}
\end{align}
where the components of $\dv u_R[i]$ are generated i.i.d. using the auxiliary random variable $U_R$.

\noindent In the beginning of block $i$, the relay has decoded message $m_{i-1}$ correctly (this will be justified below) and sends
\begin{equation}
\dv x_2[i]=\frac{\sqrt{P_2}}{\sqrt{P_2-D}}\hat{\dv x}[m_{i-1}].
\end{equation}

For the decoding arguments at the source and the relay, we give simple arguments based on intuition (the rigorous decoding uses joint typicality). Also, since all the random variables are i.i.d., we sometimes omit the time index. The relay decodes the index $m_i$ from the received $\dv y_2[i]$ at the end of block $i$. Since signal $\dv x_{1R}[m_i]$ is precoded at the source against the interference caused by the information message $w_i$, decoding at the relay can be done reliably as long as $n$ is large and
\begin{align}
\hat{R} \leq \frac{1}{2}\log\Big(1+\frac{{\gamma}P_1}{N_2}\Big).
\label{QuantizationRate__LowerBound2__GeneralGaussianChannel__ChannelCodingConstraint}
\end{align}
The destination decodes message $w_i$ from the received $\dv y_3[i]$ at the end of block $i$, considering signal $\dv x_{1R}[m_i]$ as unknown noise, with
\begin{align}
\dv y_3[i] &= \dv x_1[i] + \dv x_2[i] + \dv s[i] +\dv z_3[i]\nonumber\\
&= \Big(\sqrt{\frac{\bar{\gamma}P_1}{P_2}}\dv x[w_i]+\sqrt{\frac{P_2}{P_2-D}}\hat{\dv x}[m_{i-1}]\Big)+\dv s[i]+(\dv z_3[i]+\dv x_{1R}[m_i]).
\label{OutputDestination__LowerBound2__GeneralGaussianChannel}
\end{align}
Let now ${\dv x}'[i]$ be the optimal linear estimator of $\Big(\sqrt{\frac{\bar{\gamma}P_1}{P_2}}\dv x[w_i]+\sqrt{\frac{P_2}{P_2-D}}\hat{\dv x}[m_{i-1}]\Big)$ given $\dv x[w_i]$ under minimum mean square error criterion, and $\dv e_{\dv x}[i]$ the resulting estimation error. The estimator $\hat\hat{\dv x}[i]$ and the estimation error $\dv e_{\dv x}[i]$ are given by
\begin{align}
{\dv x}'[i] &= \mathbb{E}\Big[\sqrt{\frac{\bar{\gamma}P_1}{P_2}}\dv x[w_i]+\sqrt{\frac{P_2}{P_2-D}}\hat{\dv x}[m_{i-1}]|\dv x[i]\Big]\nonumber\\
            &= \Big(\sqrt{\frac{\bar{\gamma}P_1}{P_2}}+\sqrt{\frac{P_2-D}{P_2}}\Big)\dv x[w_i]\\
\dv e_{\dv x}[i] &= \sqrt{\frac{P_2}{P_2-D}}\hat{\dv x}[m_{i-1}]-\sqrt{\frac{P_2-D}{P_2}}\dv x[w_i].
\end{align}

\noindent We can alternatively write the output $\dv y_3[i]$ in \eqref{OutputDestination__LowerBound2__GeneralGaussianChannel} as
\begin{align}
\dv y_3[i] = \xi\dv x[w_i]+\dv s[i]+\Big(\dv z_3[i]+\dv e_{\dv x}[i]+\dv x_{1R}[m_i]\Big),
\label{EquivalentOutputDestination__LowerBound2__GeneralGaussianChannel}
\end{align}
where
\begin{equation}
\xi := \sqrt{\frac{\bar{\gamma}P_1}{P_2}}+\sqrt{\frac{P_2-D}{P_2}}
\end{equation}
and $\dv e_{\dv x}[i]$ is Gaussian with variance $D$ and is independent of $\dv x[w_i]$ and $\dv s[i]$.

\noindent Now, considering the equivalent form \eqref{EquivalentOutputDestination__LowerBound2__GeneralGaussianChannel} of the output $\dv y_3[i]$, it is easy to see that the destination can decode message $w_i$ correctly at the end of block $i$ as long as $n$ is large and
\begin{align}
R &\leq I(U;Y_3)-I(U;S)\nonumber\\
  &= \frac{1}{2}\log\Big(1+\frac{(\sqrt{\bar{\gamma}P_1}+\sqrt{P_2-D})^2}{N_3+D+{\gamma}P_1}\Big).
\label{Decoding__at__Destination__LowerBound2__GeneralGaussianChannel}
\end{align}

\noindent Furthermore, combining \eqref{QuantizationRate__LowerBound2__GeneralGaussianChannel__SourceCodingConstraint} and \eqref{QuantizationRate__LowerBound2__GeneralGaussianChannel__ChannelCodingConstraint} we get
\begin{equation}
D \geq P_2\frac{N_2}{N_2+{\gamma}P_1}.
\label{MinimumDistortion__LowerBound2__GeneralGaussianChannel}
\end{equation}
Finally, observing that the RHS of \eqref{Decoding__at__Destination__LowerBound2__GeneralGaussianChannel} decreases with $D$, we obtain \eqref{LowerBound2__GeneralGaussianChannel} by taking the equality in \eqref{MinimumDistortion__LowerBound2__GeneralGaussianChannel} and maximizing the RHS of \eqref{Decoding__at__Destination__LowerBound2__GeneralGaussianChannel} over $\gamma \in [0,1]$. This completes the proof. \begin{flushright} \qed \end{flushright}

We now turn to establish a lower bound on the capacity of the state-dependent Gaussian RC using the idea of state transmission. In this section, the source describes the channel state to only the relay. The relay guesses the information message and the transmitted state description and then transmits the message cooperatively with the source using binning against the state estimate, in a manner similar to that we described for the coding scheme for Theorem \ref{Theorem__LowerBound1__DiscreteChannel}.

For convenience we define the following quantities $\tilde{Q}_S(\cdot)$ and $R(\cdot)$ which we will use throughout the remaining sections.

\begin{definition}\label{definition1} Let
\begin{align}
\tilde{Q}_S(t,Q,D) &:=(1-t)^2Q-t(t-2)D\nonumber\\
R(\alpha,P,Q,N) &:= \frac{1}{2}\log\Big(\frac{P(P+Q+N)}{PQ(1-\alpha)^2+N(P+\alpha^2Q)}\Big)\nonumber
\end{align}
for non-negative $t, D, P, Q, N$, and $\alpha \in \mathbb{R}$.
\end{definition}

The following theorem provides a lower bound on the capacity of the state-dependent general Gaussian RC with informed source.

\begin{theorem}\label{Theorem__LowerBound1__GeneralGaussianChannel}
The capacity of the state-dependent Gaussian RC with informed source is lower-bounded by
\begin{align}
R^{\text{lo}}_{\text{G}} = \max\:\min\:\Big\{&R\Big(\alpha,(1-\rho^2_{12}-\rho^2_{1s})\bar{\theta}P_{1r},\xiup^2\tilde{Q},N_2+{\theta}P_{1r}+P_{1d}\Big), \nonumber\\
&R\Big(\alpha,(1-\rho^2_{12}-\rho^2_{1s})\bar{\theta}P_{1r},\xiup^2\tilde{Q},N_3+{\theta}P_{1r}+P_{1d}\Big)+\frac{1}{2}\log\Big(1+\frac{(\rho_{12}\sqrt{\bar{\theta}P_{1r}}+\sqrt{P_2})^2}{N_3+\xiup^2D+{\theta}P_{1r}+(1-\rho^2_{12}-\rho^2_{1s})\bar{\theta}P_{1r}+P_{1d}}\Big)\Big\}\nonumber\\
&+ \frac{1}{2}\log\big(1+\frac{P_{1d}}{N_3+{\theta}P_{1r}}\big)
\label{LowerBound1__GeneralGaussianChannel}
\end{align}
where
\begin{align}
\label{Distortion__LowerBound1__GeneralGaussianChannel}
D &= Q\frac{N_2+P_{1d}}{N_2+{\theta}P_{1r}+P_{1d}}\\
\label{Variance__EquivalentState__LowerBound1__GeneralGaussianChannel}
\tilde{Q} &= \tilde{Q}_S(\alpha_2,Q,D), \quad \xiup = 1+\rho_{1s}\sqrt{\frac{\bar{\theta}P_{1r}}{Q}}\\
\alpha_2 =& \frac{(\rho_{12}\sqrt{\bar{\theta}P_{1r}}+\sqrt{P_2})^2}{(\rho_{12}\sqrt{\bar{\theta}P_{1r}}+\sqrt{P_2})^2+(1-\rho^2_{12}-\rho^2_{1s})\bar{\theta}P_{1r}+(N_3+\xi^2D+{\theta}P_{1r}+P_{1d})}
\label{ScaleFactors__DPCs__LowerBound1__GeneralGaussianChannel}
\end{align}
and the maximization is over $P_{1r} \geq 0$, $P_{1d} \geq 0$ such that $0 \leq P_{1r}+P_{1d} \leq P_1$, $\theta \in [0,1]$, $\rho_{12} \in [0,1]$ and $\rho_{1s} \in [-1,0]$ such that $0 \leq \rho^2_{12}+\rho^2_{1s} \leq 1$ and  $\alpha \in \mathbb{R}$ such that $R((1-\rho^2_{12}-\rho^2_{1s})\bar{\theta}P_{1r},\xiup^2\tilde{Q},N_2+{\theta}P_{1r}+P_{1d}) \geq 0$ and $R((1-\rho^2_{12}-\rho^2_{1s})\bar{\theta}P_{1r},\xiup^2\tilde{Q},N_3+{\theta}P_{1r}+P_{1d}) + 1/2\log(1+P_{1d}/(N_3+{\theta}P_{1r})) \geq 0$.
\end{theorem}

\vspace{0.6cm}

\textbf{Proof:} A formal proof of Theorem \ref{Theorem__LowerBound1__GeneralGaussianChannel} appears in Appendix \ref{appendixTheorem__LowerBound1__GeneralGaussianChannel}. 

An outline of proof of Theorem \ref{Theorem__LowerBound1__GeneralGaussianChannel} is as follows. The result in Theorem \ref{Theorem__LowerBound1__DiscreteChannel} for the DM case can be extended to memoryless channels with discrete time and continuous alphabets using standard techniques \cite[Chapter 7]{G68}. For the state-dependent Gaussian relay channel \eqref{ChannelModel__StateDependent__GaussianRC}, we evaluate the rate \eqref{AchievableRate__Theorem__LowerBound1__DiscreteChannel} with the following choice of input distribution. We choose $\hat{S}_D=\O$, $U_D=\O$. Furthermore, we consider the test channel $\hat{S}_R=aS+\tilde{S}_R$, where  $a:=1-D/Q$ and $\tilde{S}_R$ is a Gaussian random variable with zero mean and variance $\sigma^2_{\tilde{S}_R}=D(1-D/Q)$, independent from $S$. The random variable $X_2$ is Gaussian with zero mean and variance $P_2$, independent of $S$ and of $\hat{S}_R$. The random variable $X_1$ is composed of three parts, $X_1=X_{SR}+X_{WR}+X_{WD}$, where $X_{SR}$ is Gaussian with zero mean and variance ${\theta}P_{1r}$, for some $\theta \in [0,1]$, is independent of $S$, $\hat{S}_R$, $X_2$; and $X_{WR}=\rho_{1s}\sqrt{\bar{\theta}P_{1r}/Q}S+\rho_{12}\sqrt{\bar{\theta}P_{1r}/P_2}X_2+X'_{WR}$, where $X'_{WR}$ is Gaussian with zero mean and variance $(1-\rho^2_{12})\bar{\theta}P_{1r}$, for some $\rho_{12} \in [0,1]$ and $\rho_{1s} \in [-1,0]$ and is independent of $X_{SR}$, $X_2$ and $(S,\hat{S}_R)$; and $X_{WD}$ is a Gaussian with zero mean and variance $P_{1d}$, chosen independently from all the other variables. The auxiliary random variables are chosen as
\begin{subequations}
\begin{align}
V &= \Big(\rho_{12}\sqrt{\frac{\bar{\theta}P_{1r}}{P_2}}+1\Big)X_2+\alpha_2\Big(\rho_{1s}\sqrt{\frac{\bar{\theta}P_{1r}}{Q}}+1\Big)\hat{S}_R \\
U &= X'_{WR} +\alpha \xiup (S-\alpha_2\hat{S}_R)\\
U_1 &= X_{WD} +\frac{P_{1d}}{P_{1d}+N_3+{\theta}P_{1r}}\xiup(1-\alpha)(S-\alpha_2\hat{S}_R)\\
U_R &= X_{SR}+\frac{{\theta}P_{1r}}{{\theta}P_{1r}+N_2+P_{1d}}(1-\alpha)S
\end{align}
\label{RandomVariables__LowerBound1__GeneralGaussianChannel}
\end{subequations}
with
\begin{subequations}
\begin{align}
\alpha_{2} &= \frac{(\rho_{12}\sqrt{\bar{\theta}P_{1r}}+\sqrt{P_2})^2}{(\rho_{12}\sqrt{\bar{\theta}P_{1r}}+\sqrt{P_2})^2+(1-\rho^2_{12}-\rho^2_{1s})\bar{\theta}P_{1r}+(N_3+\xi^2D+{\theta}P_{1r}+P_{1d})}\\
D &:= Q\frac{N_2+P_{1d}}{N_2+{\theta}P_{1r}+P_{1d}} \quad \text{and} \quad \xiup=1+\rho_{1s}\sqrt{\frac{\bar{\theta}P_{1r}}{Q}}.
\end{align}
\end{subequations}

\noindent Through straightforward algebra which is omitted for brevity, it can be shown that the evaluation of \eqref{AchievableRate__Theorem__LowerBound1__DiscreteChannel} with the aforementioned input distribution gives \eqref{LowerBound1__GeneralGaussianChannel}.

\begin{remark}\label{remark12}
The parameter $\alpha$ in Theorem \ref{Theorem__LowerBound1__GeneralGaussianChannel} stands for DPC's scale factor in precoding the information message against the interference on its way to the relay and to the destination. Because the model \eqref{ChannelModel__StateDependent__GaussianRC} has the links to the relay and to the destination corrupted  by noise terms with distinct variances, one cannot remove the effect of the interference on the two links simultaneously via one single DPC as in \cite{KELW07}. This explains why the parameter $\alpha$ is left to be optimized over in \eqref{LowerBound1__GeneralGaussianChannel}. However, in the spirit of \cite{KELW07}, one can improve the rate of Theorem \ref{Theorem__LowerBound1__GeneralGaussianChannel} by time sharing coding schemes that are similar to the one we employed for Theorem \ref{Theorem__LowerBound1__GeneralGaussianChannel} but with different inflation parameters tailored respectively for the link to the relay and the link to the destination, as in \cite{ZV09b}. 
\end{remark}

\subsection{Upper Bounds on Channel Capacity}\label{secIV_subsecC}

Similar to the general DM model, in the general Gaussian model \eqref{ChannelModel__StateDependent__GaussianRC} the relay does not know the states of the channel directly but can potentially get information about $S^n$ from the observed output sequence $Y_2^{i-1}$. Also, $Y_2^{i-1}$ may even contain information about future values of the state, and this makes establishing upper bounds on the capacity that are strictly better than the cut-set upper bound
\begin{align}
R^{\text{up}}_{\text{G}} &= \max_{p(x_1,x_2|s)} \min \Big\{ I(X_1;Y_2,Y_3|S,X_2),\:\: I(X_1,X_2;Y_3|S)\Big\}
\label{CutSetUpperBound__GeneralGaussianChannel}
\end{align}
more difficult.


\noindent Note that both $X_1$ and $X_2$ know the state $S$ in \eqref{CutSetUpperBound__GeneralGaussianChannel}. For the special case Gaussian model \eqref{GaussianChannelModel__RCwithHyperSource}, we establish an upper bound that is \textit{strictly} better than \eqref{CutSetUpperBound__GeneralGaussianChannel} by accounting for that the source input component $X_{1R,i}$ at time $i$ does not know the state $S^n$ at all and that the relay output $Y^{i-1}_2$ is function of only the strictly causal part of the state in this case. The following theorem states the corresponding result.

\begin{theorem}\label{Theorem__UpperBound__GaussianModelWithHyperSource}
The capacity of the state-dependent Gaussian relay model \eqref{GaussianChannelModel__RCwithHyperSource} is upper-bounded by
\begin{align}
R^{\text{up}}_{\text{G}} = \max \min \Bigg\{&\frac{1}{2}\log\Big(1+\frac{P_{1R}}{N_2}\Big)+\frac{1}{2}\log\Big(1+\frac{P_{1D}(1-\rho^2_{12}-\rho^2_{1s})}{N_3}\Big),\nonumber\\
&\frac{1}{2}\log\Big(1+\frac{(\sqrt{P_2}+\rho_{12}\sqrt{P_{1D}})^2}{P_{1D}(1-\rho^2_{12}-\rho^2_{1s})+(\sqrt{Q}+\rho_{1s}\sqrt{P_{1D}})^2+N_3}\Big)+\frac{1}{2}\log\Big(1+\frac{P_{1D}(1-\rho^2_{12}-\rho^2_{1s})}{N_3}\Big)\Bigg\},
\label{UpperBound__GaussianModelWithHyperSource}
\end{align}
where the maximization is over parameters $\rho_{12} \in [0,1]$, $\rho_{1s} \in [-1,0]$ such that
\begin{equation}
\rho^2_{12}+\rho^2_{1s} \leq 1.
\label{MaximizationRangeUpperBound__GaussianModelWithHyperSource}
\end{equation}
\end{theorem}

\vspace{0.6cm}

\noindent \textbf{Proof:} The proof of Theorem \ref{Theorem__UpperBound__GaussianModelWithHyperSource} appears in Appendix \ref{appendixTheorem__UpperBound__GaussianModelWithHyperSource}.

\vspace{0.2cm}

\begin{remark}\label{remark10}
Similar to in the DM case, the upper bound in Theorem~\ref{Theorem__UpperBound__GaussianModelWithHyperSource} improves upon the cut-set upper bound through the second term of the minimization. The second term of the minimization is strictly tighter than that of the cut-set upper bound because it accounts for the rate loss incurred by not knowing the state $S^n$ at all at the source encoder component $X_{1R,i}$ that is heard at the relay and that the relay output $Y^{i-1}_2$ can depend on the state only strictly-causally in this case. Further, investigating closely the proof in Appendix \ref{appendixTheorem__UpperBound__GaussianModelWithHyperSource}, it can be seen that, by opposition to the corresponding DM case, the relay ignores completely any information about the state in the multiaccess part of \eqref{UpperBound__GaussianModelWithHyperSource}. 
\end{remark}

\subsection{Capacity for Some Special Cases}\label{secIV_subsecD}

In this section, we characterize the capacity for some special Gaussian models. 

\subsubsection{Capacity for some special cases}\label{secIV_subsecD_subssubsec1}

An important special case of \eqref{GaussianChannelModel__RCwithHyperSource} is when the interference affects only the channel to the destination, i.e.,
\begin{subequations}
\begin{align}
Y_{2,i}&=X_{1R,i}+Z_{2,i}\\
Y_{3,i}&=X_{1D,i}+X_{2,i}+S_i+Z_{3,i}.
\end{align}
\label{DegenerateGaussianChannelModel1__RCwithHyperSource}
\end{subequations}
In this case, the upper bound in Theorem~\ref{Theorem__UpperBound__GaussianModelWithHyperSource} is tight. The following theorem characterizes the channel capacity in this case.

\begin{theorem}\label{Theorem__Capacity__FirstDegenerateGaussianModelWithHyperSource}
The capacity of the state-dependent Gaussian relay model \eqref{DegenerateGaussianChannelModel1__RCwithHyperSource} is given by 
\begin{align}
C_{\text{G}} = \max \min \Bigg\{&\frac{1}{2}\log\Big(1+\frac{P_{1R}}{N_2}\Big)+\frac{1}{2}\log\Big(1+\frac{P_{1D}(1-\rho^2_{12}-\rho^2_{1s})}{N_3}\Big),\nonumber\\
&\frac{1}{2}\log\Big(1+\frac{(\sqrt{P_2}+\rho_{12}\sqrt{P_{1D}})^2}{P_{1D}(1-\rho^2_{12}-\rho^2_{1s})+(\sqrt{Q}+\rho_{1s}\sqrt{P_{1D}})^2+N_3}\Big)+\frac{1}{2}\log\Big(1+\frac{P_{1D}(1-\rho^2_{12}-\rho^2_{1s})}{N_3}\Big)\Bigg\},
\label{Capacity__FirstDegenerateGaussianModelWithHyperSource}
\end{align}
where the maximization is over parameters $\rho_{12} \in [0,1]$ and $\rho_{1s} \in [-1,0]$ such that
\begin{equation}
\rho^2_{12}+\rho^2_{1s} \leq 1.
\end{equation}
\end{theorem}

\vspace{0.6cm}

\textbf{Proof:} The proof of Theorem~\ref{Theorem__Capacity__FirstDegenerateGaussianModelWithHyperSource} appears in Appendix~\ref{appendixTheorem__Capacity__FirstDegenerateGaussianModelWithHyperSource}.

\vspace{0.2cm}

\begin{remark}
The computation of the lower bound \eqref{LowerBound__DiscreteModelWithHyperSource} in the proof of Theorem~\ref{Theorem__Capacity__FirstDegenerateGaussianModelWithHyperSource} for the model \eqref{GaussianChannelModel__RCwithHyperSource} gives the following rate
\begin{align}
R^{\text{lo}}_{\text{G}} = \max \min \Bigg\{&\frac{1}{2}\log\Big(1+\frac{P_{1R}}{N_2+Q}\Big)+\frac{1}{2}\log\Big(1+\frac{P_{1D}(1-\rho^2_{12}-\rho^2_{1s})}{N_3}\Big),\nonumber\\
&\frac{1}{2}\log\Big(1+\frac{(\sqrt{P_2}+\rho_{12}\sqrt{P_{1D}})^2}{P_{1D}(1-\rho^2_{12}-\rho^2_{1s})+(\sqrt{Q}+\rho_{1s}\sqrt{P_{1D}})^2+N_3}\Big)+\frac{1}{2}\log\Big(1+\frac{P_{1D}(1-\rho^2_{12}-\rho^2_{1s})}{N_3}\Big)\Bigg\},
\label{LowerBound__GaussianModelWithHyperSource}
\end{align}
where the maximization is over parameters $\rho_{12} \in [0,1]$ and $\rho_{1s} \in [-1,0]$ such that
\begin{equation}
\rho^2_{12}+\rho^2_{1s} \leq 1.
\end{equation}
The achievable rate \eqref{LowerBound__GaussianModelWithHyperSource} differs from the upper bound \eqref{UpperBound__GaussianModelWithHyperSource} in Theorem~\ref{Theorem__UpperBound__GaussianModelWithHyperSource} only through the first logarithm term in which the state is taken as unknown noise. Substituting $\rho:=\rho_{1s}$ and $\zeta:=1-\rho^2_{12}-\rho^2_{1s}$ in \eqref{UpperBound__GaussianModelWithHyperSource} and \eqref{LowerBound__GaussianModelWithHyperSource}, it is easy to see that if $P_{1R}$, $P_{1D}$, $P_2$, $Q$, $N_2$ and $N_3$ satisfy 
\begin{equation}
N_2 \leq \max_{\zeta \in [0,1], \: \rho \in [-1,0]} \:\: \frac{P_{1R}[P_{1D}\zeta+(\sqrt{Q}+\rho\sqrt{P_{1D}})^2+N_3]}{(\sqrt{P_2}+\sqrt{1-\zeta-\rho^2}\sqrt{P_{1D}})^2}-Q
\end{equation}
then the channel capacity is given by
\begin{equation}
\mc C_{\text{G}} = \max_{\zeta \in [0,1], \: \rho \in [-1,0]} \frac{1}{2}\log\Big(1+\frac{(\sqrt{P_2}+\sqrt{1-\zeta-\rho^2}\sqrt{P_{1D}})^2}{P_{1D}\zeta+(\sqrt{Q}+\rho\sqrt{P_{1D}})^2+N_3}\Big)+\frac{1}{2}\log\Big(1+\frac{P_{1D}\zeta}{N_3}\Big).
\end{equation}
\begin{flushright} \qed \end{flushright}
\end{remark}

\vspace{0.2cm}

Let us now consider another another special case of \eqref{ChannelModel__StateDependent__GaussianRC}, in which $X_1=(X_{1R},X_{1D})$ with average power constraint $\sum_{i=1}^{n} X^2_{1,i} \leq nP_1$ on $X^n_1$, $Y_3=(Y^{(1)}_3,Y^{(2)}_3)$ and the conditional distribution $W_{Y_3|X_{1D},S,X_2}$ factorizes as $W_{Y^{(1)}_3|X_2}W_{Y^{(2)}_3|X_{1D},S}$, 
\begin{subequations}
\begin{align}
Y_{2,i} &= X_{1R,i}+S_i+Z_{2,i}\\
Y^{(1)}_{3,i} &= X_{1D,i}+S_i+Z^{(1)}_{3,i}\\
Y^{(2)}_{3,i} &= X_{2,i}+Z^{(2)}_{3,i},
\end{align}
\label{SecondDegenerateGaussianChannelModel__RCwithHyperSource}
\end{subequations}
where the noises $Z^{(1)}_{3,i}$ and $Z^{(2)}_{3,i}$ are zero mean Gaussian random variables with variances $N_3$, and are mutually independent and independent from the state sequence $S^n$, the source input $X^n_1=(X^n_{1R},X^n_{1D})$ and the relay input $X^n_2$.

\begin{corollary}\label{Corollary__Capacity__SecondDegenerateGaussianModelWithHyperSource}
The capacity of the state-dependent Gaussian relay model \eqref{SecondDegenerateGaussianChannelModel__RCwithHyperSource} is given by
\begin{align}
& \mc C_{\text{G}} = \max \: \min \big\{\frac{1}{2}\log\big(1+\frac{{\gamma}P_1}{N_2}\big),\frac{1}{2}\log(1+\frac{P_2}{N_3})\big\}+\frac{1}{2}\log\big(1+\frac{(1-\gamma)P_1}{N_3}\big),
\label{Capacity__SecondDegenerateGaussianModelWithHyperSource}
\end{align}
where the maximization is over $\gamma \in [0,1]$.
\end{corollary}

The proof of Corollary \ref{Corollary__Capacity__SecondDegenerateGaussianModelWithHyperSource} follows by specializing the cut-set upper bound to the model \eqref{SecondDegenerateGaussianChannelModel__RCwithHyperSource} and then observing that this upper bound can actually be attained using a combination of binning and generalized block Markov scheme where we let $X_{1R}$ and $X_{1D}$ to be zero-mean Gaussian with variances ${\gamma}P_1$ and $(1-\gamma)P_1$, respectively, for some $0 \leq \gamma \leq 1$, independent of $S$ and $X_2$; $X_2$ is zero-mean Gaussian with variance $P_2$ independent of $S$; and $X_{1R}$ and $X_{1D}$ obtained with standard DPCs for the links to the relay and to the receiver component $Y^{(3)}_2$, respectively. The source sends information to the receiver via the relay through the dirty paper coded $X_{1R}$, and independent information via the direct link through the dirty paper coded $X_{1D}$.

\subsubsection{Analysis of Some Extreme Cases}\label{secIV__subsecD_subsubsec2}

We now summarize the behavior of some of the developed lower and upper bounds in some extreme cases.
\begin{itemize}
\item[1)] If $N_2 \longrightarrow 0$, e.g, the relay is located spatially very close to the source, the lower bound of Theorem \ref{Theorem__LowerBound2__GeneralGaussianChannel} and the cut-set upper bound \eqref{CutSetUpperBound__GeneralGaussianChannel} tend asymptotically to the same value
\begin{equation}
\mc C_{\text{G}} =\frac{1}{2}\log\Big(1+\frac{(\sqrt{P_1}+\sqrt{P_2})^2}{N_3}\Big)-o(1)
\label{LowerBound__ExtremeCase__CleanRelay}
\end{equation}
where $o(1) \longrightarrow 0$ as $N_2 \longrightarrow 0$.

\noindent Equation \eqref{LowerBound__ExtremeCase__CleanRelay} reflects the rationale for our coding scheme for the lower bound in Theorem \ref{Theorem__LowerBound2__GeneralGaussianChannel} which is tailored to be asymptotically optimal whenever the relay can learn with negligible distortion the input that it should send. In this case, the rate \eqref{LowerBound__ExtremeCase__CleanRelay} can be interpreted as the information between two transmit antennas which both know the channel state and one receive antenna. (For comparison, note that the coding scheme of Theorem \ref{Theorem__LowerBound1__GeneralGaussianChannel} achieves rate smaller than that of Theorem \ref{Theorem__LowerBound2__GeneralGaussianChannel} if $N_2 \longrightarrow 0$, because even though with the coding scheme of Theorem \ref{Theorem__LowerBound1__GeneralGaussianChannel} as well the relay obtains the state estimate at almost no expense if $N_2$ is arbitrarily small, it also needs to know the information message to perform binning, however). 

\item[2)] \textit{Arbitrarily strong channel state:} In the asymptotic case $Q \rightarrow \infty$, the capacity of the Gaussian model \eqref{GaussianChannelModel__RCwithHyperSource} is given by 
\begin{equation}
\mc C_{\text{G}} =\frac{1}{2}\log\Big(1+\frac{P_{1D}}{N_3}\Big).
\label{Capacity__ExtremeCase__RCwithHyperSource}
\end{equation}
This can be easily seen since both the upper bound of Theorem~\ref{Theorem__Capacity__FirstDegenerateGaussianModelWithHyperSource} and the lower bound \eqref{LowerBound__GaussianModelWithHyperSource} tend to the RHS of \eqref{Capacity__ExtremeCase__RCwithHyperSource} in this case, which is also clearly achievable through standard DPC at the source and by turning the relay off.

\noindent For the Gaussian model \eqref{ChannelModel__StateDependent__GaussianRC}, the lower bound of Theorem \ref{Theorem__LowerBound1__GeneralGaussianChannel} tends to
\begin{equation}
R^{\text{lo}}_{\text{G}} =\frac{1}{2}\log\Big(1+\frac{P_1}{\max(N_2,N_3)}\Big).
\end{equation}
The lower bound of Theorem \ref{Theorem__LowerBound2__GeneralGaussianChannel} does not depend on the strength of the channel state, as we indicated previously.

\item[3)] If $N_2 \longrightarrow \infty$, i.e., the link to the relay is broken or too noisy, the cut-set upper bound \eqref{CutSetUpperBound__GeneralGaussianChannel} and the lower of Theorem \ref{Theorem__LowerBound1__GeneralGaussianChannel} agree and give the channel capacity
\begin{align}
\mc C_{\text{G}} &= \frac{1}{2}\log(1+\frac{P_1}{N_3}).
\end{align}
Also, the lower and upper bounds on the capacity of the model \eqref{GaussianChannelModel__RCwithHyperSource} agree and give the channel capacity as the RHS of \eqref{Capacity__ExtremeCase__RCwithHyperSource}.

\noindent Note that, for the Gaussian model \eqref{ChannelModel__StateDependent__GaussianRC}, the lower of Theorem \ref{Theorem__LowerBound2__GeneralGaussianChannel} is suboptimal if $N_2 \longrightarrow \infty$, and tends to
\begin{align}
R^{\text{lo}}_{\text{G}} &= \frac{1}{2}\log(1+\frac{P_1}{N_3+P_2})
\label{LowerBound__ExtremeCase__NoisyRelay}
\end{align}
This is because the distortion in Theorem \ref{Theorem__LowerBound2__GeneralGaussianChannel} is equal to its maximum value $P_2$ in this case. Equation \eqref{LowerBound__ExtremeCase__NoisyRelay} reflects a limitation of our coding scheme for the lower bound in  Theorem \ref{Theorem__LowerBound2__GeneralGaussianChannel} if the relay fails to reconstruct the input described by the source. In this case, the input from the relay acts as additional noise at the destination, thus causing the cooperative transmission to perform less good than simple direct transmission. The achievable rate \eqref{LowerBound__ExtremeCase__NoisyRelay} is, however, still better than had the state been merely treated as unknown noise if $P_2 \leq Q$. (For comparison, note that the lower bound of Theorem \ref{Theorem__LowerBound1__GeneralGaussianChannel} vanishes if $N_2 \longrightarrow \infty$).



\end{itemize}

\section{Numerical Examples and Discussion}\label{secV}
      
In this section we discuss some numerical examples, for the  general Gaussian RC with informed source \eqref{ChannelModel__StateDependent__GaussianRC}, the model \eqref{GaussianChannelModel__RCwithHyperSource} and the special case \eqref{DegenerateGaussianChannelModel1__RCwithHyperSource}.

\begin{figure}[!h]
	\begin{center}
	\includegraphics[width=0.7\linewidth]{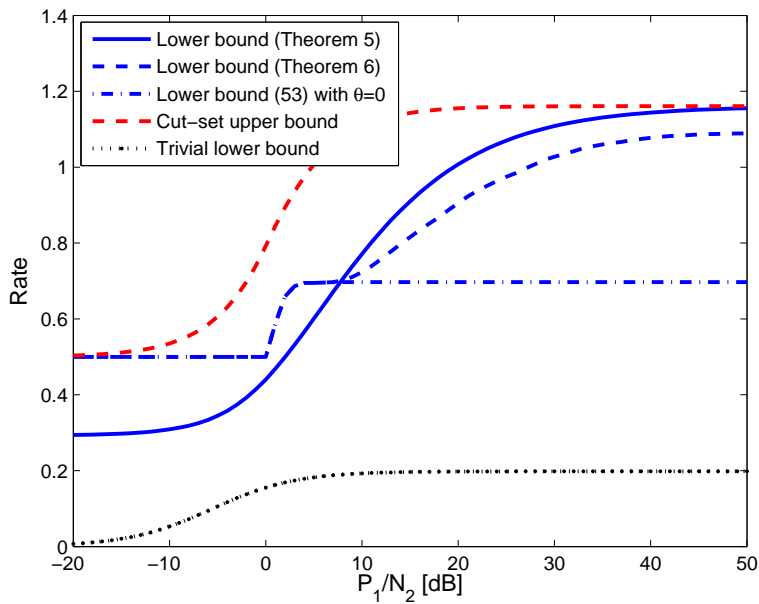}
	\end{center}
	\vspace{-1cm}
	\caption{Illustration of the lower bound of Theorem \ref{Theorem__LowerBound2__GeneralGaussianChannel} and lower bound of Theorem \ref{Theorem__LowerBound1__GeneralGaussianChannel} for the state-dependent General Gaussian RC with informed source \eqref{ChannelModel__StateDependent__GaussianRC} versus the SNR in the link source-to-relay. Numerical values are: $P_1=P_2=N_3=10$ dB and $Q=15$ dB.}
	\label{Fig1__IllustrativeExamples__General__RC}
\end{figure}


Figure \ref{Fig1__IllustrativeExamples__General__RC} illustrates the lower bound of Theorem \ref{Theorem__LowerBound2__GeneralGaussianChannel} and the lower bound of Theorem \ref{Theorem__LowerBound1__GeneralGaussianChannel} for the model \eqref{ChannelModel__StateDependent__GaussianRC}, as functions of the signal-to-noise-ratio (SNR) at the relay, i.e., $\text{SNR}=P_1/N_2$ (in decibels). Also shown for comparison are the cut-set upper bound had the state been known also at the relay and the destination and the trivial lower bound obtained by considering the channel state as unknown noise and implementing full-DF at the relay. In order to show the effect of describing the state to the relay, the figure also shows a special case of the lower bound of Theorem \ref{Theorem__LowerBound1__GeneralGaussianChannel} obtained by setting $\theta=0$ in \eqref{LowerBound1__GeneralGaussianChannel}, i.e., a Gaussian version of the achievable rate \eqref{AchievabeRatePartialDecodeAndForwardDiscreteMemorylessChannel} that we mentioned in Remark~\ref{remark__comparison-to-EURASIP}, and is a (slightly) improved version of \cite[Theorem3]{ZV09b}.

\noindent The figure shows that the lower bound of Theorem \ref{Theorem__LowerBound2__GeneralGaussianChannel} is asymptotically optimal at large $\text{SNR}$, and the lower bound of Theorem \ref{Theorem__LowerBound1__GeneralGaussianChannel} is asymptotically optimal at small $\text{SNR}$. This shows the relevance of transmitting to the relay only a description of the appropriate input that it should send upon sending to it a description of the state itself at large $\text{SNR}$. At moderate $\text{SNR}$, however, sending a description of the state to the relay may improve upon sending to it a description of the appropriate Gelf'and-Pinsker binned codeword that it should send --- (How the two bounds compare depends essentially on the strength of the state. For example, at large $\text{SNR}$, the stronger the state the larger the advantage of the lower bound of Theorem \ref{Theorem__LowerBound2__GeneralGaussianChannel} upon that of Theorem \ref{Theorem__LowerBound1__GeneralGaussianChannel}). Furthermore, the figure also shows that the lower bound of Theorem \ref{Theorem__LowerBound1__GeneralGaussianChannel} is better than that of \cite[Theorem3]{ZV09b}, thereby reflecting the utility of describing the state to the relay (recall that the coding scheme that we employed for the lower bound of Theorem \ref{Theorem__LowerBound1__GeneralGaussianChannel} involves also a partial cancellation of the state by the source to the relay, so that the relay benefits from it and the source benefits in turn). Figure \ref{Fig1__IllustrativeExamples__Degraded__RC} shows similar bounds computed for an example degraded Gaussian RC.

\begin{figure}[!htpb]
	\begin{center}
	\includegraphics[width=0.7\linewidth, height=0.5\linewidth]{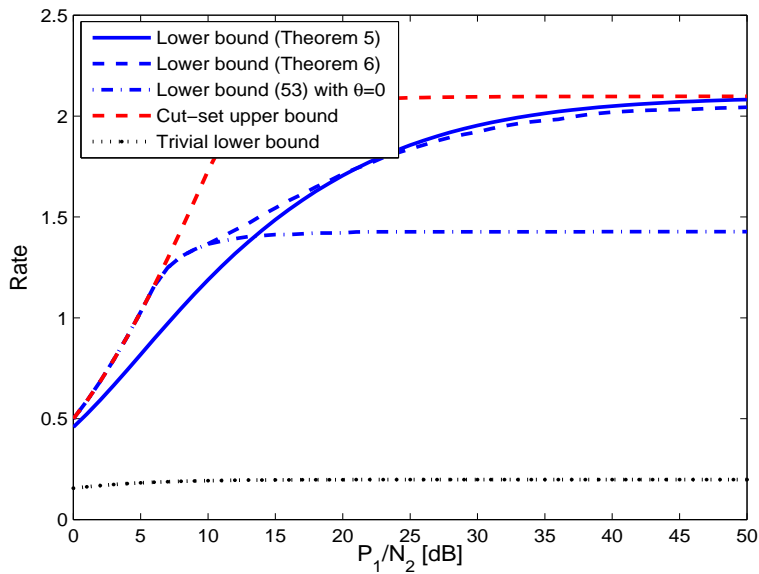}
	\end{center}
	\vspace{-1cm}
	\caption{Illustration of the lower bound of Theorem \ref{Theorem__LowerBound2__GeneralGaussianChannel} and lower bound of Theorem \ref{Theorem__LowerBound1__GeneralGaussianChannel} for an example state-dependent degraded Gaussian RC with informed source of \eqref{ChannelModel__StateDependent__GaussianRC}, versus the SNR in the link source-to-relay. Numerical values are: $P_1=10$ dB, $P_2=20$ dB, $Q=15$ dB, $N_3=10$ dB.}
	\label{Fig1__IllustrativeExamples__Degraded__RC}
\end{figure}

\begin{remark}\label{remark14}
The lower bound of Theorem \ref{Theorem__LowerBound2__GeneralGaussianChannel} is asymptotically close to optimal in $\text{SNR}$ as we mentioned in the "Extremes Cases Analysis" section and is visible from Figure \ref{Fig1__IllustrativeExamples__General__RC}. This is because the appropriate relay input, which is precoded at the source against the state and is encoded in a manner that it should combine coherently with the source transmission in next block, can be sent by the source to the relay at almost no expense in power and can be learned by the relay with negligible distortion in this case. One can be tempted to expect a similar behavior for the lower bound of Theorem \ref{Theorem__LowerBound1__GeneralGaussianChannel} since, for the latter as well, the relay can learn a "good" estimate of the state at almost no expense in source's power and with negligible distortion. This should not be, however, since our coding scheme for Theorem \ref{Theorem__LowerBound1__GeneralGaussianChannel} requires the relay to also decode the source's information message. Related to this aspect, the effect of the limitation which we mentioned in Remark \ref{remark12} is visible at large $\text{SNR}$ for this lower bound.  \hspace{0.5cm} \qed
\end{remark}

\begin{figure}[!htpb]

        \begin{minipage}[t]{\linewidth}
        \vspace{-0.5cm}
        \begin{center}
        \subfigure[]
        {
        \includegraphics[width=0.7\linewidth]{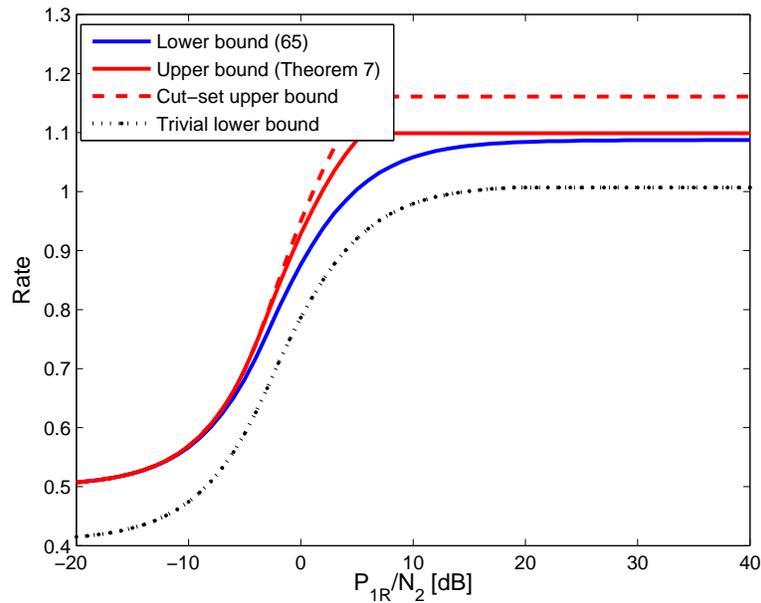}
        \label{Fig3a__IllustrativeExamples__RC__HyperSource}
        }
        \subfigure[]
        {
        \includegraphics[width=0.7\linewidth]{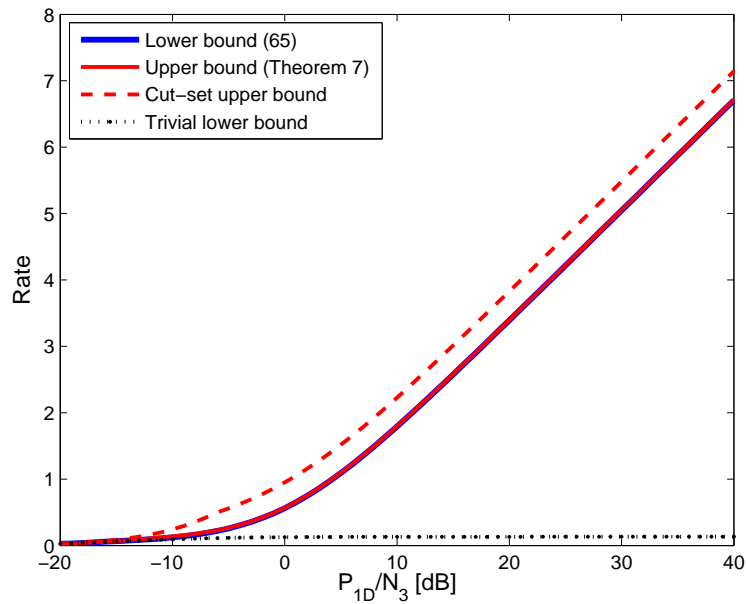}
  \label{Fig3b__IllustrativeExamples__RC__HyperSource}
        }
        \vspace{-0cm}
        \caption{Lower and upper bounds on the capacity of the state-dependent Gaussian RC with informed source \eqref{GaussianChannelModel__RCwithHyperSource}. (a) bounds versus the SNR $P_{1R}/N2$ in the link source-to-relay, for numerical values $P_{1R}=P_{1D}=P_2=N3=10$ dB, $Q=5$ and (b) bounds versus the SNR $P_{1D}/N3$ in the link source-to-destination $P_{1R}=P_{1D}=P_2=N2=10$ dB, $Q=20$ dB.}
        \label{Fig2__IllustrativeExamples__RC__HyperSource}
        \end{center}
        \end{minipage}
\end{figure}

Figure \ref{Fig2__IllustrativeExamples__RC__HyperSource} illustrates the upper bound~\eqref{UpperBound__GaussianModelWithHyperSource} of Theorem~\ref{Theorem__UpperBound__GaussianModelWithHyperSource} and the lower bound \eqref{LowerBound__GaussianModelWithHyperSource} for the special case model \eqref{GaussianChannelModel__RCwithHyperSource}. For comparison, the figure shows also the cut-set upper bound had the state been known also at the relay and the destination and the trivial lower bound obtained by considering the channel state as unknown noise and using a generalized block Markov coding scheme as in \cite{GZ05}. The curves are plotted against the signal-to-noise-ratio (SNR) at the relay, i.e., $\text{SNR}=P_{1R}/N_2$ (in decibels). Observe that the upper bound \eqref{UpperBound__GaussianModelWithHyperSource} is strictly better than the cut-set upper bound. The improvement is due to that the upper bound \eqref{UpperBound__GaussianModelWithHyperSource} accounts for some inevitable rate loss which is caused by not knowing the state at the relay, as we mentioned previously. Also, the improvement is visible mainly at small to relatively large values of $\text{SNR}$.  

\begin{figure}[!htpb]
	\begin{center}
	\includegraphics[width=0.7\linewidth, height=0.5\linewidth]{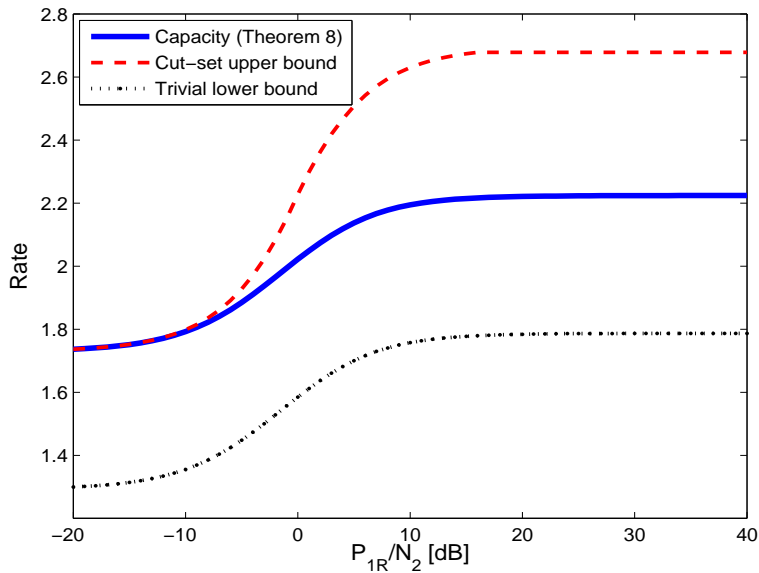}
	\end{center}
	\caption{Capacity of the state-dependent Gaussian RC model \eqref{DegenerateGaussianChannelModel1__RCwithHyperSource}, versus the SNR in the link source-to-relay. Numerical values are: $P_{1R}=10$ dB, $P_{1D}=P_2=20$ dB, $Q=10$ dB, $N_3=10$ dB.}
	\label{Fig__IllustrativeExamples__Degenerate__RC__HyperSource}
\end{figure}

Figure \ref{Fig__IllustrativeExamples__Degenerate__RC__HyperSource} illustrates the capacity result of \eqref{DegenerateGaussianChannelModel1__RCwithHyperSource} as given by Theorem \ref{Theorem__Capacity__FirstDegenerateGaussianModelWithHyperSource}, as function the SNR in the link source-to-relay of $P_{1R}/N_2$ (in decibels). Also shown for comparison are the cut-set upper bound and the trivial lower bound obtained by considering the channel state as unknown noise and using a generalized block Markov coding scheme as in \cite{GZ05}.



\vspace{-0.3cm}

\section{Conclusion}\label{secVI}

In this paper, we consider a state-dependent relay channel with the channel states available noncausally at only the source, i.e., neither at the relay nor at the destination. We refer to this communication model as \textit{state-dependent RC with informed source}. This setup may model some scenarios of node cooperation over wireless networks with some of the terminals equipped with cognition capabilities that enable estimating to high accuracy the states of the channel.

We investigate this problem in the discrete memoryless (DM) case and in the Gaussian case. For both cases, we derive lower and upper bounds on the channel capacity. A key feature of the model we study is that, assuming decode-and-forward relaying, the input of the relay should be generated using binning against the state that controls the channel in order to combat its effect and, at the same time, combine coherently with the source transmission. We develop two lower bounds on the capacity by using coding schemes which achieve this goal differently. In the first coding scheme, the source describes the channel state to the relay and to the destination, through a combined coding for multiple descriptions, binning and decode-and-forward scheme. The relay guesses an estimate of the transmitted information message and of the channel state and then utilizes the state estimate to perform cooperative binning with the source for sending the information message. The destination utilizes its output and the already recovered state to guess an estimate of the currently transmitted message and state description. In the second coding scheme, the source describes to the relay the appropriate input that the relay would send had the relay known the channel state. The relay then simply guesses this input and sends it in the appropriate subsequent block. The lower bound obtained with this scheme achieves close to optimal for some special cases. 

Furthermore, the upper bounds that we establish in the discrete memoryless and the memoryless Gaussian cases are not trivial and account for not knowing the state at the relay and destination. Also, considering a special case in which the source input has two components one of which is independent of the channel state, we show that our upper bound is strictly tighter than that obtained by assuming that the channel state is also available at the relay and the destination, i.e., the max-flow min-cut or cut-set upper bound, and it helps characterizing the rate loss due to the asymmetry caused by having the channel state available at only one source encoder component. Also, we characterize the channel capacity fully in some cases, including when the state does not affect the channel to the relay.

\vspace{-0.3cm}

\appendix
Throughout this section we denote the set of strongly jointly $\epsilon$-typical sequences \cite[Chapter 14.2]{CT91} with respect to the distribution $P_{X,Y}$ as $T_{\epsilon}^n(P_{X,Y})$.

\renewcommand{\theequation}{A-\arabic{equation}}
\setcounter{equation}{0}  
\subsection{Proof of Theorem~\ref{Theorem__LowerBound1__DiscreteChannel}}\label{appendixTheorem__LowerBound1__DiscreteChannel}

Consider the random coding scheme that we outlined in Section \ref{secIII}. We now analyse the average probability of error.

\textbf{Analysis of Probability of Error:} The average probability of error is given by
\begin{align}
\text{Pr}(\text{Error}) &=\sum_{\dv s \in \mc S^n}\text{Pr}(\dv s)\text{Pr}(\text{error}|\dv s)\nonumber\\
& \leq \sum_{\dv s\notin T_{\epsilon}^n(Q_S)}\text{Pr}(\dv s) + \sum_{\dv s \in T_{\epsilon}^n(Q_S)}\text{Pr}(\dv s)\text{Pr}(\text{error}|\dv s).
\label{AverageProbabilityOfError}
\end{align}
The first term, $\text{Pr}(\dv s \notin T_{\epsilon}^n(Q_S))$, on the RHS of \eqref{AverageProbabilityOfError} goes to zero as $n \rightarrow +\infty$, by the strong asymptotic equipartition property (AEP) \cite[p. 384]{CT91}. Thus, it is sufficient to upper bound the second term on the RHS of \eqref{AverageProbabilityOfError}.

We now examine the probabilities of the error events associated with the encoding and decoding procedures. The error event is contained in the union of the following error events; where the events $E_{1i}$ and $E_{2i}$ correspond to encoding errors at block $i$; the events $E_{ki}$, $k=3,\hdots,6$, correspond to decoding errors at the relay at block $i$; and the events $E_{ki}$, $k=7,\hdots,13$, correspond to decoding errors at the destination at block $i$.

\begin{itemize}
\item Let  $E_{1i}=E^{(1)}_{1i} \cup E^{(2)}_{1i} \cup E^{(3)}_{1i}$, with
\begin{align}
E^{(1)}_{1i} &= \Big\{(\dv s[i+2],\hat{\dv s}_R[\iota_{Ri}]) \notin T^n_{\epsilon}(P_{S,\hat{S}_R}), \:\:\:\textit{for all}\:\:\: \iota_{Ri} \in[1,2^{n\hat{R}_R}] \Big\}\nonumber\\
E^{(2)}_{1i} &= \Big\{(\dv s[i+2],\hat{\dv s}_D[\iota_{Di}]) \notin T^n_{\epsilon}(P_{S,\hat{S}_D}), \:\:\:\textit{for all}\:\:\: \iota_{Di} \in[1,2^{n\hat{R}_D}] \Big\}\nonumber\\
E^{(3)}_{1i} &= \Big\{(\dv s[i+2],\hat{\dv s}_R[\iota_{Ri}],\hat{\dv s}_D[\iota_{Di}]) \notin T^n_{\epsilon}(P_{S,\hat{S}_R,\hat{S}_D}), \:\:\:\textit{for all}\:\:\: (\iota_{Ri},\iota_{Di}) \in[1,2^{n\hat{R}_R}]\times[1,2^{n\hat{R}_D}] \Big\}.
\end{align}

From known results in rate distortion theory \cite[p. 336]{CT91}, it follows that $P(E^{(1)}_{1i}) \longrightarrow 0$ exponentionally with $n$ if $\hat{R}_R > I(S;\hat{S}_R)$. Similarly, $P(E^{(2)}_{1i}) \longrightarrow 0$ exponentionally with $n$ if $\hat{R}_D > I(S;\hat{S}_D)$. It remains to show that $P(E^{(3)}_{1i}) \longrightarrow 0$ exponentionally with $n$ if $\hat{R}_R + \hat{R}_D > I(S;\hat{S}_R,\hat{S}_D)+I(\hat{S}_R;\hat{S}_D)$, which we prove by following the arguments in \cite{GC82}. 

Define the random set,
\begin{align}
\mc A_{\dv s[i+2]} = \Big\{(\iota_{Ri},\iota_{Di}) \in[1,2^{n\hat{R}_R}]\times[1,2^{n\hat{R}_D}]\:s.t.:\Big(\dv s[i+2],\hat{\dv s}_R[\iota_{Ri}],\hat{\dv s}_D[\iota_{Di}]\Big) \in T^n_{\epsilon}(P_{S,\hat{S}_R,\hat{S}_D})\Big\}. 
\label{Encoding__Step__ElGamal__Binning}
\end{align}
Then, we have
\begin{align}
P(E^{(3)}_{1i}) & \leq \max_{\dv s[i+2] \in T^n_{\epsilon}(Q_S)} P(\|\mc A_{\dv s[i+2]}\|=0)
\end{align}
and, using Chebychev's inequality, 
\begin{align}
P(\|\mc A_{\dv s[i+2]}\|=0) &\leq P\Big(|\|\mc A_{\dv s[i+2]}\|-\mathbb{E}[\mc A_{\dv s[i+2]}]| > \epsilon \mathbb{E}[\mc A_{\dv s[i+2]}]\Big)\nonumber\\
&\leq \frac{\text{var}(\|\mc A_{\dv s[i+2]}\|)}{\epsilon^2(\mathbb{E}[\mc A_{\dv s[i+2]}])^2}.
\end{align}
Now, to obtain bounds on $\mathbb{E}[\mc A_{\dv s[i+2]}]$ and $\text{var}(\|\mc A_{\dv s[i+2]}\|)$, we define the indicator functions,
\begin{align}
\mathds{1}\Big((\iota_{Ri},\iota_{Di})\in \mc A_{\dv s[i+2]}\Big)=
\left\{
\begin{array}{rl}
1, & \text{if}\:\: (\iota_{Ri},\iota_{Di})\in \mc A_{\dv s[i+2]}\\
0, & \text{otherwise}.
\end{array}
\right.
\end{align}

The cardinality of the set $\mc A_{\dv s[i+2]}$ is given by
\begin{equation}
\|\mc A_{\dv s[i+2]}\| = \sum_{\iota_{Ri}=1}^{2^{n\hat{R}_R}} \sum_{\iota_{Di}=1}^{2^{n\hat{R}_D}} \mathds{1}\Big((\iota_{Ri},\iota_{Di})\in \mc A_{\dv s[i+2]}\Big).
\end{equation}
Or,
\begin{align}
\mathbb{E}\mathds{1}\Big((\iota_{Ri},\iota_{Di})\in \mc A_{\dv s[i+2]}\Big) & \geq 2^{-n[H(\hat{S}_R)+H(\hat{S}_D)-H(\hat{S}_R;\hat{S}_D|S)+\epsilon+2\delta(\epsilon)]}.
\end{align}
Thus,
\begin{align}
\mathbb{E}[\|\mc A_{\dv s[i+2]}\|] &= \sum_{\iota_{Ri}=1}^{2^{n\hat{R}_R}} \sum_{\iota_{Di}=1}^{2^{n\hat{R}_D}} \mathbb{E}\mathds{1}\Big((\iota_{Ri},\iota_{Di})\in \mc A_{\dv s[i+2]}\Big)\nonumber\\
&\geq 2^{n[\hat{R}_R+\hat{R}_D-H(\hat{S}_R)-H(\hat{S}_D)+H(\hat{S}_R;\hat{S}_D|S)-\epsilon-2\delta(\epsilon)]}.
\end{align}
Similarly, one can show that
\begin{align}
\text{var}(\|\mc A_{\dv s[i+2]}\|) &\leq 2^{n[\hat{R}_R+\hat{R}_D-H(\hat{S}_R)-H(\hat{S}_D)+H(\hat{S}_R;\hat{S}_D|S)-\epsilon+2\delta(\epsilon)]}.
\end{align}
Thus, the above Chebychev's inequality yields
\begin{align}
P(\|\mc A_{\dv s[i+2]}\|=0) &\leq \frac{1}{\epsilon^2}2^{-n[\hat{R}_R+\hat{R}_D-H(\hat{S}_R)-H(\hat{S}_D)+H(\hat{S}_R;\hat{S}_D|S)-\epsilon-6\delta(\epsilon)]} 
\end{align}
Then, $P(\|\mc A_{\dv s[i+2]}\|=0) \longrightarrow 0$ as $n \longrightarrow \infty$ if
\begin{align}
\hat{R}_R+\hat{R}_D &> H(\hat{S}_R)+H(\hat{S}_D)-H(\hat{S}_R;\hat{S}_D|S)-\epsilon-6\delta(\epsilon)\nonumber\\
&=I(S;\hat{S}_R,\hat{S}_D)+I(\hat{S}_R;\hat{S}_D)-\epsilon-6\delta(\epsilon).
\end{align}
\item Let $E_{2i}$ be the event that there is no pair $\Big(\dv u_R(w_{i-1},j^{\star}_{Vi},w_i,j^{\star}_{Ui},k_i,j_{Ri}),\dv u_D(w_{i-1},j^{\star}_{Vi},w_i,j^{\star}_{Ui},l_i,j_{Di})\Big)$ satisfying \eqref{Encoding__Step__Marton__Binning}, i.e., the set $\mc D_{\iota_{Ri}\iota_{Di}}$ is empty. 

Again, using Chebychev's inequality, it is easy to see that 
\begin{align}
P\Big(\|\mc D_{\iota_{Ri}\iota_{Di}}\|=0\Big) &\leq P\Big(|\|\mc D_{\iota_{Ri}\iota_{Di}}\|-\mathbb{E}[\mc D_{\iota_{Ri}\iota_{Di}}]| > \epsilon \mathbb{E}[\mc D_{\iota_{Ri}\iota_{Di}}]\Big)\nonumber\\
& \leq \frac{\text{var}(\|\mc D_{\iota_{Ri}\iota_{Di}}\|)}{\epsilon^2(\mathbb{E}[\mc D_{\iota_{Ri}\iota_{Di}}])^2}.
\end{align}
We obtain bounds on $\mathbb{E}[\mc D_{\iota_{Ri}\iota_{Di}}]$ and $\text{var}(\|\mc D_{\iota_{Ri}\iota_{Di}}\|)$ by proceeding in a way similar to that for the event $E_{1i}$. We define the indicator functions,
\begin{align}
\mathds{1}\Big(\Big(\dv u_R(w_{i-1},j^{\star}_{Vi},w_i,j^{\star}_{Ui},k_i,j_{Ri}),&\dv u_D(w_{i-1},j^{\star}_{Vi},w_i,j^{\star}_{Ui},l_i,j_{Di})\Big) \in \mc D_{\iota_{Ri}\iota_{Di}}\Big)=\nonumber\\
&{
\left\{
\begin{array}{rl}
1, & \text{if}\:\: \Big(\dv u_R(w_{i-1},j^{\star}_{Vi},w_i,j^{\star}_{Ui},k_i,j_{Ri}),\dv u_D(w_{i-1},j^{\star}_{Vi},w_i,j^{\star}_{Ui},l_i,j_{Di})\Big) \in \mc D_{\iota_{Ri}\iota_{Di}}\\
0, & \text{otherwise}. 
\end{array}
\right.
}
\end{align}
The cardinality of the set $\mc D_{\iota_{Ri}\iota_{Di}}$ is given by
\begin{equation}
\|\mc D_{\iota_{Ri}\iota_{Di}}\| = \sum_{k_i \in \mc B_{\iota_{Ri}},l_i \in \mc C_{\iota_{Di}}}\sum_{j_{Ri} \in J_R, j_{Di} \in J_D} \mathds{1}\Big(\Big(\dv u_R(w_{i-1},j^{\star}_{Vi},w_i,j^{\star}_{Ui},k_i,j_{Ri}),\dv u_D(w_{i-1},j^{\star}_{Vi},w_i,j^{\star}_{Ui},l_i,j_{Di})\Big) \in \mc D_{\iota_{Ri}\iota_{Di}}\Big).
\end{equation}
Thus, 
\begin{align}
\mathbb{E}[\mc D_{\iota_{Ri}\iota_{Di}}]&= \sum_{k_i \in \mc B_{\iota_{Ri}},l_i \in \mc C_{\iota_{Di}}}\sum_{j_{Ri} \in J_R, j_{Di} \in J_D} \mathbb{E}\mathds{1}\Big(\Big(\dv u_R(w_{i-1},j^{\star}_{Vi},w_i,j^{\star}_{Ui},k_i,j_{Ri}),\dv u_D(w_{i-1},j^{\star}_{Vi},w_i,j^{\star}_{Ui},l_i,j_{Di})\Big) \in \mc D_{\iota_{Ri}\iota_{Di}}\Big)\nonumber\\
& \geq \|\mc B_{\iota_{Ri}}\| \|\mc C_{\iota_{Di}}\|J_RJ_D2^{-n[I(U_R;S,\hat{S}_R,\hat{S}_D|U,V)+I(U_D;S,\hat{S}_R,\hat{S}_D|U,V)-I(U_R;U_D|U,VS,\hat{S}_R,\hat{S}_D)+o(1)]}\nonumber\\
&= 2^{n[R_R+R_D-\hat{R}_R-\hat{R}_D-I(U_R;U_D|U,V,S,\hat{S}_R,\hat{S}_D)-o(1)]}
\end{align}
where $o(1) \longrightarrow 0$ as $n \longrightarrow \infty$.

Evaluating the variance, we obtain that
\begin{equation}
\text{var}(\|\mc D_{\iota_{Ri}\iota_{Di}}\|) \leq 2^{n[R_R+R_D-\hat{R}_R-\hat{R}_D-I(U_R;U_D|U,V,S,\hat{S}_R,\hat{S}_D)+o(1)]}.
\end{equation}
Therefore, for sufficiently large $n$
\begin{equation}
P\Big(\|\mc D_{\iota_{Ri}\iota_{Di}}\|=0\Big) \leq \epsilon
\end{equation}
provided that \eqref{Condition__on__Sum__Rate__Encoding__Step__Marton__Binning} is true.
\item Let $E_{3i}$ be the event that $\dv u(w_{i-1},j^{\star}_{Vi},w_i,j^{\star}_{Ui})$, $\dv u_R(w_{i-1},j^{\star}_{Vi},w_i,j^{\star}_{Ui},k_i,j^{\star}_{Ri})$ are not jointly typical with $(\dv y_2[i],\hat{\dv s}_R[\iota_{Ri-2}])$ given $\dv v(w_{i-1},j^{\star}_{Vi})$. That is
\begin{align}
&E_{3i} = \Big\{\Big(\dv v(w_{i-1},j^{\star}_{Vi}),\dv u(w_{i-1},j^{\star}_{Vi},w_i,j^{\star}_{Ui}),\dv u_R(w_{i-1},j^{\star}_{Vi},w_i,j^{\star}_{Ui},k_i,j^{\star}_{Ri}),\dv y_2[i],\hat{\dv s}_R[\iota_{Ri-2}]\Big) \notin T_{\epsilon}^n(P_{V,U,U_R,Y_2,\hat{S}_R})\Big\}.
\end{align}
For $\dv v(w_{i-1},j^{\star}_{Vi})$, $\dv u(w_{i-1},j^{\star}_{Vi},w_i,j^{\star}_{Ui})$, $\dv u_R(w_{i-1},j^{\star}_{Vi},w_i,j^{\star}_{Ui},k_i,j^{\star}_{Ri})$, $\dv u_D(w_{i-1},j^{\star}_{Vi},w_i,j^{\star}_{Ui},l_i,j^{\star}_{Di})$ jointly typical with $\dv s[i]$, $\hat{\dv s}_R[\iota_{Ri-2}]$, $\hat{\dv s}_D[\iota_{Di-2}]$ and with the source input $\dv x_1[i]$ and the relay input $\dv x_2[i]$, we have $\text{Pr}(E_{3i}|E_{1i}^c,E_{2i}^c) \longrightarrow 0$ as $n \longrightarrow \infty$ by the Markov Lemma \cite[p. 436]{CT91}. 
\item  Let $E_{4i}$ be the event that $\dv u(w_{i-1},j^{\star}_{Vi},w'_i,j_{Ui})$, $\dv u_R(w_{i-1},j^{\star}_{Vi},w'_i,j_{Ui},k_i,j_{Ri})$ are jointly typical with $(\dv y_2[i],\hat{\dv s}_R[\iota_{Ri-2}])$ given $\dv v(w_{i-1},j^{\star}_{Vi})$, for some $w'_i \in [1,M]$, $j_{Ui} \in J_U$, $k_i \in [1,M_R]$ and $j_{Ri} \in J_R$, with $w'_i \neq w_i$. That is,
\begin{align}
E_{4i} = \Big\{&\exists \; w'_i \in [1,M],\;  j_{Ui} \in J_U, k_i \in [1,M_R], j_{Ri} \in J_R \; \text{s.t.:}\; w'_i \neq w_i, \nonumber\\
&\Big(\dv v(w_{i-1},j^{\star}_{Vi}),\dv u(w_{i-1},j^{\star}_{Vi},w'_i,j_{Ui}),\dv u_R(w_{i-1},j^{\star}_{Vi},w'_i,j_{Ui},k_i,j_{Ri}),\dv y_2[i],\hat{\dv s}_R[\iota_{Ri-2}]\Big) \in T_{\epsilon}^n(P_{V,U,U_R,Y_2,\hat{S}_R})\Big\}.
\end{align}
Using the union bound and standard arguments on jointly typical sequences, the probability of the event $E_{4i}$ conditioned on $E_{1i}^c$, $E_{2i}^c$, $E_{3i}^c$ can be easily bounded as
\begin{align}
\text{Pr}(E_{4i}|E_{1i}^c,E_{2i}^c,E_{3i}^c)  & \leq MJ_UM_RJ_R2^{-n[I(U,U_R;Y_2,\hat{S}_R|V)-\epsilon]}\nonumber\\
                            & =  2^{-n[I(U;Y_2|V,\hat{S}_R)-I(U;S,\hat{S}_D|V,\hat{S}_R)-R+4\epsilon]}.
\label{ProbabilityErrorEvent3ForAchievabilityInTheoreom1}
\end{align}
 Thus, $\text{Pr}(E_{4i}|E_{1i}^c,E_{2i}^c,E_{3i}^c) \longrightarrow 0$ as $n \longrightarrow \infty$ if $R< I(U;Y_2|V,\hat{S}_R)-I(U;S,\hat{S}_D|V,\hat{S}_R)$.
\item  Let $E_{5i}$ be the event that $\dv u(w_{i-1},j^{\star}_{Vi},w_i,j'_{Ui})$, $\dv u_R(w_{i-1},j^{\star}_{Vi},w_i,j'_{Ui},k_i,j_{Ri})$ are jointly typical with $(\dv y_2[i],\hat{\dv s}_R[\iota_{Ri-2}])$ given $\dv v(w_{i-1},j^{\star}_{Vi})$, for some $j'_{Ui} \in J_U$, $k_i \in [1,M_R]$, $j_{Ri} \in J_R$ with $j'_{Ui} \neq j^{\star}_{Ui}$. That is, 
\begin{align}
E_{5i} = \Big\{&\exists \; j'_{Ui} \in J_U, k_i \in [1,M_R], j_{Ri} \in J_R \; \text{s.t.}\; j'_{Ui} \neq j^{\star}_{Ui}, \nonumber\\
&\Big(\dv v(w_{i-1},j^{\star}_{Vi}),\dv u(w_{i-1},j^{\star}_{Vi},w_i,j'_{Ui}),\dv u_R(w_{i-1},j^{\star}_{Vi},w_i,j'_{Ui},k_i,j_{Ri}),\dv y_2[i],\hat{\dv s}_R[\iota_{Ri-2}]\Big) \in T_{\epsilon}^n(P_{V,U,U_R,Y_2,\hat{S}_R})\Big\}.
\end{align}
Conditioned on the events $E_{1i}^c$, $E_{2i}^c$, $E_{3i}^c$, $E_{4i}^c$,  the probability of the event $E_{5i}$ can be bounded using the union bound, as   
\begin{align}
\text{Pr}(E_{5i}|E_{1i}^c,E_{2i}^c,E_{3i}^c,E_{4i}^c)  & \leq J_UM_RJ_R2^{-n[I(U,U_R;Y_2,\hat{S}_R|V)-\epsilon]}\nonumber\\
                            & =  2^{-n[I(U;Y_2|V,\hat{S}_R)-I(U;S,\hat{S}_D|V,\hat{S}_R)+3\epsilon]}.
\label{ProbabilityErrorEvent3ForAchievabilityInTheoreom1}
\end{align}
Thus, $\text{Pr}(E_{5i}|E_{1i}^c,E_{2i}^c,E_{3i}^c,E_{4i}^c) \longrightarrow 0$ as $n \longrightarrow \infty$.

\item Let $E_{6i}$ be the event that $\dv u_R(w_{i-1},j^{\star}_{Vi},w_i,j^{\star}_{Ui},k'_i,j_{Ri})$ is jointly typical with $(\dv y_2[i],\hat{\dv s}_R[\iota_{Ri-2}])$ given $\dv v(w_{i-1},j^{\star}_{Vi})$, $\dv u(w_{i-1},j^{\star}_{Vi},w_i,j^{\star}_{Ui})$, for some $k'_i \in [1,M_R]$, $j_{Ri} \in J_R$ with $k'_i \neq k_i$. That is,
\begin{align}
E_{6i} = \Big\{&\exists \; k'_i \in [1,M_R], j_{Ri} \in J_R \; \text{s.t.}\; k'_i \neq k_i, \nonumber\\
&\Big(\dv v(w_{i-1},j^{\star}_{Vi}),\dv u(w_{i-1},j^{\star}_{Vi},w_i,j^{\star}_{Ui}),\dv u_R(w_{i-1},j^{\star}_{Vi},w_i,j^{\star}_{Ui},k'_i,j_{Ri}),\dv y_2[i],\hat{\dv s}_R[\iota_{Ri-2}]\Big) \in T_{\epsilon}^n(P_{V,U,U_R,Y_2,\hat{S}_R})\Big\}.
\end{align}
Conditioned on the events $E_{1i}^c$, $E_{2i}^c$, $E_{3i}^c$, $E_{4i}^c$, $E_{5i}^c$, the probability of the event $E_{6i}$ can be bounded using the union bound, as
\begin{align}
\text{Pr}(E_{6i}|E_{1i}^c,E_{2i}^c,E_{3i}^c,E_{4i}^c,E_{5i}^c)  & \leq M_RJ_R2^{-n[I(U_R;Y_2,\hat{S}_R|U,V)-\epsilon]}\nonumber\\
                            & =  2^{-n(4\epsilon)}.
\label{ProbabilityErrorEvent3ForAchievabilityInTheoreom1}
\end{align}
Thus, $\text{Pr}(E_{6i}|E_{1i}^c,E_{2i}^c,E_{3i}^c,E_{4i}^c,E_{5i}^c) \longrightarrow 0$ as $n \longrightarrow \infty$.

\item For decoding the triple $(\hat{w}_{i-1},\hat{j}_{Ui-1},\hat{l}_{i-1})$ and the index $\hat{j}_{Vi}$ at the destination, let $E_{7i}$ be the union of the following two events
\begin{align}
&E_{7i}^{(1)} = \Big\{\Big(\dv v(w_{i-2},j_V(\hat{\dv s}_R[\iota_{Ri-3}],w_{i-2})),\dv u(w_{i-2},j_V(\hat{\dv s}_R[\iota_{Ri-3}],w_{i-2}),w_{i-1},j^{\star}_{Ui-1}),\nonumber\\
&\hspace{1cm}\dv u_D(w_{i-2},j^{\star}_{Vi-1},w_{i-1},j^{\star}_{Ui-1},l_{i-1},j^{\star}_{Di-1}), \dv y_3[i-1],\hat{\dv s}_D[\iota_{Di-3}]\Big) \notin T_{\epsilon}^n(P_{V,U,U_D,Y_3,\hat{S}_D})\Big\}\nonumber\\
&E_{7i}^{(2)} = \Big\{\Big(\dv v(w_{i-1},j_V(\hat{\dv s}_R[\iota_{Ri-2}],w_{i-1})),\dv y_3[i],\hat{\dv s}_D[\iota_{Di-2}]\Big) \notin T_{\epsilon}^n(P_{V,Y_3,\hat{S}_D})\Big\}\nonumber.
\end{align}

For $\dv v(w_{i-2},j^{\star}_{Vi-1})$, $\dv u(w_{i-2},j^{\star}_{Vi-1},w_{i-1},j^{\star}_{Ui-1})$, $\dv u_R(w_{i-2},j^{\star}_{Vi-1},w_{i-1},j^{\star}_{Ui-1},k_{i-1},j^{\star}_{Ri-1})$, $\dv u_D(w_{i-2},j^{\star}_{Vi-1},w_{i-1},j^{\star}_{Ui-1},l_{i-1},j^{\star}_{Di-1})$ jointly typical with $\dv s[i-1]$, $\hat{\dv s}_R[\iota_{Ri-3}]$, $\hat{\dv s}_D[\iota_{Di-3}]$ and with the source input $\dv x_1[i-1]$ and the relay input $\dv x_2[i-1]$, we have $\text{Pr}(E_{7i}^{(1)}| \cap_{k=1}^{6} E_{ki}^c) \longrightarrow 0$ as $n \longrightarrow \infty$ by the Markov Lemma. Similarly, $\text{Pr}(E_{7i}^{(2)}|\cap_{k=1}^{6} E_{ki}^c) \longrightarrow 0$ as $n \longrightarrow \infty$. Thus, $\text{Pr}(E_{7i}|\cap_{k=1}^{6} E_{ki}^c) \longrightarrow 0$ as $n \longrightarrow \infty$.

\item For decoding the triple $(\hat{w}_{i-1},\hat{j}_{Ui-1},\hat{l}_{i-1})$ and the index $\hat{j}_{Vi}$ at the destination, let $E_{8i}$ be the event
\begin{align}
E_{8i} = \Big\{&\exists \; w'_{i-1} \in [1,M], j_{Ui-1} \in J_U, l_{i-1} \in [1,M_D], j_{Di-1} \in J_D, j_{Vi} \in J_V \; \text{s.t.:}\; w'_{i-1} \neq w_{i-1}, \nonumber\\
&\Big(\dv v(w_{i-2},j^{\star}_{Vi-1}),\dv u(w_{i-2},j^{\star}_{Vi-1},w'_{i-1},j_{Ui-1}),\dv u_D(w_{i-2},j^{\star}_{Vi-1},w'_{i-1},j_{Ui-1},l_{i-1},j_{Di-1}), \dv y_3[i-1],\hat{\dv s}_D[\iota_{Ri-3}]\Big) \in T_{\epsilon}^n(P_{V,U,U_D,Y_3,\hat{S}_D}),\nonumber\\
&\Big(\dv v(w'_{i-1},j_{Vi}),\dv y_3[i],\hat{\dv s}_D[\iota_{Di-2}]\Big) \in T_{\epsilon}^n(P_{V,Y_3,\hat{S}_D})\Big\}\nonumber.
\end{align}

Conditioned on $\cap_{k=1}^{7} E_{ki}^c$, the probability of the event $E_{8i}$ can be bounded using the union bound, as
\begin{align}
\text{Pr}(E_{8i}|\cap_{k=1}^{7} E_{ki}^c)  & \leq MJ_UM_DJ_DJ_V2^{-n[I(U,U_D;Y_3,\hat{S}_D|V)-\epsilon]}2^{-n[I(V;Y_3,\hat{S}_D)-\epsilon]}\nonumber\\
                            & =  2^{-n[I(V,U;Y_3|\hat{S}_D)-I(V,U;S,\hat{S}_R|\hat{S}_D)-R-[I(U;Y_3,\hat{S}_D|V)-I(U;S,\hat{S}_R,\hat{S}_D|V)]_{-}+2\epsilon]}.
\label{ProbabilityErrorEvent3ForAchievabilityInTheoreom1}
\end{align}
 Thus, $\text{Pr}(E_{8i}|\cap_{k=1}^{7} E_{ki}^c) \longrightarrow 0$ as $n \longrightarrow \infty$ if $R < I(U,V;Y_3,\hat{S}_D)-I(U,V;S,\hat{S}_R,\hat{S}_D)$.

\item For decoding the triple $(\hat{w}_{i-1},\hat{j}_{Ui-1},\hat{l}_{i-1})$ and the index $\hat{j}_{Vi}$ at the destination, let $E_{9i}$ be the event
\begin{align}
E_{9i} = \Big\{&\exists\; j_{Vi} \in J_V \; \text{s.t.:}\;  j_{Vi} \neq  j^{\star}_{Vi}, \nonumber\\
&\Big(\dv v(w_{i-2},j^{\star}_{Vi-1}),\dv u(w_{i-2},j^{\star}_{Vi-1},w_{i-1},j^{\star}_{Ui-1}),\dv u_D(w_{i-2},j^{\star}_{Vi-1},w_{i-1},j^{\star}_{Ui-1},l_{i-1},j_{Di-1}), \dv y_3[i-1],\hat{\dv s}_D[\iota_{Ri-3}]\Big) \in T_{\epsilon}^n(P_{V,U,U_D,Y_3,\hat{S}_D}),\nonumber\\
&\Big(\dv v(w_{i-1},j_{Vi}),\dv y_3[i],\hat{\dv s}_D[\iota_{Di-2}]\Big) \in T_{\epsilon}^n(P_{V,Y_3,\hat{S}_D})\Big\}\nonumber.
\end{align}

Conditioned on $\cap_{k=1}^{8} E_{ki}^c$, the probability of the event $E_{9i}$ can be bounded using the union bound, as
\begin{align}
\text{Pr}(E_{9i}|\cap_{k=1}^{8} E_{ki}^c)  & \leq J_V2^{-n[I(V;Y_3,\hat{S}_D)-\epsilon]}\nonumber\\
                            & =  2^{-n[I(V;Y_3,\hat{S}_D)-I(V;S,\hat{S}_R,\hat{S}_D)-2\epsilon]}.
\label{ProbabilityErrorEvent3ForAchievabilityInTheoreom1}
\end{align}
 Thus, $\text{Pr}(E_{9i}|\cap_{k=1}^{8} E_{ki}^c) \longrightarrow 0$ as $n \longrightarrow \infty$ if $I(V;Y_3,\hat{S}_D)-I(V;S,\hat{S}_R,\hat{S}_D) > 2\epsilon$.

\item For decoding the triple $(\hat{w}_{i-1},\hat{j}_{Ui-1},\hat{l}_{i-1})$ and the index $\hat{j}_{Vi}$ at the destination, let $E_{10i}$ be the event
\begin{align}
E_{10i} = \Big\{&\exists \; j'_{Ui-1} \in J_U, l_{i-1} \in [1,M_D], j_{Di-1} \in J_D, j_{Vi} \in J_V \; \text{s.t.:}\; j'_{Ui-1} \neq j^{\star}_{Ui-1},\; j_{Vi} \neq j^{\star}_{Vi} \nonumber\\
&\Big(\dv v(w_{i-2},j^{\star}_{Vi-1}),\dv u(w_{i-2},j^{\star}_{Vi-1},w_{i-1},j'_{Ui-1}),\dv u_D(w_{i-2},j^{\star}_{Vi-1},w_{i-1},j'_{Ui-1},l_{i-1},j_{Di-1}), \dv y_3[i-1],\hat{\dv s}_D[\iota_{Ri-3}]\Big) \in T_{\epsilon}^n(P_{V,U,U_D,Y_3,\hat{S}_D}),\nonumber\\
&\Big(\dv v(w_{i-1},j_{Vi}),\dv y_3[i],\hat{\dv s}_D[\iota_{Di-2}]\Big) \in T_{\epsilon}^n(P_{V,Y_3,\hat{S}_D})\Big\}\nonumber.
\end{align}

Conditioned on $\cap_{k=1}^{9} E_{ki}^c$, the probability of the event $E_{10i}$ can be bounded using the union bound, as
\begin{align}
\text{Pr}(E_{10i}|\cap_{k=1}^{9} E_{ki}^c)  & \leq J_UM_DJ_DJ_V2^{-n[I(U,U_D;Y_3,\hat{S}_D|V)-\epsilon]}2^{-n[I(V;Y_3,\hat{S}_D)-\epsilon]}\nonumber\\
                            & =  2^{-n[I(U,V;Y_3|\hat{S}_D)-I(U,V;S,\hat{S}_R|\hat{S}_D)-[I(U;Y_3,\hat{S}_D|V)-I(U;S,\hat{S}_R,\hat{S}_D|V)]_{-}+\epsilon]}.
\label{ProbabilityErrorEvent3ForAchievabilityInTheoreom1}
\end{align}
 Thus, $\text{Pr}(E_{10i}|\cap_{k=1}^{9} E_{ki}^c) \longrightarrow 0$ as $n \longrightarrow \infty$.
 
\item For decoding the triple $(\hat{w}_{i-1},\hat{j}_{Ui-1},\hat{l}_{i-1})$ and the index $\hat{j}_{Vi}$ at the destination, let $E_{11i}$ be the event
\begin{align}
E_{11i} = \Big\{&\exists \; j'_{Ui-1} \in J_U, l_{i-1} \in [1,M_D], j_{Di-1} \in J_D, j_{Vi} \in J_V \; \text{s.t.:}\; j'_{Ui-1} \neq j^{\star}_{Ui-1},\nonumber\\
&\Big(\dv v(w_{i-2},j^{\star}_{Vi-1}),\dv u(w_{i-2},j^{\star}_{Vi-1},w_{i-1},j'_{Ui-1}),\dv u_D(w_{i-2},j^{\star}_{Vi-1},w_{i-1},j'_{Ui-1},l_{i-1},j_{Di-1}), \dv y_3[i-1],\hat{\dv s}_D[\iota_{Ri-3}]\Big) \in T_{\epsilon}^n(P_{V,U,U_D,Y_3,\hat{S}_D}),\nonumber\\
&\Big(\dv v(w_{i-1},j^{\star}_{Vi}),\dv y_3[i],\hat{\dv s}_D[\iota_{Di-2}]\Big) \in T_{\epsilon}^n(P_{V,Y_3,\hat{S}_D})\Big\}\nonumber.
\end{align}

Conditioned on $\cap_{k=1}^{10} E_{ki}^c$, the probability of the event $E_{11i}$ can be bounded using the union bound, as
\begin{align}
\text{Pr}(E_{11i}|\cap_{k=1}^{10} E_{ki}^c)  & \leq J_UM_DJ_D2^{-n[I(U,U_D;Y_3,\hat{S}_D|V)-\epsilon]}\nonumber\\
                            & =  2^{-n[I(U;Y_3|V,\hat{S}_D)-I(U;S,\hat{S}_R|V,\hat{S}_D)-[I(U;Y_3,\hat{S}_D|V)-I(U;S,\hat{S}_R,\hat{S}_D|V)]_{-}+3\epsilon]}.
\label{ProbabilityErrorEvent3ForAchievabilityInTheoreom1}
\end{align}
 Thus, $\text{Pr}(E_{11i}|\cap_{k=1}^{10} E_{ki}^c) \longrightarrow 0$ as $n \longrightarrow \infty$.

\item For decoding the triple $(\hat{w}_{i-1},\hat{j}_{Ui-1},\hat{l}_{i-1})$ and the index $\hat{j}_{Vi}$ at the destination, let $E_{12i}$ be the event
\begin{align}
E_{12i} = \Big\{&\exists \; l'_{i-1} \in [1,M_D], j_{Di-1} \in J_D, j_{Vi} \in J_V \; \text{s.t.:}\; l'_{i-1} \neq l_{i-1},\; j_{Vi} \neq j^{\star}_{Vi}, \nonumber\\
&\Big(\dv v(w_{i-2},j^{\star}_{Vi-1}),\dv u(w_{i-2},j^{\star}_{Vi-1},w_{i-1},j^{\star}_{Ui-1}),\dv u_D(w_{i-2},j^{\star}_{Vi-1},w_{i-1},j^{\star}_{Ui-1},l'_{i-1},j_{Di-1}), \dv y_3[i-1],\hat{\dv s}_D[\iota_{Ri-3}]\Big) \in T_{\epsilon}^n(P_{V,U,U_D,Y_3,\hat{S}_D}),\nonumber\\
&\Big(\dv v(w_{i-1},j_{Vi}),\dv y_3[i],\hat{\dv s}_D[\iota_{Di-2}]\Big) \in T_{\epsilon}^n(P_{V,Y_3,\hat{S}_D})\Big\}\nonumber.
\end{align}

Conditioned on $\cap_{k=1}^{11} E_{ki}^c$, the probability of the event $E_{12i}$ can be bounded using the union bound, as
\begin{align}
\text{Pr}(E_{12i}|\cap_{k=1}^{11} E_{ki}^c)  & \leq M_DJ_DJ_V2^{-n[I(U_D;Y_3,\hat{S}_D|U,V)-\epsilon]}2^{-n[I(V;Y_3,\hat{S}_D)-\epsilon]}\nonumber\\
                            & =  2^{-n[I(V;Y_3,\hat{S}_D)-I(V;S,\hat{S}_R,\hat{S}_D)-[I(U;Y_3,\hat{S}_D|V)-I(U;S,\hat{S}_R,\hat{S}_D|V)]_{-}+2\epsilon]}.
\label{ProbabilityErrorEvent3ForAchievabilityInTheoreom1}
\end{align}
 Thus, $\text{Pr}(E_{12i}|\cap_{k=1}^{11} E_{ki}^c) \longrightarrow 0$ as $n \longrightarrow \infty$.

\item For decoding the triple $(\hat{w}_{i-1},\hat{j}_{Ui-1},\hat{l}_{i-1})$ and the index $\hat{j}_{Vi}$ at the destination, let $E_{13i}$ be the event
\begin{align}
E_{13i} = \Big\{&\exists \; l'_{i-1} \in [1,M_D], j_{Di-1} \in J_D, j_{Vi} \in J_V \; \text{s.t.:}\; l'_{i-1} \neq l_{i-1},\nonumber\\
&\Big(\dv v(w_{i-2},j^{\star}_{Vi-1}),\dv u(w_{i-2},j^{\star}_{Vi-1},w_{i-1},j^{\star}_{Ui-1}),\dv u_D(w_{i-2},j^{\star}_{Vi-1},w_{i-1},j^{\star}_{Ui-1},l'_{i-1},j_{Di-1}), \dv y_3[i-1],\hat{\dv s}_D[\iota_{Ri-3}]\Big) \in T_{\epsilon}^n(P_{V,U,U_D,Y_3,\hat{S}_D}),\nonumber\\
&\Big(\dv v(w_{i-1},j^{\star}_{Vi}),\dv y_3[i],\hat{\dv s}_D[\iota_{Di-2}]\Big) \in T_{\epsilon}^n(P_{V,Y_3,\hat{S}_D})\Big\}\nonumber.
\end{align}

Conditioned on $\cap_{k=1}^{12} E_{ki}^c$, the probability of the event $E_{13i}$ can be bounded using the union bound, as
\begin{align}
\text{Pr}(E_{13i}|\cap_{k=1}^{12} E_{ki}^c)  & \leq M_DJ_D2^{-n[I(U_D;Y_3,\hat{S}_D|U,V)-\epsilon]}\nonumber\\
                            & =  2^{-n[-|I(U;Y_3,\hat{S}_D|V)-I(U;S,\hat{S}_R,\hat{S}_D|V)+4\epsilon]}.
\label{ProbabilityErrorEvent3ForAchievabilityInTheoreom1}
\end{align}
 Thus, $\text{Pr}(E_{13i}|\cap_{k=1}^{12} E_{ki}^c) \longrightarrow 0$ as $n \longrightarrow \infty$.

\end{itemize}
 
This concludes the proof of Theorem~\ref{Theorem__LowerBound1__DiscreteChannel}.
\renewcommand{\theequation}{B-\arabic{equation}}
\setcounter{equation}{0}  
\subsection{Proof of Theorem~\ref{Theorem__LowerBound2__DiscreteChannel}}\label{appendixTheorem__LowerBound2__DiscreteChannel}

First we generate a random codebook that we use to obtain the lower bound in Theorem~\ref{Theorem__LowerBound2__DiscreteChannel}. This scheme is based on a combination of block Markov coding \cite{CG79}, Gel'fand-Pinsker binning \cite{GP80}, and classic rate distortion theory \cite[Chapter 13]{CT91}. Next, we outline the encoding and decoding procedures.

We transmit in $B$ blocks, each of length $n$. During each of the first $B$ blocks, the source encodes a message $w_i \in [1,2^{nR}]$ and sends it over the channel, where $i=1,\hdots,B$ denotes the index of the block. For convenience we let $w_{B+1}=1$. For fixed $n$, the average rate $R\frac{B}{B+1}$ over $B+1$ blocks approaches $R$ as $B \longrightarrow +\infty$.

\textbf{Codebook generation:} Fix a measure $P_{S,U,U_R,X_1,X_2,X,\hat{X},Y_2,Y_3}$ of the form \eqref{Measure__AchievableRate__Theorem__LowerBound2__DiscreteChannel}. Calculate the marginal $P_{\hat{X}}$ induced by this measure. Fix $\epsilon > 0$ and let
\begin{subequations}
\begin{align}
J &= 2^{n[I(U;S)+2\epsilon]} & J_R &= 2^{n[I(U_R;U,S)+2\epsilon]} \\
M &= 2^{n[R-4\epsilon]} & M_R &= 2^{n[\hat{R}-4\epsilon]}.
\end{align}
\label{ValuesForBinningVariablesInTheorem__LowerBound2__DiscreteChannel}
\end{subequations}

\begin{enumerate}
\item[1)] We generate $JM$ independent and identically distributed (i.i.d.) codewords  $\{\dv u(w,j)\}$ indexed by $w=1,\hdots,M$, $j=1,\hdots,J$, each with i.i.d. components drawn according to $P_{U}$.

\item[2)] We generate $J_RM_R$ i.i.d. codewords  $\{\dv u_R(m,j_R)\}$ indexed by $m=1,\hdots,M_R$, $j_R=1,\hdots,J_R$, each with i.i.d. components drawn according to $P_{U_R}$.

\item[3)] Independently, we randomly generate a rate distortion codebook consisting of $M_R$ sequences $\hat{\dv x}$ drawn i.i.d. according to the $n-$product of the marginal $P_{\hat{X}}$. We index these sequences as $\hat{\dv x}[m], m=1,\hdots,M_R$.
\end{enumerate}

\textbf{Encoding:} We pick up the story in block $i$. Let $w_i \in \{1,\hdots,M\}$ be the new message to be sent from the source node at the beginning of block $i$, and  $w_{i+1} \in \{1,\hdots,M\}$ the message to be sent in the next block $i+1$ (note that we can assume that $w_i \neq w_{i+1}$, as the indices $\{w_k\}$ are assumed i.i.d. on $\{1,\hdots,2^{nR}\}$, and so $\text{Pr}(w_i=w_{i+1})=2^{-2nR} \rightarrow 0$ as $n \rightarrow +\infty$). The encoding at the beginning of block $i$ is as follows.

\begin{itemize}
\item[i)] The source searches for the smallest $j \in \{1,\cdots,J\}$ such that $\dv u(w_i,j)$ is jointly typical with $\dv s[i]$. (The properties of strongly typical sequences guarantee that there exists one such $j$). Denote this $j$ by $j^{\star}_i=j(\dv s[i],w_i)$. 

\item[ii)] Similarly, the source finds  $j^{\star}_{i+1}=j(\dv s[i+1],w_{i+1})$ such that $\dv u(w_{i+1},j^{\star}_{i+1})$ is jointly typical with $\dv s[i+1]$ and then  generates a vector $\dv x[w_{i+1}]$ with i.i.d. components given $\dv u(w_{i+1}, j^{\star}_{i+1})$ and $\dv s[i+1]$, drawn according to the marginal $P_{X|U,S}$.

\item[iii)] Then, the source indices $\dv x[w_{i+1}]$ by $m_i$ if there exists an $m_i \in \{1,\hdots,M_R\}$ such that $\dv x[w_{i+1}]$ and $\hat{\dv x}[m_i]$ are jointly strongly typical. If there is more than one such $m_i$, the source selects the first in lexicographic order. If there is no such $m_i$, let $m_i=1$. Shannon's rate-distortion theory \cite[Chapter 13]{CT91} ensures that the encoding of $\dv x[w_{i+1}]$ is accomplished successfully with high probability provided that $n$ is sufficiently large and
\begin{align}
\hat{R} &> I(X;\hat{X}).
\label{Step1__Constraint__AchievableRate__Theorem__LowerBound2__DiscreteChannel}
\end{align}
\item[iv)] Next, the source looks for the smallest $j_R \in \{1,\cdots,J_R\}$ such that $\dv u_R(m_i,j_R)$ is jointly typical with $(\dv s[i],\dv u(w_i,j^{\star}_i))$. (Again, the properties of strongly typical sequences guarantee that there exists one such $j_R$). Denote this $j_R$ by $j^{\star}_{Ri}=j_R(\dv s[i],\dv u(w_i,j^{\star}_i))$.
\end{itemize}

\noindent Continuing with the strategy. Let $m_0=1$. The encoding at the beginning of block $i$ is as follows.
\begin{itemize}
\item[1)] The relay knows $m_{i-1}$ (this will be justified below), and sends $\dv x_2[i]=\hat{\dv x}[m_{i-1}]$.

\item[2)]  The source transmits the pair $(w_i,m_i)$. It sends a vector $\dv x_1[i]$ with i.i.d. components given the vectors $\dv u(w_i,j^{\star}_i)$, $\dv u_R(m_i,j^{\star}_{Ri})$ and $\dv s[i]$, drawn according to the marginal $P_{X_1|U,U_R,S}$ induced by the distribution~\eqref{Measure__AchievableRate__Theorem__LowerBound2__DiscreteChannel}.

\textbf{Decoding:} The reconstruction of the vector $\dv x[w_{i+1}]$ at the relay and the decoding procedure at destination at the end of block $i$, are as follows.

\item[1)] The relay knows $m_{i-1}$ and estimates $m_i$ from the received  $\dv y_2[i]$. It declares that  $\hat{m}_i$ is sent if there is a unique $\hat{m}_i \in \{1,\hdots,M_R\}$ such that $\dv u_R(\hat{m}_i,j_{Ri})$ and $\dv y_2[i]$ are jointly typical for some $j_{Ri} \in \{1,\hdots,J_R\}$. One can show that the decoding error in this step is small for sufficiently large $n$ if
\begin{align}
\hat{R} &< I(U_R;Y_2)-I(U_R;U,S)\nonumber\\
        &= I(U_R;Y_2)-I(U_R;S)-I(U_R;U|S).
\label{Step3__Constraint__AchievableRate__Theorem__LowerBound2__DiscreteChannel}
\end{align}

\item[2)] The destination estimates $w_i$ from the received $\dv y_3[i]$. It declares that  $\hat{w}_i$ is sent if there is a unique $\hat{w}_i \in \{1,\hdots,M\}$ such that $\dv u(\hat{w}_i,j_i)$ and $\dv y_3[i]$ are jointly typical for some $j_i \in \{1,\hdots,J\}$. One can show that the decoding error in this step is small for sufficiently large $n$ if
\begin{align}
R &< I(U;Y_3)-I(U;S).
\label{Step4__Constraint__AchievableRate__Theorem__LowerBound2__DiscreteChannel}
\end{align}

\end{itemize}

\textbf{Analysis of Probability of Error:} Fix a probability distribution $P_{S,U,U_R,X_1,X_2,X,\hat{X},Y_2,Y_3}$ satisfying \eqref{Measure__AchievableRate__Theorem__LowerBound2__DiscreteChannel}. Let $\dv s[i]$ and $(w_i,m_i)$ be the state sequence in block $i$ and the message pair sent from the source node in block $i$, respectively. As we already mentioned above, at the beginning of block $i$ the source transmits $\dv x_1(w_i,m_i)$ and the relay transmits $\dv x_2[i]=\hat{\dv x}[m_{i-1}]$.

The average probability of error is such that
\begin{align}
\text{Pr}(\text{Error}) & \leq \sum_{\dv s \notin T_{\epsilon}^n(Q_S)}\text{Pr}(\dv s) + \sum_{\dv s \in T_{\epsilon}^n(Q_S)}\text{Pr}(\dv s)\text{Pr}(\text{error}|\dv s).
\label{AverageProbabilityOfError}
\end{align}
The first term, $\text{Pr}(\dv s \notin T_{\epsilon}^n(Q_S))$, on the RHS of \eqref{AverageProbabilityOfError} goes to zero as $n \rightarrow \infty$, by the asymptotic equipartition property (AEP) \cite[p. 384 ]{CT91}. Thus, it is sufficient to upper bound the second term on the RHS of \eqref{AverageProbabilityOfError}.

We now examine the probabilities of the error events associated with the encoding and decoding procedures. The error event is contained in the union of the following error events; where the events $E_{1i}$, $E_{2i}$ and $E_{3i}$ correspond to encoding errors at block $i$; the events $E_{4i}$ and $E_{5i}$ correspond to decoding errors at the relay at block $i$; and the events $E_{6i}$ and $E_{7i}$ correspond to decoding errors at the destination at block $i$.

\begin{itemize}
\item Let $E_{1i}$ be the event that there is no sequence $\dv u(w_i,j)$ jointly typical with $\dv s[i]$, i.e.,
\begin{equation*}
E_{1i} = \Big\{\nexists \:j \in \{1,\hdots,J\}\:\text{s.t.}\: \Big(\dv u(w_i,j),\dv s[i]\Big) \in T_{\epsilon}^n(P_{U,S})\Big\}.
\end{equation*}
To bound the probability of the event $E_{1i}$, we use a standard argument \cite{GP80}. More specifically, for $\dv u(w_i,j)$ and $\dv s[i]$ generated independently with i.i.d. components drawn according to $P_{U}$ and $Q_S$, respectively, the probability that $\dv u(w_i,j)$ is jointly typical with $\dv s[i]$ is greater than $(1-\epsilon)2^{-n(I(U;S)+\epsilon)}$ for sufficiently large $n$. There is a total of $J$ such $\dv u$'s in each bin. The probability of the event $E_{1i}$, the probability that there is no such $\dv u$, is therefore bounded as
\begin{equation}
\text{Pr}(E_{1i}) \leq [1-(1-\epsilon)2^{-n(I(U;S)+\epsilon)}]^J.
\label{BoundingProbabilityErrorEventE1}
\end{equation}
Taking the logarithm on both sides of \eqref{BoundingProbabilityErrorEventE1} and substituting $J$ using \eqref{ValuesForBinningVariablesInTheorem__LowerBound2__DiscreteChannel} we obtain $\ln(\text{Pr}(E_{1i})) \leq -(1-\epsilon)2^{n\epsilon}$. Thus, $\text{Pr}(E_{1i}) \rightarrow 0 \quad \text{as}\quad  n \rightarrow \infty$.
\item Let $E_{2i}$ be the event that there is no sequence $\dv u(w_{i+1},j)$ jointly typical with $\dv s[i+1]$, and $E_{3i}$ the event that there is no sequence $\dv u_R(m_i,j_R)$ jointly typical with $(\dv s[i],\dv u(w_i,j^{\star}_i))$. Proceeding similarly to for the event $E_{1i}$, it can be easily shown that, conditioned on $E^c_{1i}$ and $E^c_{1i} \cap E^c_{2i}$, respectively, these tow events have vanishing probabilities as $n \rightarrow +\infty$.
\item  For the decoding at the relay, let $E_{4i}$ be the event that $\dv u_R(m_i,j^{\star}_{Ri})$ is not jointly typical with $\dv y_2[i]$. That is
\begin{align}
&E_{4i} = \Big\{\big(\dv u_R(m_i,j^{\star}_{Ri}),\dv y_2[i]\big) \notin T_{\epsilon}^n(P_{U_R,Y_2,\hat{X}})\Big\}.
\end{align}
For $\dv u(w_i,j^{\star}_i)$, $\dv u_R(m_i,j^{\star}_{Ri})$ jointly typical with $\dv s[i]$, and with the source input $\dv x_1[i]$ and the relay input $\dv x_2[i]$, we have $\text{Pr}(E_{4i}|E_{1i}^c,E_{2i}^c,E_{3i}^c) \longrightarrow 0$ as $n \longrightarrow \infty$ by the Markov Lemma \cite[p. 436]{CT91}.
\item  For the decoding at the relay, let $E_{5i}$ be the event that $\dv u_R(m'_i,j_{Ri})$ is jointly typical with $\dv y_2[i]$ for some $m'_i \in [1,M_R]$ and $j_{Ri} \in J_R$, with $m'_i \neq m_i$. That is,
\begin{align}
E_{5i} = \Big\{&\exists \; m'_i \in [1,M_R], j_{Ri} \in J_R \; \text{s.t.}\; m'_i \neq m_i, \nonumber\\
&\big(\dv u_R(m'_i,j_{Ri}),\dv y_2[i]\big) \in T_{\epsilon}^n(P_{U_R,Y_2,\hat{X}})\Big\}.
\end{align}
Conditioned on the events $E_{1i}^c$, $E_{2i}^c$, $E_{3i}^c$ and $E_{4i}^c$, the probability of the event $E_{5i}$ can be bounded using the union bound, as
\begin{align}
\text{Pr}(E_{5i}|E_{1i}^c,E_{2i}^c,E_{3i}^c,E_{4i}^c)  & \leq M_RJ_R2^{-n[I(U_R;Y_2)-\epsilon]}\nonumber\\
                            & =  2^{-n[I(U_R;Y_2)-I(U_R;U,S)-\hat{R}+\epsilon]}.
\label{ProbabilityErrorEvent3ForAchievabilityInTheoreom1}
\end{align}
Thus, $\text{Pr}(E_{3i}|E_{1i}^c,E_{2i}^c,E_{3i}^c,E_{4i}^c) \longrightarrow 0$ as $n \longrightarrow \infty$ if $R < I(U_R;Y_2)-I(U_R;S)-I(U_R;U|S)$.
\item  For the decoding at the destination, let $E_{6i}$ be the event that $\dv u(w_i,j^{\star}_i)$ is not jointly typical with $\dv y_3[i]$. That is
\begin{align}
&E_{6i} = \Big\{\big(\dv u(w_i,j^{\star}_i),\dv y_3[i]\big) \notin T_{\epsilon}^n(P_{U,Y_3})\Big\}.
\end{align}
For $\dv u(w_i,j^{\star}_i)$, $\dv u_R(m_i,j^{\star}_{Ri})$ jointly typical with $\dv s[i]$, and with the source input $\dv x_1[i]$ and the relay input $\dv x_2[i]$, we have $\text{Pr}(E_{6i}|E_{1i}^c,E_{2i}^c,E_{3i}^c,E_{4i}^c,E_{5i}^c) \longrightarrow 0$ as $n \longrightarrow \infty$ by the Markov Lemma \cite[p. 436]{CT91}.
\item For the decoding at the destination, let $E_{7i}$ be the event that $\dv u(w'_i,j_i)$ is jointly typical with $\dv y_3[i]$ for some $w'_i \in [1,M]$ and $j_i \in J$, with $w'_i \neq k_i$. That is,
\begin{align}
E_{7i} = \Big\{&\exists \; w'_i \in [1,M], j_i \in J \; \text{s.t.}\; w'_i \neq k_i, \nonumber\\
&\big(\dv u(w'_i,j_i),\dv y_3[i]\big) \in T_{\epsilon}^n(P_{U,Y_3})\Big\}.
\end{align}
Conditioned on the events $E_{1i}^c$, $E_{2i}^c$, $E_{3i}^c$, $E_{4i}^c$, $E_{5i}^c$ and $E_{6i}^c$, the probability of the event $E_{7i}$ can be bounded using the union bound, as
\begin{align}
\text{Pr}(E_{7i}|E_{1i}^c,E_{2i}^c,E_{3i}^c,E_{4i}^c,E_{5i}^c,E_{6i}^c)  & \leq MJ2^{-n[I(U;Y_3)-\epsilon]}\nonumber\\
                            & =  2^{-n[I(U;Y_3)-I(U;S)-R+\epsilon]}
\label{ProbabilityErrorEvent3ForAchievabilityInTheoreom1}
\end{align}
Thus, $\text{Pr}(E_{7i}|E_{1i}^c,E_{2i}^c,E_{3i}^c,E_{4i}^c,E_{5i}^c,E_{6i}^c) \longrightarrow 0$ as $n \longrightarrow +\infty$ if $R < I(U;Y_3)-I(U;S)$.
\end{itemize}
This concludes the proof of Theorem~\ref{Theorem__LowerBound2__DiscreteChannel}.

\renewcommand{\theequation}{C-\arabic{equation}} 
\setcounter{equation}{0}  
\subsection{Proofs of Theorem~\ref{Theorem_UpperBound__DiscreteChannel}}\label{appendixTheorem__UpperBound__DiscreteChannel} 

Let an $(\epsilon_n,n,R)$ code be given. By Fano's inequality, we have
\begin{align}
nR &= H(W)\nonumber\\
   &\leq I(W;Y^n_3)+1+nR\epsilon_n.
\end{align}

\noindent Let us define $\bar{U}_i=(S^n_{i+1},Y^{i-1}_2,Y^{i-1}_3)$ and $\bar{V}_i=(W,S^n_{i+1},Y^{i-1}_3)$, $i=1,\hdots,n$.  

\noindent We have
{\allowdisplaybreaks
\begin{align}
I(W;Y^n_3) &\leq I(W;Y^n_2,Y^n_3)\nonumber\\
      &\stackrel{(a)}{=} I(W;Y^n_2,Y^n_3)-I(W;S^n)\\
      &= \sum_{i=1}^{n} I(W;Y_{2,i},Y_{3,i}|Y^{i-1}_2,Y^{i-1}_3)-I(W;S_i|S^n_{i+1})\nonumber\\
      &= \sum_{i=1}^{n} I(W,S^n_{i+1};Y_{2,i},Y_{3,i}|Y^{i-1}_2,Y^{i-1}_3)-I(S^n_{i+1};Y_{2,i},Y_{3,i}|W,Y^{i-1}_2,Y^{i-1}_3)-I(W;S_i|S^n_{i+1})\nonumber\\
      &\stackrel{(b)}{=} \sum_{i=1}^{n} I(W,S^n_{i+1};Y_{2,i},Y_{3,i}|Y^{i-1}_2,Y^{i-1}_3)-I(S_i;Y^{i-1}_2,Y^{i-1}_3|W,S^n_{i+1})-I(W;S_i|S^n_{i+1})\nonumber\\
      &= \sum_{i=1}^{n} I(W,S^n_{i+1};Y_{2,i},Y_{3,i}|Y^{i-1}_2,Y^{i-1}_3)-I(S_i;W,Y^{i-1}_2,Y^{i-1}_3|S^n_{i+1})\nonumber\\
      &= \sum_{i=1}^{n} I(W;Y_{2,i},Y_{3,i}|S^n_{i+1},Y^{i-1}_2,Y^{i-1}_3)+I(S^n_{i+1};Y_{2,i},Y_{3,i}|Y^{i-1}_2,Y^{i-1}_3)-I(S_i;Y^{i-1}_2,Y^{i-1}_3|S^n_{i+1})-I(S_i;W|S^n_{i+1},Y^{i-1}_2,Y^{i-1}_3)\nonumber\\
      &\stackrel{(c)}{=} \sum_{i=1}^{n} I(W;Y_{2,i},Y_{3,i}|S^n_{i+1},Y^{i-1}_2,Y^{i-1}_3)-I(S_i;W|S^n_{i+1},Y^{i-1}_2,Y^{i-1}_3)\nonumber\\
      &\stackrel{(d)}{=} \sum_{i=1}^{n} I(W;Y_{2,i},Y_{3,i}|S^n_{i+1},Y^{i-1}_2,Y^{i-1}_3,X_{2,i})-I(S_i;W|S^n_{i+1},Y^{i-1}_2,Y^{i-1}_3,X_{2,i})\nonumber\\
      &= \sum_{i=1}^{n} I(\bar{V}_i;Y_{2,i},Y_{3,i}|\bar{U}_i,X_{2,i})-I(\bar{V}_i;S_i|\bar{U}_i,X_{2,i})
\end{align}
where: $(a)$ follows since message $W$ is independent of the state $S^n$; $(b)$ follows from  Csiszar and Korner's ``summation by parts"-lemma \cite{CK78}
\begin{align}
\sum_{i=1}^{n} I(S^n_{i+1};Y_{2,i},Y_{3,i}|W,Y^{i-1}_2,Y^{i-1}_3) = \sum_{i=1}^{n} I(S_i;Y^{i-1}_2,Y^{i-1}_3|W,S^n_{i+1})
\end{align}
$(c)$ follows similarly, from  Csiszar and Korner's ``summation by parts"
\begin{align}
\sum_{i=1}^{n}  I(S^n_{i+1};Y_{2,i},Y_{3,i}|Y^{i-1}_2,Y^{i-1}_3) = \sum_{i=1}^{n} I(S_i;Y^{i-1}_2,Y^{i-1}_3|S^n_{i+1})
\end{align}
$(d)$ follows from the fact that $X_{2i}$ is a deterministic function of $Y^{i-1}_2$.

\noindent Similarly, 
{\allowdisplaybreaks
\begin{align}
I(W;Y^n_3) &\stackrel{(e)}{\leq} \sum_{i=1}^{n} I(W,S^n_{i+1},Y^{i-1}_3;Y_{3,i})-I(W,S^n_{i+1},Y^{i-1}_3;S_i)\nonumber\\
           &= \sum_{i=1}^{n} I(\bar{V}_i;Y_{3,i})-I(\bar{V}_i;S_i)
\end{align}
where $(e)$ follows exactly as in the converse part of the proof of the capacity of Gel'fand-Pinsker channel \cite{GP80} by replacing $Y^n$ with $Y^n_3$.

From the above, we have
\begin{align}
R &\leq \frac{1}{n} \sum_{i=1}^{n} I(\bar{V}_i;Y_{2,i},Y_{3,i}|\bar{U}_i,X_{2,i})-I(\bar{V}_i;S_i|\bar{U}_i,X_{2,i})+1+nR\epsilon_n\nonumber\\
R &\leq \frac{1}{n} \sum_{i=1}^{n} I(\bar{V}_i;Y_{3,i})-I(\bar{V}_i;S_i)+1+nR\epsilon_n
\label{MultiLetter__UpperBound__DiscreteChannel}
\end{align}

We introduce a random variable $T$ which is uniformly  distributed over $\{1,\cdots,n\}$. Set $S=S_T$, $\bar{U}=\bar{U}_T$, $\bar{V}=\bar{V}_T$, $X_1=X_{1,T}$, $X_2=X_{2,T}$, $Y_2=Y_{2,T}$, and $Y_3=Y_{3,T}$. We substitute $T$ into the above bounds. Considering the first bound in \eqref{MultiLetter__UpperBound__DiscreteChannel}, we have
\begin{align}
\frac{1}{n} &\sum_{i=1}^{n} I(\bar{V}_i;Y_{2,i},Y_{3,i}|\bar{U}_i,X_{2,i})-I(\bar{V}_i;S_i|\bar{U}_i,X_{2,i})\nonumber\\
  &= I(\bar{V};Y_2,Y_3|\bar{U},X_2,T)-I(\bar{V};S|\bar{U},X_2,T)\nonumber\\
  &= I(T,\bar{V};Y_2,Y_3|\bar{U},X_2)-I(T;Y_2,Y_3|\bar{U},X_2)-I(T,\bar{V};S|\bar{U},X_2)+I(T;S|\bar{U},X_2)\nonumber\\
  &\leq I(T,\bar{V};Y_2,Y_3|\bar{U},X_2)-I(T,\bar{V};S|\bar{U},X_2)+I(T;S|\bar{U},X_2)\nonumber\\
  &= I(T,\bar{V};Y_2,Y_3|\bar{U},X_2)-I(T,\bar{V};S|\bar{U},X_2)
\label{FirstTermUpperBound__WithTimeSharingVariable__DiscreteChannel}
\end{align}
where in the last equality we used the fact that $T$ is independent of all the other variables.

\noindent Similarly, considering the second bound in \eqref{MultiLetter__UpperBound__DiscreteChannel}, we obtain
\begin{align}
\frac{1}{n} &\sum_{i=1}^{n} I(\bar{V}_i;Y_{3,i})-I(\bar{V}_i;S_i)\nonumber\\
            &= I(\bar{V};Y_3|T)-I(\bar{V};S|T)\nonumber\\
            &= I(T,\bar{V};Y_3)-I(T;Y_3)-I(T,\bar{V};S)+I(T;S)\nonumber\\
	    &\leq I(T,\bar{V};Y_3)-I(T,\bar{V};S).
\label{SecondTermUpperBound__WithTimeSharingVariable__DiscreteChannel}
\end{align}
\noindent Let us now define $U=\bar{U}$ and $V=(T,\bar{V})$. Using \eqref{MultiLetter__UpperBound__DiscreteChannel}, \eqref{FirstTermUpperBound__WithTimeSharingVariable__DiscreteChannel} and \eqref{SecondTermUpperBound__WithTimeSharingVariable__DiscreteChannel}, we then get
\begin{align}
R &\leq I(V;Y_2,Y_3|U,X_2)-I(V;S|U,X_2)+1+nR\epsilon_n\nonumber\\
R &\leq I(V;Y_3)-I(V;S)+1+nR\epsilon_n.
\end{align}

So far we have shown that, for a given sequence of $(\epsilon_n,n,R)-$codes with $\epsilon_n$ going to zero as $n$ goes to infinity, there exists a probability distribution of the form \eqref{Measure__UpperBound__DiscreteChannel} such that the rate $R$ essentially satisfies \eqref{UpperBound__DiscreteChannel}. This completes the proof of Theorem \ref{Theorem_UpperBound__DiscreteChannel}.

It remains to show that the rate \eqref{UpperBound__DiscreteChannel} is not altered if one restricts the random variables $U$ and $U$ to have their alphabet sizes limited as indicated in \eqref{BoundsOnCardinalityOfAuxiliaryRandonVariables__UpperBound__DiscreteMemorylessChannel}. This is done by invoking the support lemma \cite[p. 310]{CK81}. Fix a distribution $\mu$ of $(S,U,V,X_1,X_2,Y_2,Y_3)$ on $\mc P({\mc S}{\times}{\mc U}{\times}{\mc V}{\times}{\mc X_1}{\times}{\mc X_2}{\times}{\mc Y_2}{\times}{\mc Y_3})$ that has the form \eqref{Measure__UpperBound__DiscreteChannel}.

\noindent To prove the bound \eqref{BoundsOnCardinalityOfAuxiliaryRandonVariableU__UpperBound__DiscreteMemorylessChannel} on $|\mc U|$, note that we have
\begin{align}
&I_{\mu}(V;Y_2,Y_3|U,X_2)-I_{\mu}(V;S|U,X_2) \nonumber\\
&\hspace{1cm}= I_{\mu}(V,X_2;Y_2,Y_3|U)-I_{\mu}(X_2;Y_2,Y_3|U)-I_{\mu}(V,X_2;S|U)+I_{\mu}(X_2;S|U)\nonumber\\
&\hspace{1cm}= H_{\mu}(Y_2,Y_3|U)-H_{\mu}(V,X_2,Y_2,Y_3|U)+H_{\mu}(V,X_2,S|U)+H_{\mu}(X_2|U)-H_{\mu}(X_2,S|U).
\end{align}

\noindent Hence, it suffices to show that the following functionals of $\mu(S,U,V,X_1,X_2,Y_2,Y_3)$
\begin{subequations}
\begin{align}
\label{Functional1__BoundsOnCardinalityOfAuxiliaryRandonVariableU__UpperBound__DiscreteMemorylessChannel}
r_{s,x,x'}(\mu) &= \mu(s,x,x') \quad \forall \: (s,x,x') \in {\mc S}{\times}{\mc X_1}{\times}{\mc X_2}\\
r_1(\mu) &= \int_{u}d_{\mu}(u)[H_{\mu}(Y_2,Y_3|u)-H_{\mu}(V,X_2,Y_2,Y_3|u)+H_{\mu}(V,X_2,S|u)+H_{\mu}(X_2|u)-H_{\mu}(X_2,S|u)]
\label{Functional2__BoundsOnCardinalityOfAuxiliaryRandonVariableU__UpperBound__DiscreteMemorylessChannel}
\end{align}
\label{Functionals__BoundsOnCardinalityOfAuxiliaryRandonVariableU__UpperBound__DiscreteMemorylessChannel}
\end{subequations}
can be preserved with another measure $\mu'$ that has the form \eqref{Measure__UpperBound__DiscreteChannel}. Observing that  there is a total of $|\mc S||\mc X_1||\mc X_2|$ functionals in \eqref{Functionals__BoundsOnCardinalityOfAuxiliaryRandonVariableU__UpperBound__DiscreteMemorylessChannel}, this is ensured by a standard application of the support lemma; and this shows that the cardinality of the alphabet of the auxiliary random variable $U_1$ can be limited as indicated in \eqref{BoundsOnCardinalityOfAuxiliaryRandonVariableU__UpperBound__DiscreteMemorylessChannel} without altering the rate \eqref{UpperBound__DiscreteChannel}.

\noindent Once the alphabet of $U$ is fixed, we apply similar arguments to bound the alphabet of $V$, where this time $(|\mc S||\mc X_1||\mc X_2|)^2-1$ functionals must be satisfied in order to preserve the joint distribution of $(S, U, X_1, X_2)$, and one more functional to preserve
\begin{align}
I_{\mu}(V;Y_3)-I_{\mu}(V;S)&=H_{\mu}(Y_3)-H_{\mu}(S)-H_{\mu}(Y_3|V)+H_{\mu}(S|V),
\end{align}
yielding the bound indicated in \eqref{BoundsOnCardinalityOfAuxiliaryRandonVariableV__UpperBound__DiscreteMemorylessChannel}. This completes the proof of Theorem~\ref{Theorem_UpperBound__DiscreteChannel}.

\renewcommand{\theequation}{D-\arabic{equation}}
\setcounter{equation}{0}  
\subsection{Proof of Theorem~\ref{Theorem__UpperBound__DiscreteModelWithHyperSource}}\label{appendixTheorem__UpperBound__DiscreteModelWithHyperSource}

We prove that for any $(\epsilon,n,R)$ code consisting of a mapping $\phi^n_1=(\phi^n_{1R},\phi^n_{1D})$ at the hyper source with $\phi^n_{1R}:\mc W \longrightarrow \mc X^n_{1R}$ and $\phi^n_{1D}:\mc W{\times}\mc S^{n} \longrightarrow \mc X^n_{1D}$ , a sequence of mappings $\phi_{2,i}: \mc Y^{i-1}_2 \longrightarrow \mc X_2$, $i=1,\hdots,n$, at the relay, and a mapping $\psi^n : \mc Y^n \longrightarrow W$ at the decoder with average error probability $P_e^n \rightarrow 0$ as $n \rightarrow 0$, the rate $R$ must satisfy \eqref{UpperBound__DiscreteModelWithHyperSource}.

By Fano's inequality, we have
\begin{equation}
H(W|Y_3^n) \leq nR\epsilon_n+1 \triangleq n\delta_n.
\end{equation}
Thus,
\begin{align}
nR &= H(W) \leq I(W;Y_3^n)+n\delta_n\nonumber\\
\label{FanosInequality}
\end{align}

We now upper bound $I(W;Y^n_3)$ as in the following lemma, the proof of which follows.

\begin{lemma}\label{LemmaProofTheoremUpperBoundDegenerateRelayModel}
\begin{subequations}
\begin{align}
\label{Equation1__LemmaUpperBoundDegenerateRelayModel}
& \text{i)} \:\:\: I(W;Y_3^n) \leq \sum_{i=1}^{n}I(X_{1R,i};Y_{2,i}|S_i,X_{2,i})+I(X_{1D,i};Y_{3,i}|S_i,X_{2,i})\\
& \text{ii)} \:\:\: I(W;Y_3^n) \leq \sum_{i=1}^{n}I(X_{1D,i};Y_{3,i}|S_i,X_{2,i})+I(X_{2,i};Y_{3,i}).
\label{Equation2__LemmaUpperBoundDegenerateRelayModel}
\end{align}
\label{EquationsLemmaUpperBoundDegenerateRelayModel}
\end{subequations}
\end{lemma}

\begin{proof}
To simplify the notation, we use $S^i=(S_1,S_2,\cdots,S_i)$, $Y_k^i=(Y_{k,1},Y_{k,2},\cdots,Y_{k,i})$, $k=2,3$, and $X_j^i=(X_{j,1},X_{j,2},\cdots,X_{j,i})$, $j=1R,1D,2$.

\vspace{0.2cm}

1) The proof of the bound on $I(W;Y^n_3)$ given in i) follows straightforwardly by  revealing the state to the destination and using the channel structure~\eqref{ConditionaProbability__ModelWithHyperSource}. 

{\allowdisplaybreaks
\begin{align}
I(W;Y^n_3) & \stackrel{(a)}{\leq} \sum_{i=1}^{n} I(X_{1R,i},X_{1D,i};Y_{2,i},Y_{3,i}|X_{2,i},S_i) \\
&= \sum_{i=1}^{n} I(X_{1R,i},X_{1D,i};Y_{2,i}|X_{2,i},S_i)+ I(X_{1R,i},X_{1D,i};Y_{3,i}|X_{2,i},S_i,Y_{2,i})\\
&= \sum_{i=1}^{n} I(X_{1R,i};Y_{2,i}|X_{2,i},S_i)+ I(X_{1D,i};Y_{2,i}|X_{1R,i},X_{2,i},S_i)\nonumber\\
&\quad + I(X_{1R,i},X_{1D,i};Y_{3,i}|X_{2,i},S_i,Y_{2,i})\\
&\stackrel{(b)}{=}  \sum_{i=1}^{n} I(X_{1R,i};Y_{2,i}|X_{2,i},S_i) + I(X_{1R,i},X_{1D,i};Y_{3,i}|X_{2,i},S_i,Y_{2,i})\\
&=  \sum_{i=1}^{n} I(X_{1R,i};Y_{2,i}|X_{2,i},S_i) + H(Y_{3,i}|X_{2,i},S_i,Y_{2,i})-H(Y_{3,i}|X_{1R,i},X_{1D,i},X_{2,i},S_i,Y_{2,i})\\
&\stackrel{(c)}{=}  \sum_{i=1}^{n} I(X_{1R,i};Y_{2,i}|X_{2,i},S_i) + H(Y_{3,i}|X_{2,i},S_i,Y_{2,i})-H(Y_{3,i}|X_{1D,i},X_{2,i},S_i)\\
&\stackrel{(d)}{\leq}  \sum_{i=1}^{n} I(X_{1R,i};Y_{2,i}|X_{2,i},S_i) + H(Y_{3,i}|X_{2,i},S_i)-H(Y_{3,i}|X_{1D,i},X_{2,i},S_i)\\
&=  \sum_{i=1}^{n} I(X_{1R,i};Y_{2,i}|X_{2,i},S_i) + I(X_{1D,i};Y_{3,i}|X_{2,i},S_i)
\label{FirstTerm__TheoremUpperBoundDegenerateRelayModel}
\end{align}}
where:\\
$(a)$ follows trivially by revealing the state to the destination; $(b)$ follows since $X_{1D,i} \leftrightarrow (X_{1R,i},X_{2,i},S_i) \leftrightarrow Y_{2,i}$; $(c)$ follows since $(X_{1R,i},Y_{2,i}) \leftrightarrow (X_{1D,i},X_{2,i},S_i) \leftrightarrow Y_{3,i}$; and $(d)$ follows since conditioning reduces entropy.

2) The proof of the bound on $I(W;Y^n_3)$ given in ii) follows as follows. 

{\allowdisplaybreaks
\begin{align}
I(W;Y^n_3)   & = I(W,S^n;Y^n_3)-I(S^n;Y^n_3|W)\nonumber\\        
   & = \Big(\sum_{i=1}^{n}I(W,S^n;Y_{3,i}|Y^{i-1}_3)\Big)-H(S^n|W)+H(S^n|W,Y^n_3)\nonumber\\
   & \stackrel{(e)}{=} \sum_{i=1}^{n} H(Y_{3,i}|Y^{i-1}_3)-H(Y_{3,i}|W,S^n,Y^{i-1}_3)-H(S_i)+H(S_i|W,Y^n_3,S^{i-1})\nonumber\\
   & \stackrel{(f)}{\leq} \sum_{i=1}^{n} H(Y_{3,i})-H(Y_{3,i}|X_{1D,i},X_{2,i},S_i)- H(S_i)+H(S_i|W,Y^n_3,S^{i-1})\nonumber\\
   & \stackrel{(g)}{=} \sum_{i=1}^{n} H(Y_{3,i})-H(Y_{3,i}|X_{1D,i},X_{2,i},S_i)- H(S_i)+H(S_i|W,Y^n_3,S^{i-1},Y^{i-1}_2)\nonumber\\
   & \stackrel{(h)}{=} \sum_{i=1}^{n} H(Y_{3,i})-H(Y_{3,i}|X_{1D,i},X_{2,i},S_i)- H(S_i)+H(S_i|W,Y^n_3,S^{i-1},Y^{i-1}_2,X_{2,i})\nonumber\\
   & \stackrel{(i)}{\leq} \sum_{i=1}^{n} I(X_{1D,i},X_{2,i},S_i;Y_{3,i})-H(S_i)+H(S_i|X_{2,i},Y_{3,i})\nonumber\\
   & \stackrel{(i)}{\leq} \sum_{i=1}^{n} I(X_{1D,i},X_{2,i},S_i;Y_{3,i})-H(S_i)+H(S_i|X_{2,i},Y_{3,i})\nonumber\\
   & = \sum_{i=1}^{n} I(X_{1D,i},X_{2,i},S_i;Y_{3,i})-I(S_i;X_{2,i},Y_{3,i}) \nonumber\\
   & = \sum_{i=1}^{n} I(X_{1D,i};Y_{3,i}|S_i,X_{2,i})+I(X_{2,i};Y_{3,i})-I(X_{2,i};S_i)\nonumber\\
   & \stackrel{(j)}{=} \sum_{i=1}^{n} I(X_{1D,i};Y_{3,i}|S_i,X_{2,i})+I(X_{2,i};Y_{3,i}),
\label{SecondTerm__TheoremUpperBoundDegenerateRelayModel}
\end{align}}
where: $(e)$ follows from the fact that the state $S^n$ is i.i.d. and is independent of the message $W$; $(f)$ follows from $(W,S^n,Y^{i-1}_3) \leftrightarrow (X_{1D,i},X_{2,i},S_i) \leftrightarrow Y_{3,i}$ is a Markov chain; $(g)$ follows from $Y^{i-1}_2 \leftrightarrow (W,S^{i-1},Y^{n}_3)  \leftrightarrow S_i$ is a Markov chain; $(h)$ follows from the fact that $X_{2,i}$ is a deterministic function of $Y^{i-1}_2$; $(i)$ follows from the fact that conditioning reduces entropy; and $(j)$ holds since $X_{2,i}$ is independent of $S_i$.
\end{proof}

We introduce a random variable $T$ which is uniformly  distributed over $\{1,\cdots,n\}$. Set $S=S_T$, $X_{1R}=X_{1R,T}$, $X_{1D}=X_{1D,T}$, $X_2=X_{2,T}$, $Y_2=Y_{2,T}$, and $Y_3=Y_{3,T}$. We substitute $T$ into the above bounds. Considering the bound \eqref{SecondTerm__TheoremUpperBoundDegenerateRelayModel}, we obtain
\begin{align}
&\frac{1}{n} \sum_{i=1}^{n} I(X_{1D,i};Y_{3,i}|S_i,X_{2,i})+I(X_{2,i};Y_{3,i})\nonumber\\
&=I(X_{1D};Y_3|S,X_2,T)+I(X_2;Y_3|T)\nonumber\\
&=I(X_{1D},X_2,S;Y_3|T)-I(S;X_2,Y_3|T)
\label{UpperBoundWithTimeSharingRandomVariablePart2}
\end{align}
and, similarly, 
\begin{align}
&\frac{1}{n} \sum_{i=1}^{n} I(X_{1R,i};Y_{2,i}|X_{2,i},S_i) + I(X_{1D,i};Y_{3,i}|X_{2,i},S_i)\nonumber\\
&=I(X_{1R};Y_2|S,X_2,T)+I(X_{1D};Y_3|S,X_2,T)
\label{UpperBoundWithTimeSharingRandomVariablePart1}
\end{align}
where the distribution on $(T,S,X_{1R},X_{1D},X_2,Y_2,Y_3)$  from a given code is of the form
\begin{align}
P_{T,S,X_{1R},X_{1D},X_2,Y_2,Y_3} &= Q_SP_TP_{X_2|T}P_{X_{1R}|X_2,T}P_{X_{1D}|S,X_2,T}\nonumber\\
&\hspace{1cm}{\times}W_{Y_2|S,X_{1R}}W_{Y_3|S,X_{1D},X_2}.
\label{AllowedDistributionUpperBoundWithTimeSharingRandomVariable}
\end{align}

We now eliminate the variable $T$ from \eqref{UpperBoundWithTimeSharingRandomVariablePart2} and \eqref{UpperBoundWithTimeSharingRandomVariablePart1} as follows. The right-hand side of \eqref{UpperBoundWithTimeSharingRandomVariablePart2} can be bounded as
{\allowdisplaybreaks
\begin{align}
& I(X_{1D},X_2,S;Y_3|T)-I(S;X_2,Y_3|T)\nonumber\\
& \stackrel{(k)}{\leq} H(Y_3)-H(Y_3|X_{1D},X_2,S)-H(S|T)+H(S|X_2,Y_3,T)\nonumber\\
& = I(X_{1D},X_2,S;Y_3)-H(S|T)+H(S|X_2,Y_3,T)\nonumber \\
& \stackrel{(l)}{\leq} I(X_{1D},X_2,S;Y_3)-H(S)+H(S|X_2,Y_3)\nonumber \\
& = I(X_{1D},X_2,S;Y_3)-I(S;X_2,Y_3)\nonumber\\
& = I(X_{1D};Y_3|S,X_2)+I(X_2;Y_3),
\label{UpperBoundTheoremDegenerateRelayModelPart2}
\end{align}}
where:\\
$(k)$ holds since $H(Y_3|T) \leq H(Y_3)$ and $H(Y_3|X_{1D},X_2,S,T)=H(Y_3|X_{1D},X_2,S)$ (by the Markovian relation $T \leftrightarrow (X_{1D},X_2,S) \leftrightarrow Y_3$); and\\
$(l)$ holds  since $S$ is independent of $T$ and $H(S|X_{1D},Y_3,T) \leq H(S|X_{1D},Y_3)$.\\
Similarly,  right-hand side of \eqref{UpperBoundWithTimeSharingRandomVariablePart2} can be bounded as
\begin{align}
I(X_{1R};Y_2|S,X_2,T)+I(X_{1D};Y_3|S,X_2,T) &\leq I(X_{1R};Y_2|S,X_2)+I(X_{1D};Y_3|S,X_2).
\label{UpperBoundTheoremDegenerateRelayModelPart1} 
\end{align}
Finally, combining \eqref{FanosInequality}, \eqref{SecondTerm__TheoremUpperBoundDegenerateRelayModel}, \eqref{UpperBoundTheoremDegenerateRelayModelPart2} at one hand, and \eqref{FanosInequality}, \eqref{FirstTerm__TheoremUpperBoundDegenerateRelayModel}, \eqref{UpperBoundTheoremDegenerateRelayModelPart1} at the other hand, we get
\begin{subequations}
\begin{align}
 R & \:\: \leq \:\: I(X_{1D};Y_3|S,X_2)+I(X_2;Y_3)\\
 R & \:\: \leq \:\: I(X_{1R};Y_2|S,X_2)+I(X_{1D};Y_3|S,X_2),
\end{align}
\label{UpperBound__Theorem__Degenerate__DiscreteModel}
\end{subequations}
where the distribution on $(S,X_{1R},X_{1D},X_2,Y_2,Y_3)$, obtained by  marginalizing \eqref{AllowedDistributionUpperBoundWithTimeSharingRandomVariable} over the variable $T$, has the form given in \eqref{MeasureForUpperBound__DiscreteModelWithHyperSource}.

We conclude that, for a given sequence of $(\epsilon_n,n,R)-$codes with $\epsilon_n$ going to zero as $n$ goes to infinity, there exists a probability distribution of the form \eqref{MeasureForUpperBound__DiscreteModelWithHyperSource} such that the rate $R$ satisfies \eqref{UpperBound__Theorem__Degenerate__DiscreteModel}. This completes the proof of Theorem~\ref{Theorem__UpperBound__DiscreteModelWithHyperSource}.

\renewcommand{\theequation}{E-\arabic{equation}}
\setcounter{equation}{0}  
\subsection{Proof of Theorem~\ref{Theorem__LowerBound1__GeneralGaussianChannel}}\label{appendixTheorem__LowerBound1__GeneralGaussianChannel}

The encoding and transmission scheme is as follows. Let $P_{1r} \geq 0$, $P_{1d} \geq 0$ and $D \geq 0$ be given such that $P_{1r}+P_{1d} \leq P_1$ and $0 \leq D \leq Q$. Also, consider the test channel $\hat{S}_R=aS+\tilde{S}_R$, where  $a:=1-D/Q$ and $\tilde{S}_R$ is a Gaussian random variable with zero mean and variance $\sigma^2_{\tilde{S}_R}=D(1-D/Q)$, independent from $S$. Using this test channel, we calculate $\mathbb{E}[(S-\hat{S}_R)^2]=D$ and $\mathbb{E}[\hat{S}^2_R]=Q-D$. Let $X_2 \sim \mc N(0,P_2)$ be jointly Gaussian with $\hat{S}_R$ with $\mathbb{E}[X_2\hat{S}_R]=0$ and independent from $S$, and $X_{SR} \sim \mc N(0,{\theta}P_{1r})$ jointly Gaussian with $(S,\hat{S}_R)$ with $\mathbb{E}[X_{SR}S]=0$ and $\mathbb{E}[X_{SR}\hat{S}_R]=0$, where $0 \leq \theta \leq 1$. Also, let $X_{WR} \sim \mc N(0,\bar{\theta}P_{1r})$ be jointly Gaussian with $(X_2,S)$ and independent of $X_{SR}$, with $\mathbb{E}[X_{WR}S]=\sigma_{1s}$ and $\mathbb{E}[X_{WR}X_2]=\sigma_{12}$; and $X_{WD} \sim \mc N(0,P_{1d})$ jointly Gaussian with and independent of $(X_{WR},X_{SR},X_2,S,\hat{S}_R)$. In what follows, we use the random variables $V$, $U$, $U_1$ and $U_R$ given by \eqref{RandomVariables__LowerBound1__GeneralGaussianChannel} to generate the auxiliary codewords $V_i$, $U_i$, $U_{1i}$ and $U_{Ri}$ which we will use in the sequel. Also, recall the definition of $\tilde{Q}$, $\xiup$ and $\alpha_2$ in \eqref{Variance__EquivalentState__LowerBound1__GeneralGaussianChannel} and \eqref{ScaleFactors__DPCs__LowerBound1__GeneralGaussianChannel}, respectively, which we will use in the rest of this proof.

We decompose the message $W$ to be sent from the source into two parts $W_r$ and $W_d$. The input $X^n_1$ from the source is divided into three independent parts, i.e., $X^n_1=X^n_{SR}+X^n_{wr}+X^n_{wd}$, where $X^n_{SR}$ carries a description $\hat{S}^n_R$ of the state $S^n$ that is intended to be recovered only at the relay and has power constraint $n{\theta}P_{1r}$, $X_{wr}^n$ carries message $W_r$ and has power constraint $n\bar{\theta}P_{1r}$ and $X_{wd}^n$ carries message $W_d$ and has power constraint $nP_{1d}$, with $P_1=P_{1r}+P_{1d}$. The message $W_r$ is sent through the relay at rate $R_r$ and the message $W_d$ is sent directly to the destination at rate $R_d$. The total rate is $R=R_r+R_d$.

\vspace{0.2cm}

As in the discrete case, a block Markov encoding is used. Let $w_i=(w_{ri},w_{di}) \in [1,2^{nR_r}]{\times}[1,2^{nR_d}]$ denote the message to be transmitted in block $i$ and $\dv s[i]$ denote the state controlling the channel in block $i$. The source quantizes $\dv s[i]$ into $\hat{\dv s}_R[\iota_{Ri-1}]$, where $\iota_{Ri-1} \in [1,2^{n\hat{R}_R}]$. Using the aforementioned test channel, the source can encode $\dv s[i]$ successfully at the quantization rate
\begin{align}
\hat{R}_R &=I(S;\hat{S}_R)\nonumber\\\
        &= \frac{1}{2}\log(\frac{Q}{D}).
\label{QuantizationRate__LowerBound1__GeneralGaussianChannel__SourceCodingConstraint}
\end{align}

In the beginning of block $i$, the relay has decoded correctly message $w_{ri-1}$ and the index $\iota_{Ri-1}$ of the description $\hat{\dv s}_{R}[\iota_{Ri-1}]$ sent by the source in the previous block $i-1$ (this will be justified below) and sends a Gaussian signal $\dv x_2[w_{ri-1}]$ which carries message $w_{ri-1}$ and is obtained via a DPC considering $\hat{\dv s}_R[\iota_{Ri-1}]$ as noncausal channel state information at the transmitter, as
\begin{equation}
\dv x_2[w_{ri-1}] = \frac{\sqrt{P_2}}{\rho_{12}\sqrt{\bar{\theta}P_{1r}}+\sqrt{P_2}}\Big(\dv v[i] - {\alpha}_2\xiup\hat{\dv s}_R[\iota_{Ri-1}]\Big),
\label{Precoding__AuxiliaryRandomVariable_V}
\end{equation}
where the components of $\dv v[i]$ are generated i.i.d. using the auxiliary random variable $V$.

\noindent Let $\iota_{Ri}$ be the index associated with the state $\dv s[i+1]$ of the next block $i+1$. In the beginning of block $i$, the source sends a superposition of three Gaussian vectors,
\begin{align}
\dv x_1[i] &= \dv x_{SR}[\iota_{Ri}]+\dv x_{wr}[w_{ri-1},w_{ri}]+\dv x_{wd}[w_{di}]\nonumber\\
\dv x_{wr}[w_{ri-1},w_{ri}] &= \rho_{1s}\sqrt{\frac{\bar{\theta}P_{1r}}{Q}}\dv s[i]+\rho_{12}\sqrt{\frac{\bar{\theta}P_{1r}}{P_2}}\dv x_2[w_{ri-1}]+\dv x'_{wr}[w_{ri}].
\label{SourceInput__LowerBound1__GeneralGaussianChannel}
\end{align}
In \eqref{SourceInput__LowerBound1__GeneralGaussianChannel}, the vectors $\dv x_{SR}[\iota_{Ri}]$ and $\dv x_{wd}[w_{di}]$ are generated i.i.d. using the auxiliary random variables $X_{SR}$ and $X_{WD}$, respectively; and the vector $\dv x'_{wr}[w_{ri}]$ has power $n(1-\rho^2_{12}-\rho^2_{1s})\bar{\theta}P_{1r}$ and is independent of $\dv s[i]$, $\dv x_2[w_{ri-1}]$, $\dv x_{SR}[\iota_{Ri}]$ and $\dv x_{wd}[w_{di}]$. Furthermore, the vector $\dv x_{SR}[\iota_{Ri}]$ carries a description $\hat{\dv s}_R[\iota_{Ri}]$ of the state $\dv s[i+1]$ that affects transmission in the next block $i+1$, intended to be recovered only at the relay; the vector $\dv x_2[w_{ri-1}]$ carries cooperative information $w_{ri-1}$, and the vector $\dv x'_{wr}[w_{ri}]$ carries new information $w_{ri}$. The vectors  $\dv x_{SR}[\iota_{Ri}]$, $\dv x_{wd}[w_{di}]$ and $\dv x'_{wr}[w_{ri}]$ are obtained via DPCs considering $(s[i],\hat{\dv s}_R[\iota_{Ri-1}])$ as noncausal channel state information at the transmitter, as
\begin{subequations}
\begin{align}
\label{Precoding__AuxiliaryRandomVariable_UR}
\dv x_{SR}[\iota_{Ri}] &= \dv u_R[i]-\frac{{\theta}P_{1r}}{{\theta}P_{1r}+N_2+P_{1d}}(1-\alpha)\dv s[i]\\
\label{Precoding__AuxiliaryRandomVariable_U1}
\dv x_{wd}[w_{di}] &= \dv u_1[i]-\frac{P_{1d}}{P_{1d}+N_3+{\theta}P_{1r}}\xiup(1-\alpha)\Big(\dv s[i]-\alpha_2\hat{\dv s}_R[\iota_{Ri-1}]\Big)\\
\dv x'_{wr}[w_{ri}] &= \dv u[i]-\alpha\xiup\Big(\dv s[i]-\alpha_2\hat{\dv s}_R[\iota_{Ri-1}]\Big)
\label{Precoding__AuxiliaryRandomVariable_U}
\end{align}
\end{subequations}
where the components of $\dv u_R[i]$, $\dv u_1[i]$ and $\dv u[i]$ are generated i.i.d. using the auxiliary random variables $U_R$, $U_1$ and $U$ respectively.

We now describe the decoding operations (we give simple arguments; the rigorous decoding uses joint typicality testing). Consider first the decoding at the relay. In block $i$, the relay receives
\begin{equation}
\dv y_2[i]=\dv x_{SR}[\iota_{Ri}]+\rho_{12}\sqrt{\frac{\bar{\theta}P_{1r}}{P_2}}\dv x_2[w_{ri-1}]+\dv x'_{wr}[w_{ri}]+\Big(1+\rho_{1s}\sqrt{\frac{\bar{\theta}P_{1r}}{Q}}\Big)\dv s[i]+(\dv z_2[i]+\dv x_{wd}[w_{di}]).
\end{equation}
The relay knows $w_{ri-1}$ and $\iota_{Ri-1}$ and decodes the pair $(w_{ri},\iota_{Ri})$ from $\dv y_2[i]$. The relay decodes $w_{ri}$ and $\iota_{Ri}$ successively, starting by $w_{ri}$. To decode $w_{ri}$, the relay subtracts out the quantity $\big(\rho_{12}\sqrt{\bar{\theta}P_{1r}/P_2}\dv x_2[w_{ri-1}] + \alpha_2\xiup\hat{\dv s}_R[\iota_{Ri-1}]\big)$ from $\dv y_2[i]$ to make the channel equivalent to
\begin{equation}
\tilde{\dv y}_2[i]=\dv x'_{wr}[w_{ri}] + \xiup\Big(\dv s[i]-\alpha_2\hat{\dv s}_R[\iota_{Ri-1}]\Big) + (\dv z_2[i]+\dv x_{SR}[\iota_{Ri}]+\dv x_{wd}[w_{di}]).
\label{Received__at__Relay__LowerBound1__GeneralGaussianChannel}
\end{equation}
The relay decodes message $w_{ri}$ from $\tilde{\dv y}_2[i]$ treating signals $\dv x_{SR}[\iota_{Ri}]$ and $\dv x_{wd}[w_{di}]$ as unknown independent noises. This can be done reliably as long as $n$ is large and
\begin{align}
R_r &\leq I(U;\tilde{Y}_2)-I(U;S-\alpha_2\hat{S}_R)\nonumber\\
  &= R\Big(\alpha,(1-\rho^2_{12}-\rho^2_{1s})\bar{\theta}P_{1r},\xiup^2\tilde{Q},N_2+{\theta}P_{1r}+P_{1d}\Big)
\label{RateConstraint__Decoding__at__Relay__LowerBound1__GeneralGaussianChannel}
\end{align}
where the equality follows through straightforward algebra which we omit here for brevity (note that the variance of the additive state $\xiup(S-\alpha_2\hat{S}_R)$  in \eqref{Received__at__Relay__LowerBound1__GeneralGaussianChannel} is $\xiup^2\mathbb{E}[(S-\alpha_2\hat{S}_R)^2]=\xiup^2[(1-\alpha_2)^2Q-\alpha_2(\alpha_2-2)D]:=\xiup^2\tilde{Q}$). Next, for the decoding of $\iota_{Ri}$, the relay subtracts out the quantity $\big(\dv u[i]-(1-\alpha)\alpha_2\xiup\hat{\dv s}_R[\iota_{Ri-1}]\big)$ from $\tilde{\dv y}_2[i]$ to make the channel equivalent to
\begin{equation}
\breve{\dv y}_2[i]=\dv x_{SR}[\iota_{Ri}]+ (1-\alpha)\dv s[i] +(\dv z_2[i]+\dv x_{wd}[w_{di}]).
\end{equation}
The relay decodes the index $\iota_{Ri}$ from $\breve{\dv y}_2[i]$ correctly as long as $n$ is large and
\begin{align}
\hat{R}_R &\leq I(U_R;\breve{Y}_2)-I(U_R;S)\nonumber\\
  &= \frac{1}{2}\log\Big(1+\frac{{\theta}P_{1r}}{N_2+P_{1d}}\Big).
\label{QuantizationRate__LowerBound1__GeneralGaussianChannel__ChannelCodingConstraint}
\end{align}

We now turn to the decoding at the destination at the end of block $i$. In block $i$, the destination receives
\begin{align}
\dv y_3[i] &= \dv x_1[i] + \dv x_2[w_{ri-1}] + \dv s[i] +\dv z_3[i]\nonumber\\
&= \Big(\rho_{12}\sqrt{\frac{\bar{\theta}P_{1r}}{P_2}}+1\Big)\dv x_2[w_{ri-1}]+\dv x'_{wr}[w_{ri}]+\dv x_{wd}[w_{di}]+\Big(\rho_{1s}\sqrt{\frac{\bar{\theta}P_{1r}}{Q}}+1\Big)\dv s[i]+(\dv z_3[i]+\dv x_{SR}[\iota_{Ri}]).
\label{OutputDestination__LowerBound1__GeneralGaussianChannel}
\end{align}
At the end of block $i$, the destination knows message $w_{ri-2}$ and decodes the pair $(w_{ri-1},w_{di-1})$ successively, treating the signal that carries the state description as unknown independent noise. It starts by decoding message  $w_{ri-1}$, using $(\dv y_3[i-1],\dv y_3[i])$. Note that $w_{ri-1}$ is carried by both auxiliary vectors $\dv v[i]$ and $\dv u[i-1]$. If $n$ is large, it can do so reliably at rate
\begin{align}
R_r &\leq I(V,U;Y_3)-I(V,U;S,\hat{S}_R)\nonumber\\
    &= [I(V;Y_3)-I(V;\hat{S}_R)]+[I(U;Y_3|V)-I(U;S,\hat{S}_R|V)]
\end{align}
where the equality follows since the choice of $(V,\hat{S}_R)$ in \eqref{RandomVariables__LowerBound1__GeneralGaussianChannel} satisfies $V \leftrightarrow \hat{S}_R \leftrightarrow S$ is a Markov chain.  

\noindent We first compute the term $[I(V;Y_3)-I(V;\hat{S}_R)]$. Let $\tilde{\dv s}[i]$ be the estimation error of $\xiup\dv s[i]$ given $\hat{\dv s}_R[\iota_{Ri-1}]$ under minimum mean square error criterion. Since $\dv s[i]$ and $\hat{\dv s}_R[\iota_{Ri-1}]$ are jointly Gaussian, $\tilde{\dv s}[i]$ is i.i.d. Gaussian with variance $\mathbb{E}[({\xiup}S-{\xiup}\hat{S}_R)^2]=\xiup^2D$ per element and is independent from $\hat{\dv s}_R[\iota_{Ri-1}]$. Thus, we can alternatively write the output $\dv y_3[i]$ as
\begin{align}
\dv y_3[i] &= \Big(\rho_{12}\sqrt{\frac{\bar{\theta}P_{1r}}{P_2}}+1\Big)\dv x_2[w_{ri-1}]+\dv x'_{wr}[w_{ri}]+\dv x_{wd}[w_{di}]+\xiup\hat{\dv s}_R[\iota_{Ri-1}]+(\dv z_3[i]+\dv x_{SR}[\iota_{Ri}]+\tilde{\dv s}[i]).
\label{EquivalentOutputDestination__LowerBound1__GeneralGaussianChannel}
\end{align}
With the choice of the auxiliary random variable $V$ as in \eqref{RandomVariables__LowerBound1__GeneralGaussianChannel} and that of the associated Costa's scale factor $\alpha_2$ set to its optimal value as in \eqref{ScaleFactors__DPCs__LowerBound1__GeneralGaussianChannel}, the destination decodes the vector $\dv v[i]$  correctly from $\dv y_3[i]$ at rate 
\begin{equation}
I(V;Y_3)-I(V;\hat{S}_R) = \frac{1}{2}\log\Big(1+\frac{(\rho_{12}\sqrt{\bar{\theta}P_{1r}}+\sqrt{P_2})^2}{N_3+\xiup^2D+{\theta}P_{1r}+(1-\rho^2_{12}-\rho^2_{1s})\bar{\theta}P_{1r}+P_{1d}}\Big)
\label{Rate__of__CooperativeInformation__LowerBound1__GeneralGaussianChannel}
\end{equation}
where the equality follows through straightforward algebra. 
\noindent Let us now compute the term $[I(U;Y_3|V)-I(U;S,\hat{S}_R|V)]$. Observing that the destination can peel off $\dv v[i-1]$ from $\dv y_3[i-1]$ to make the channel equivalent to
\begin{align}
\tilde{\dv y}_3[i-1] &= \dv y_3[i-1]-\Big(\Big(\rho_{12}\sqrt{\frac{\bar{\theta}P_{1r}}{P_2}}+1\Big)\dv x_2[w_{ri-2}]+\alpha_2\xiup\hat{\dv s}_R[\iota_{Ri-2}]\Big)\nonumber\\
&= \dv x'_{wr}[w_{ri-1}]+\xiup\dv s[i-1]-\alpha_2\xiup\hat{\dv s}_R[\iota_{Ri-2}]+(\dv z_3[i-1]+\dv x_{SR}[\iota_{Ri-1}]+\dv x_{wd}[w_{di-1}]),
\end{align}
it is easy to see that, if $n$ is large and with the choice of the auxiliary random variable $U$ as in \eqref{RandomVariables__LowerBound1__GeneralGaussianChannel}, the destination obtains the vector $\dv u[i-1]$ correctly from $\dv y_3[i-1]$ at rate 
\begin{align}
I(U;Y_3|V)-I(U;S,\hat{S}_R|V) &= I(U;\tilde{Y}_3)-I(U;{\xiup}(S-\alpha_2\hat{S}_R))\nonumber\\
&= R\Big(\alpha,(1-\rho^2_{12}-\rho^2_{1s})\bar{\theta}P_{1r},\xiup^2\tilde{Q},N_3+{\theta}P_{1r}+P_{1d}\Big)
\label{Rate__of__NewInformation__LowerBound1__GeneralGaussianChannel}
\end{align}
where the last equality follows through straightforward algebra.

Finally, the destination can peel off $\dv u[i-1]$ from $\tilde{\dv y}_3[i-1]$ to make the channel equivalent to
\begin{align}
\breve{\dv y}_3[i-1] &= \tilde{\dv y}_3[i-1]-\Big(\dv x'_{wr}[w_{ri-1}]+\alpha\xiup(\dv s[i-1]-\alpha_2\hat{\dv s}_R[\iota_{Ri-2}])\Big)\nonumber\\
&= \dv x_{wd}[w_{di-1}]+\xiup(1-\alpha)(\dv s[i-1]-\alpha_2\xiup\hat{\dv s}_R[\iota_{Ri-2}])+(\dv z_3[i-1]+\dv x_{SR}[\iota_{Ri-1}]).
\label{EquivalentChannel__DecodingMessageWd}
\end{align}
From \eqref{EquivalentChannel__DecodingMessageWd}, it is easy to see that if $n$ is large, and with the choice of the auxiliary random variable $U_1$ as in \eqref{RandomVariables__LowerBound1__GeneralGaussianChannel}, the destination obtains the vector $\dv u_1[i-1]$ (which carries message $w_{di-1}$) correctly at rate
\begin{align}
 R_d &\leq I(U_1;\breve{Y}_3)-I(U_1;\xiup(1-\alpha)(S-\alpha_2\hat{S}_R)) \nonumber\\
&=\frac{1}{2}\log(1+\frac{P_{1d}}{N_3+{\theta}P_{1r}}).
\label{Rate__of__MessageWd__LowerBound1__GeneralGaussianChannel}
\end{align}

\noindent Finally, for given $D$, adding \eqref{RateConstraint__Decoding__at__Relay__LowerBound1__GeneralGaussianChannel} and \eqref{Rate__of__MessageWd__LowerBound1__GeneralGaussianChannel}, we obtain the first term of the minimization in \eqref{LowerBound1__GeneralGaussianChannel}; and adding \eqref{Rate__of__CooperativeInformation__LowerBound1__GeneralGaussianChannel}, \eqref{Rate__of__NewInformation__LowerBound1__GeneralGaussianChannel} and \eqref{Rate__of__MessageWd__LowerBound1__GeneralGaussianChannel}, we obtain the second term of the minimization in \eqref{LowerBound1__GeneralGaussianChannel}. Also, similar to in the proof of Theorem~\ref{Theorem__LowerBound2__GeneralGaussianChannel}, observing that the rate terms in \eqref{LowerBound1__GeneralGaussianChannel} decrease with $D$, we obtain the lower bound in Theorem~\ref{Theorem__LowerBound1__GeneralGaussianChannel} by taking the equality in \eqref{QuantizationRate__LowerBound1__GeneralGaussianChannel__ChannelCodingConstraint} and maximizing the minimization in \eqref{LowerBound1__GeneralGaussianChannel} over $P_{1r} \geq 0$, $P_{1d} \geq 0$ such that $0 \leq P_{1r}+P_{1d} \leq P_1$, $\theta \in [0,1]$, $\rho_{12} \in [0,1]$ and $\rho_{1s} \in [-1,0]$ such that $0 \leq \rho^2_{12}+\rho^2_{1s} \leq 1$ and $\alpha \in \mathbb{R}$ such that the RHS of \eqref{RateConstraint__Decoding__at__Relay__LowerBound1__GeneralGaussianChannel} is non-negative and the sum of the RHS of \eqref{Rate__of__NewInformation__LowerBound1__GeneralGaussianChannel} and the RHS of \eqref{Rate__of__MessageWd__LowerBound1__GeneralGaussianChannel} is non-negative. This completes the proof.

\renewcommand{\theequation}{F-\arabic{equation}}
\setcounter{equation}{0}  
\subsection{Proof of Theorem~\ref{Theorem__UpperBound__GaussianModelWithHyperSource}}\label{appendixTheorem__UpperBound__GaussianModelWithHyperSource}

In this section, we first use the upper bound for the DM case in Theorem~\ref{Theorem__UpperBound__DiscreteModelWithHyperSource} to obtain a new upper bound on the capacity of the state-dependent additive Gaussian model \eqref{GaussianChannelModel__RCwithHyperSource}. Then, we show that this new upper bound is maximized by jointly Gaussian $(S,X_{1R},X_{1D},X_2,Z_2,Z_3)$.

From Theorem~\ref{Theorem__UpperBound__DiscreteModelWithHyperSource}, we have that, given any $(\epsilon_n,n,R)$ sequence of codes with average error probability $P^n_e \longrightarrow 0$ as $n \longrightarrow +\infty$, the transmission rate $R$ satisfies 
\begin{align}
R &\leq \min \:\Big\{I(X_{1R};Y_2|X_2,S),\: I(X_2;Y_3)\Big\}+I(X_{1D};Y_3|X_2,S)
\end{align}
for some joint measure of the form
\begin{align}
P_{S,X_{1R},X_{1D},X_2,Y_2,Y_3} &= Q_SP_{X_2}P_{X_{1R}|X_2}P_{X_{1D}|X_2,S}W_{Y_2|X_{1R},S}W_{Y_3|X_{1D},X_2,S}.
\label{Proof__MeasureForUpperBound__DiscreteModelWithHyperSource}
\end{align}

Since the channel structure~\eqref{GaussianChannelModel__RCwithHyperSource} satisfies $W_{Y_2|X_{1R},X_2,S}=W_{Y_2|X_{1R},S}$, it follows that
\begin{align}
I(X_{1R};Y_2|S,X_2)&=H(Y_2|S,X_2)-H(Y_2|S,X_2,X_{1R})\nonumber\\
&= H(Y_2|S,X_2)-H(Y_2|S,X_{1R})\nonumber\\
&\leq H(Y_2|S)-H(Y_2|S,X_{1R})\nonumber\\
&= I(X_{1R};Y_2|S).
\end{align}

An upper bound on the capacity of the channel~\eqref{GaussianChannelModel__RCwithHyperSource} is then given by 
\begin{align}
R &\leq \min \:\Big\{I(X_{1R};Y_2|S),\: I(X_2;Y_3)\Big\}+I(X_{1D};Y_3|X_2,S)
\label{Proof__UpperBound__GaussianModelWithHyperSource}
\end{align}
for some joint measure of the form
\begin{align}
P_{S,X_{1R},X_{1D},X_2,Y_2,Y_3} &= Q_SP_{X_2}P_{X_{1R}}P_{X_{1D}|X_2,S}W_{Y_2|X_{1R},S}W_{Y_3|X_{1D},X_2,S}.
\label{Proof__MeasureForUpperBound__GaussianModelWithHyperSource}
\end{align}
(Note that, in contrast to in Theorem~\ref{Theorem__UpperBound__DiscreteModelWithHyperSource} and \eqref{Proof__MeasureForUpperBound__DiscreteModelWithHyperSource}, the inputs $X_{1R}$ and $X_2$ are independent in \eqref{Proof__MeasureForUpperBound__GaussianModelWithHyperSource}).

Fix a joint distribution on $(S,X_{1R},X_{1D},X_2,Y_2,Y_3)$ of the form~\eqref{Proof__MeasureForUpperBound__GaussianModelWithHyperSource} satisfying
\begin{align}
&\mathbb{E}[X^2_{1R}]=\tilde{P}_{1R} \leq P_{1R}, \quad \mathbb{E}[X^2_{1D}]=\tilde{P}_{1D} \leq P_{1D}, \quad \mathbb{E}[X^2_2]=\tilde{P}_2 \leq P_2,\nonumber\\
&\mathbb{E}[X_{1D}X_2]=\sigma_{12},\quad \mathbb{E}[X_{1D}S]=\sigma_{1s}.
\label{FixedSecondMomentsOuterBoundFullDuplexGaussianCase}
\end{align}
We shall also use the correlation coefficients $\rho_{12} \in [-1,1]$, $\rho_{1s} \in [-1,1]$ defined as
\begin{equation}
\rho_{12}=\frac{\sigma_{12}}{\sqrt{\tilde{P}_{1D}\tilde{P}_2}},\quad \rho_{1s}=\frac{\sigma_{1s}}{\sqrt{\tilde{P}_{1D}Q}}.
\label{CorrelationCoefficientsDefinition__UpperBound__GaussianChannel}
\end{equation}

We first compute the first term in the minimization on the RHS of \eqref{Proof__UpperBound__GaussianModelWithHyperSource}. We have

\begin{align}
\label{UpperBound__RC__Orth__Components__FirstTerm__Step8}
R & \leq  I(X_{1R};Y_2|S) + I(X_{1D};Y_3|X_2,S) \\
&= h(X_{1R}+Z_2|S)-h(Z_2)+h(X_{1D}+Z_3|X_2,S)-h(Z_3)\\
&\stackrel{(a)}{\leq} h(X_{1R}+Z_2)-h(Z_2)+h(X_{1D}+Z_3|X_2,S)-h(Z_3)\\
&\stackrel{(b)}{\leq} \frac{1}{2}\log\Big(1+\frac{\tilde{P}_{1R}}{N_2}\Big)+\frac{1}{2}\log\Big(1+\frac{\tilde{P}_{1D}(1-\rho^2_{12}-\rho^2_{1s})}{N_3}\Big), 
\label{FirstTermUpperBoundGaussianFullDuplex}
\end{align}
where: $(a)$ holds since conditioning reduces entropy; and $(b)$ holds since the conditional differential entropy $h(X_{1R}+Z_2)$ is maximized if $(X_{1R},Z_2)$ are jointly Gaussian, and the conditional differential entropy $h(X_{1D}+Z_3|X_2,S)$ is maximized if $(S,X_{1D},X_2,Z_3)$ are jointly Gaussian.

We now compute the term $[I(X_2;Y_3)+I(X_{1D};Y_3|X_2,S)]$. We have
\begin{align}
I(X_2;Y_3)+I(X_{1D};Y_3|X_2,S) &\stackrel{(c)}{=} I(X_{1D};Y_3|X_2,S) + I(X_2;Y_3)-I(X_2;S)\nonumber\\
&= I(X_{1D};Y_3|X_2,S) + I(X_2;Y_3|S)-I(X_2;S|Y_3)\nonumber\\
&= h(Y_3|S)-h(Y_3|S,X_{1D},X_2)-h(S|Y_3)+h(S|X_1,Y_3)\nonumber\\
&= h(Y_3)-h(S)+h(S|X_2,Y_3)-h(Z_3)
\label{SecondTermUpperBoundFullDuplex}
\end{align}
where: $(c)$ follows since $X_2$ and $S$ are independent. 

For fixed second moments \eqref{FixedSecondMomentsOuterBoundFullDuplexGaussianCase}, we have
\begin{align}
h(Y_3) & \leq \frac{1}{2}\log(2\pi{e})(\tilde{P}_{1D}+\tilde{P}_2+2\sigma_{12}+2\sigma_{1s}+Q+N_3),
\label{EntropyOfReceivedForSecondTermUpperBoundFullDuplex}
\end{align}
where equality is attained if $Y_3$ is Gaussian. Similarly, the term $h(S|X_2,Y_3)$ is maximized if $(S,X_2,Y_3)$ are jointly Gaussian. Let $\hat{S}(X_2,Y_3)=\mathbb{E}[S|X_2,Y_3]$ be the MMSE estimator of $S$ given $(X_2,Y_3)$, i.e.,
\begin{align}
\hat{S}(X_2,Y_3) &= \mathbb{E}[S|X_2,X_{1D}+S+Z_3]\nonumber\\
&= \gamma_1X_2+\gamma_2(X_{1D}+S+Z_3)
\end{align}
with
\begin{align}
\gamma_1 &= -\frac{\sigma_{12}(Q+\sigma_{1s})}{\tilde{P}_2(\tilde{P}_{1D}+2\sigma_{1s}+Q+N_3)-\sigma^2_{12}}\nonumber\\
\gamma_2 &= \frac{\tilde{P}_2(Q+\sigma_{1s})}{\tilde{P}_2(\tilde{P}_{1D}+2\sigma_{1s}+Q+N_3)-\sigma^2_{12}}.
\end{align}
\begin{align}
 h(S|X_2,Y_3) & = h(S-\hat{S}(X_2,Y_3)|X_2,Y_3)\nonumber\\
              &\leq h(S-\gamma_1X_2-\gamma_2(X_{1D}+S+Z_3))\nonumber\\
	      &= \frac{1}{2}\log(2\pi{e})\mathbb{E}\Big[\Big(S-\gamma_1X_2-\gamma_2(X_{1D}+S+Z_3)\Big)^2\Big]\nonumber\\
	      &= \frac{1}{2}\log\Big((2\pi{e})\frac{Q\tilde{P}_{1D}\tilde{P}_2+\tilde{P}_2N_3Q-\sigma^2_{1s}\tilde{P}_2-\sigma^2_{12}Q}{\tilde{P}_2(\tilde{P}_{1D}+2\sigma_{1s}+Q+N_3)-\sigma^2_{12}}\Big),
\label{ConditionalEntropyForSecondTermUpperBoundFullDuplex}
\end{align}
where the inequality is attained with equality if $S,X_{1D},X_2,Y_3$ are jointly Gaussian. Then, from \eqref{SecondTermUpperBoundFullDuplex}, \eqref{EntropyOfReceivedForSecondTermUpperBoundFullDuplex} and \eqref{ConditionalEntropyForSecondTermUpperBoundFullDuplex} and straightforward algebra, we obtain
\begin{align}
 I(X_2;Y_3)+I(X_{1D};Y_3|S,X_2) 
 &=\frac{1}{2}\log\Big(1+\frac{(\sqrt{\tilde{P}_2}+\rho_{12}\sqrt{\tilde{P}_{1D}})^2}{\tilde{P}_{1D}(1-\rho^2_{12}-\rho^2_{1s})+(\sqrt{Q}+\rho_{1s}\sqrt{\tilde{P}_{1D}})^2+N_3}\Big)\nonumber\\
 &\hspace{2.5cm}+\frac{1}{2}\log\Big(1+\frac{\tilde{P}_{1D}(1-\rho^2_{12}-\rho^2_{2s})}{N_3}\Big).
 \label{SecondTermUpperBoundGaussianFullDuplex}
 \end{align}

For convenience, let us now define the function $\Theta_1(\tilde{P}_{1R},\tilde{P}_{1D},\rho_{12},\rho_{1s})$ as the RHS of \eqref{FirstTermUpperBoundGaussianFullDuplex} and the function $\Theta_2(\tilde{P}_{1D},\tilde{P}_2,\rho_{12},\rho_{2s})$ as the RHS of \eqref{SecondTermUpperBoundGaussianFullDuplex}. From the above analysis, the capacity of the channel is upper-bounded as
\begin{align}
C \leq \max\:\min \{\Theta_1(\tilde{P}_{1R},\tilde{P}_{1D},\rho_{12},\rho_{1s}),\Theta_2(\tilde{P}_{1D},\tilde{P}_2,\rho_{12},\rho_{1s})\}
\label{IntermediaryStepOuterBoundGaussianFullDuplex}
\end{align}
where the maximization is over all covariance matrices of $(X_{1R},X_{1D},X_2,S)$ of the form 
\begin{align}
&\Lambda_{X_{1R},X_{1D},X_2,S}=
{\left(
\begin{array}{ccccc}
\tilde{P}_{1R} &    0    &   0   & 0\\
0  &  \tilde{P}_{1R} & \rho_{12}\sqrt{\tilde{P}_{1D}\tilde{P}_2} & \rho_{1s}\sqrt{\tilde{P}_{1D}Q}\\
0  &  \rho_{12}\sqrt{\tilde{P}_{1D}\tilde{P}_2} & \tilde{P}_2 &  0 \\
0  & \rho_{1s}\sqrt{\tilde{P}_{1D}Q} & 0 & Q
\end{array}
\right),}
\label{AllowableCovarianceMatrixOuterBoundFullDuplex}
\end{align}
that satisfy
\begin{equation}
\tilde{P}_{1R} \leq P_{1R}, \quad \tilde{P}_{1D} \leq P_{1D}, \quad \tilde{P}_2 \leq P_2
\end{equation}
and have non-negative discriminant,
\begin{equation}
Q\tilde{P}_{1R}\tilde{P}_{1D}\tilde{P}_2(1-\rho^2_{12}-\rho^2_{2s})\geq 0,
\end{equation}
i.e., for $Q > 0$,
\begin{equation}
\rho^2_{12}+\rho^2_{2s} \leq 1.
\label{ConditionOnRhosOuterBoundGaussianFullDuplex}
\end{equation}

Investigating $\Theta_1(\tilde{P}_{1R},\tilde{P}_{1D},\rho_{12},\rho_{1s})$ and $\Theta_2(\tilde{P}_{1D},\tilde{P}_2,\rho_{12},\rho_{1s})$, it can be seen that it suffices to consider $\rho_{12} \in [0,1]$ and $\rho_{1s} \in [-1,0]$ for the maximization in \eqref{IntermediaryStepOuterBoundGaussianFullDuplex}.

Also, it is easy to see that, for fixed $\tilde{P}_{1D}$, the functions $\Theta_1(\tilde{P}_{1R},\tilde{P}_{1D},\rho_{12},\rho_{1s})$ and $\Theta_2(\tilde{P}_{1D},\tilde{P}_2,\rho_{12},\rho_{1s})$ increase monotonically with $\tilde{P}_{1R}$ and $\tilde{P}_2$. So, for fixed $\tilde{P}_{1D}$, they are maximized at $\tilde{P}_{1R}=P_{1R}$ and $\tilde{P}_2=P_2$. To complete the proof, we should show that $\Theta_1(P_{1R},\tilde{P}_{1D},\rho_{12},\rho_{1s})$ and $\Theta_2(\tilde{P}_{1D},P_2,\rho_{12},\rho_{1s})$ are also maximized at $\tilde{P}_{1D}=P_{1D}$. 

\noindent It is clear that the function $\Theta_1(P_{1R},\tilde{P}_{1D},\rho_{12},\rho_{1s})$ increases with $\tilde{P}_{1D}$. The term $\Theta_2(\tilde{P}_{1D},P_2,\rho_{12},\rho_{1s})$ can be seen as the sum rate of a two-user state-dependent MAC with state information $\Delta_S^n$ known to one encoder, both encoders sending a common message and the informed encoder sending, in addition, an individual message \cite{SBSV07a}. As argued in \cite{SBSV07a}, this sum rate increases with the power of the informed encoder \cite[Appendix E]{SBSV07a}, i.e., $\tilde{P}_{1D}$ here. This concludes the proof of Theorem~\ref{Theorem__UpperBound__GaussianModelWithHyperSource}.

\renewcommand{\theequation}{G-\arabic{equation}}
\setcounter{equation}{0}  
\subsection{Proof of Theorem~\ref{Theorem__Capacity__FirstDegenerateGaussianModelWithHyperSource}}\label{appendixTheorem__Capacity__FirstDegenerateGaussianModelWithHyperSource}

1) \textit{Converse Part:} the proof of the converse part of Theorem~\ref{Theorem__Capacity__FirstDegenerateGaussianModelWithHyperSource} follows by noticing that the computation of the upper bound \eqref{Proof__UpperBound__GaussianModelWithHyperSource} in the proof of  Theorem~\ref{Theorem__UpperBound__GaussianModelWithHyperSource} for the special case \eqref{DegenerateGaussianChannelModel1__RCwithHyperSource}, and using the same jointly Gaussian distribution as in Appendix~\ref{appendixTheorem__UpperBound__GaussianModelWithHyperSource}, gives the RHS of \eqref{Capacity__FirstDegenerateGaussianModelWithHyperSource}.

2) \textit{Achievability Part:} Recall the lower bound in Corollary~\ref{Corollary__LowerBound1__DiscreteChannel}. With the choice $\hat{S}_R=\hat{S}_D=\O$,  $U_R=U_D=\O$, $U=X_{1R}$ independent of $S$ and  $V=X_2$ independent of $S$, we obtain
\begin{align}
R^{\text{lo}} &= \max \min \big\{\:I(X_{1R};Y_2|X_2), \:\: I(X_{1R},X_2;Y_3)\:\big\}+[I(U_1;Y_3|X_{1R},X_2)-I(U_1;S|X_{1R},X_2)]^{+}
\label{LowerBound__DiscreteModelWithHyperSource}
\end{align}
where $[x]^{+} := \max(x,0)$ and the maximization is over all measures of the form
\begin{align}
P_{S,U_1,X_{1R},X_{1D},X_2,Y_2,Y_3} &= Q_SP_{X_2}P_{X_{1R}|X_2}P_{U_1,X_{1D}|S,X_2}W_{Y_2|S,X_{1R}}W_{Y_3|X_{1D},X_2,S}.
\label{Measure__CorollaryLowerBound__DiscreteModelWithHyperSource}
\end{align}

In the proof of the direct part of Theorem~\ref{Theorem__Capacity__FirstDegenerateGaussianModelWithHyperSource} we compute the rate \eqref{LowerBound__DiscreteModelWithHyperSource} using an appropriate jointly Gaussian distribution on $(S,U_1,X_{1R},X_{1D},X_2)$. The algebra in this section is similar to that in the proof of \cite[Theorem 3]{ZKLV10} and \cite[Theorem 6]{SBSV07a}.

\noindent We first compute the term $[I(U_1;Y_3|X_{1R},X_2)-I(U_1;S|X_{1R},X_2)]$ in the RHS of \eqref{LowerBound__DiscreteModelWithHyperSource} because this gives insights about the distribution that we should use to compute the lower bound. We assume that $X_{1R}$, $X_{1D}$ and $X_2$ are jointly Gaussian random variables with zero-mean and variance $P_{1R}$, $P_{1D}$ and $P_2$, respectively. The random variables $X_{1R}$ and $X_2$  are independent and independent of the state $S$. The random variable $X_{1D}$ is independent of $X_{1R}$ and jointly Gaussian with $(S,X_2)$, with $\mathbb{E}[X_{1D}X_2]=\rho_{12}\sqrt{P_{1D}P_2}$ and $\mathbb{E}[X_{1D}S]=\rho_{1s}\sqrt{P_{1D}Q}$, for some correlation coefficients $\rho_{12} \in [-1,1]$ and $\rho_{1s} \in [-1,1]$.

Let $\hat{X}_{1D}=\mathbb{E}[X_{1D}|S,X_{1R},X_2]$ be the optimal linear estimator of $X_{1D}$ given $(S,X_{1R},X_2)$ under minimum mean square error criterion, and $X'_{1D}$ be the resulting estimation error (note that $\mathbb{E}[X_{1D}|S,X_{1R},X_2]=\mathbb{E}[X_{1D}|S,X_2]$). The estimator $\hat{X}_{1D}$ and the estimation error $X'_{1D}$ are given by
\begin{align}
\hat{X}_{1D} &=\rho_{12}\sqrt{\frac{P_{1D}}{P_2}}X_2+\rho_{1s}\sqrt{\frac{P_{1D}}{Q}}S\\
X'_{1D} &=X_{1D}-\hat{X}_{1D}.
\end{align}

We can then write $Y_3$ in \eqref{DegenerateGaussianChannelModel1__RCwithHyperSource} alternatively as
\begin{align}
Y_3 &=X'_{1D}+(1+\rho_{12}\sqrt{\frac{P_{1D}}{P_2}})X_2+(1+\rho_{1s}\sqrt{\frac{P_{1D}}{Q}})S+Z_3.
\end{align}
Let now
\begin{equation}
Y'_3:=Y_3-\mathbb{E}[Y_3|X_{1R},X_2]=X'_{1D}+(1+\rho_{1s}\sqrt{\frac{P_{1D}}{Q}})S+Z_3.
\label{EquivalentForm__FirstDegenerateGaussianChannelModel__RCwithHyperSource}
\end{equation}
Noticing now that $X'_{1D}$ is independent of the state $S$ in \eqref{EquivalentForm__FirstDegenerateGaussianChannelModel__RCwithHyperSource}, it is clear that an optimal choice of the associated auxiliary random variable $U_1$ is 
\begin{align}
U_1 &=X'_{1D}+{\alpha}(1+\rho_{1s}\sqrt{\frac{P_{1D}}{Q}})S,
\label{DPCforEquivalentFictitiousFirstDegenerateGaussianChannelModel__RCwithHyperSource}
\end{align}
where $\alpha$ is Costa's parameter given by
\begin{align}
\alpha &= \frac{\mathbb{E}[X'^2_{1D}]}{\mathbb{E}[X'^2_{1D}]+\mathbb{E}[Z^2_3]} = \frac{P_{1D}(1-\rho^2_{12}-\rho^2_{1s})}{P_{1D}(1-\rho^2_{12}-\rho^2_{1s})+N_3}.
\label{CostasParameterIntermediaryStep}
\end{align}
Then we can easily show that
\begin{equation}
I(U_1;Y_3|X_{1R},X_2)-I(U_1;S|X_{1R},X_2) = I(U_1;Y'_3)-I(U_1;S)
\end{equation}
By substituting $X'_{1D}$ in \eqref{DPCforEquivalentFictitiousFirstDegenerateGaussianChannelModel__RCwithHyperSource}, we get
\begin{align}
U_1 &= X_{1D}-\rho_{12}\sqrt{\frac{P_{1D}}{P_2}}X_2 +\alpha_{\text{opt}}S
\label{AuxiliaryRandomVariable__FirstDegenerateGaussianChannelModel__RCwithHyperSource}
\end{align}
with
\begin{align}
\alpha_{\text{opt}} &= \alpha(1+\rho_{1s}\sqrt{\frac{P_{1D}}{Q}})-\rho_{1s}\sqrt{\frac{P_{1D}}{Q}}\nonumber\\
&= \Big[\frac{P_{1D}(1-\rho^2_{12}-\rho^2_{1s})}{P_{1D}(1-\rho^2_{12}-\rho^2_{1s})+N_3}\Big(1+\rho_{1s}\sqrt{\frac{P_{1D}}{Q}}\Big)-\rho_{1s}\sqrt{\frac{P_{1D}}{Q}}\Big].
\label{OptimalAlpha__FirstDegenerateGaussianChannelModel__RCwithHyperSource}
\end{align}
Now, it is easy to see that, with the choice \eqref{AuxiliaryRandomVariable__FirstDegenerateGaussianChannelModel__RCwithHyperSource}, we have
\begin{align}
I(U_1;Y_3|X_{1R},X_2)-I(U_1;S|X_{1R},X_2) &= I(U_1;Y'_3)-I(U_1;S)\nonumber\\
&= \frac{1}{2}\log\left(1+\frac{\mathbb{E}[X'^2_{1D}]}{N_3}\right)\nonumber\\
&= \frac{1}{2}\log\left(1+\frac{P_{1D}(1-\rho^2_{12}-\rho^2_{1s})}{N_3}\right).
\label{ConditionalTerm__LowerBound__FirstDegenerateGaussianChannelModel__RCwithHyperSource}
\end{align}
 
 We now compute the terms $I(X_{1R};Y_2|X_2)$ and $I(X_2;Y_3)$. It is easy to see that, with the aforementioned jointly Gaussian input distribution,
 \begin{align}
I(X_{1R};Y_2|X_2) &= I(X_{1R};Y_2)\nonumber\\
                  &= \frac{1}{2}\log(1+\frac{P_{1R}}{N_2}).
\label{FirstTerm__Minimum__LowerBound__FirstDegenerateGaussianChannelModel__RCwithHyperSource}
\end{align}
Also, we have 
\begin{align}
I(X_{1R},X_2;Y_3) &\stackrel{(a)}{=} I(X_2;Y_3)\nonumber\\
		&= h(Y_3)-h(Y_3|X_2)\nonumber\\
           &= h(Y_3)-h(X'_{1D}+\mathbb{E}[X_{1D}|X_2]+\mathbb{E}[X_{1D}|S]+S+Z_3|X_2)\nonumber\\
           &\stackrel{(b)}{=} h(Y_3)-h(X'_{1D}+\mathbb{E}[X_{1D}|S]+S+Z_3)\nonumber\\
	   &\stackrel{(c)}{=} \frac{1}{2}\log\Big(\frac{\mathbb{E}[(X_{1D}+X_2+S)^2]+\mathbb{E}[Z^2_3]}{\mathbb{E}[X'^2_{1D}]+\mathbb{E}[(S+\mathbb{E}[X_{1D}|S])^2]+\mathbb{E}[Z^2_3]}\Big)\nonumber\\
	   &= \frac{1}{2}\log\Big(1+\frac{(\sqrt{P_2}+\rho_{12}\sqrt{P_{1D}})^2}{P_{1D}(1-\rho^2_{12}-\rho^2_{1s})+(\sqrt{Q}+\rho_{1s}\sqrt{P_{1D}})^2+N_3}\Big).
\label{SecondTerm__Minimum__LowerBound__FirstDegenerateGaussianChannelModel__RCwithHyperSource}
\end{align}
where: $(a)$ holds since $X_{1R}$ is independent of $(X_2,Y_3)$, $(b)$ holds since $X'_{1D}$ and $S$ are independent of $X_2$, and $(c)$ follows through straightforward algebra.

Adding \eqref{ConditionalTerm__LowerBound__FirstDegenerateGaussianChannelModel__RCwithHyperSource} and \eqref{FirstTerm__Minimum__LowerBound__FirstDegenerateGaussianChannelModel__RCwithHyperSource} we obtain the first term of the minimization in \eqref{Capacity__FirstDegenerateGaussianModelWithHyperSource}; and adding \eqref{ConditionalTerm__LowerBound__FirstDegenerateGaussianChannelModel__RCwithHyperSource} and \eqref{SecondTerm__Minimum__LowerBound__FirstDegenerateGaussianChannelModel__RCwithHyperSource} we obtain the second term of the minimization in \eqref{Capacity__FirstDegenerateGaussianModelWithHyperSource}. 

Finally, we obtain the capacity in Theorem~\ref{Theorem__Capacity__FirstDegenerateGaussianModelWithHyperSource} by maximizing the RHS of \eqref{Capacity__FirstDegenerateGaussianModelWithHyperSource} over all possible values of $\rho_{12} \in [-1,1]$ and $\rho_{1s} \in [-1,1]$. Investigating the two terms of the minimization, we can easily see that it suffices to consider $\rho_{12} \in [0,1]$ and $\rho_{1s} \in [-1,0]$. This concludes the proof of Theorem~\ref{Theorem__Capacity__FirstDegenerateGaussianModelWithHyperSource}.

\bibliographystyle{IEEEtran}
\bibliography{Draft__InitialSumission__RCwithInformedSource__OneColumn}
\end{document}